\documentclass[12pt]{article}
\makeatletter \@addtoreset{equation}{section} \makeatother
\renewcommand{\theequation}{\thesection.\arabic{equation}}
\newcommand{\ba}{\begin{array}}
\newcommand{\ea}{\end{array}}
\newcommand{\beq}{\begin{equation}}
\newcommand{\eeq}{\end{equation}}
\newcommand{\bea}{\begin{eqnarray}}
\newcommand{\eea}{\end{eqnarray}}

\def\bce{\begin{center}}
\def\ece{\end{center}}

\def\nonu{\nonumber}

\def\pa{\partial}

\def\be{\beta}

\def\eps6{{\displaystyle \mathop{\epsilon}^{6}}{}}
\def\g6{{\displaystyle \mathop{g}^{6}}{}}

\def\nab6{{\displaystyle \mathop{\nabla}^{6}}{}}

\def\0{{\sst{(0)}}}
\def\1{{\sst{(1)}}}
\def\2{{\sst{(2)}}}
\def\3{{\sst{(3)}}}
\def\4{{\sst{(4)}}}
\def\5{{\sst{(5)}}}
\def\6{{\sst{(6)}}}
\def\7{{\sst{(7)}}}
\def\8{{\sst{(8)}}}

\def\ba{\begin{array}}
\def\ea{\end{array}}
\def\beq{\begin{equation}}
\def\eeq{\end{equation}}
\def\be{\begin{equation}}
\def\ee{\end{equation}}

\def\eps{\epsilon}

\def\ba{\begin{array}}
\def\ea{\end{array}}
\def\beq{\begin{equation}}
\def\eeq{\end{equation}}
\def\be{\begin{equation}}
\def\ee{\end{equation}}

\def\eps{\epsilon}

\def\eps6{{\displaystyle \mathop{\epsilon}^{6}}{}}

\def\nab6{{\displaystyle \mathop{\nabla}^{6}}{}}

\newcommand{\bean}{\begin{eqnarray*}}
\newcommand{\eean}{\end{eqnarray*}}

\begin{document}
\thispagestyle{empty} \addtocounter{page}{-1}
   \begin{flushright}
\end{flushright}

\vspace*{1.3cm}
  
\centerline{ \Large \bf   Spin-5 Casimir Operator
and }
\centerline{ \Large \bf Its Three-Point Functions with Two Scalars
 }
\vspace*{1.5cm}
\centerline{{\bf Changhyun Ahn }  and {\bf Hyunsu Kim}
}
\vspace*{1.0cm} 
\centerline{\it 
Department of Physics, Kyungpook National University, Taegu
702-701, Korea} 
\vspace*{0.8cm} 
\centerline{\tt ahn@knu.ac.kr, \qquad kimhyun@knu.ac.kr
} 
\vskip2cm

\centerline{\bf Abstract}
\vspace*{0.5cm}

By calculating the second-order pole in the operator product expansion (OPE) 
between 
the spin-$3$ Casimir operator and the spin-$4$ Casimir operator known 
previously, the spin-$5$ Casimir operator is obtained in the
coset model  
based on $(A_{N-1}^{(1)} \oplus A_{N-1}^{(1)}, A_{N-1}^{(1)})$ at level $(k,1)$.  
This spin-$5$ Casimir
operator  
consisted of the quintic, quartic (with one derivative) 
and cubic (with two derivatives) WZW currents contracted with 
$SU(N)$ invariant tensors. 
The three-point functions 
with two scalars  for all values of 't Hooft
coupling in the large $N$ limit were obtained 
by analyzing the zero-mode eigenvalue equations carefully.
These three-point functions were dual to those 
in $AdS_3$ higher spin gravity theory with matter. 
Furthermore, the exact three-point functions that hold for any 
finite $N$  and $k$ are obtained. 
The zero mode eigenvalue equations for the spin-$5$ current in CFT 
coincided with those of the spin-$5$ field in asymptotic symmetry 
algebra of the higher spin theory on the $AdS_3$. 
This paper also
describes the structure constant appearing in the spin-$4$ Casimir operator 
from the OPE between the spin-$3$ Casimir operator and itself for $N=4, 5$
in the more general coset minimal model with two arbitrary levels 
$(k_1, k_2)$.   

\baselineskip=18pt
\newpage
\renewcommand{\theequation}
{\arabic{section}\mbox{.}\arabic{equation}}

\section{Introduction}

Simple examples of the AdS/CFT correspondence allow
a detailed study of holography that would not be 
possible in a string theory setup. 
Recent progress along these lines is based on 
an examination of higher spin 
theories of gravity in AdS space, which include  a large number of fields 
with spins $s=2, 3, \cdots, N$. In three bulk dimensions,
the higher spin theory is quite simple. The massless sector is 
given semiclassically by the Chern-Simons action. Therefore, 
the graviton and its higher spin fields have no propagating modes.
Two-dimensional conformal field theories are well-established quantum field 
theories and have a range of applications in different areas of 
physics.  A high degree of analytical control over these CFTs
can provide a rich source of CFTs with nontrivial bulk AdS duals. 
The simple theory is described using large $N$ vector models.  
This will give rise to a controlled environment 
in which to study the puzzle of the emergence of a gravity dual.      

The $W_N$ 
minimal model conformal field theory (CFT) 
is dual in the 't Hooft $\frac{1}{N}$ expansion
to the higher spin theory of Vasiliev on the $AdS_3$ coupled 
with one complex scalar \cite{GG,GG1}.
The CFTs have the following diagonal coset, $\frac{G}{H}$, which is expressed as
\bea
\frac{G}{H} =\frac{\widehat{SU}(N)_k \oplus \widehat{SU}(N)_1}
{\widehat{SU}(N)_{k+1}}.
\label{cosetexp}
\eea
The higher spin-$s$ Casimir 
currents of degree $s$ with $s=3, 
\cdots, N$  can be obtained from polynomial  combinations 
of the individual numerator $SU(N)$ currents of spin-$1$ 
\cite{BS}. 
The diagonal denominator 
$SU(N)$ currents of spin-$1$ 
commute with these higher spin-$s$ Casimir currents.
One of the levels for the spin-$1$ current is fixed by the positive 
integer $k$
and the other is fixed by $1$  in the numerator 
of the coset CFT (\ref{cosetexp}).
One of the specialties of the above coset model (\ref{cosetexp})
is that the additional currents generated in the OPEs for the general
level $l$ in the second numerator $SU(N)$ current 
become null fields for the level $l=1$ and hence decouple \cite{BBSS2}. 
See recent review papers \cite{GG2,AGKP} for the above duality.

The spin-$3$ (Casimir) current was reported in \cite{BBSS2}.
The four independent cubic terms were made from the two spin-$1$ currents 
in the numerator of the coset (\ref{cosetexp}).
The spin-$4$ (Casimir) current was generated \cite{Ahn1111} by calculating the OPE between this spin-$3$ Casimir current and itself, 
which consists of 
quartic terms and quadratic terms with two derivatives in the above spin-$1$ 
currents. 
The next step is to determine how one can construct the next higher spin-$5$ Casimir 
current.
Because the complete form of the spin-$3$ and spin-$4$ Casimir currents is 
known, the OPE between the spin-$3$ current and  
spin-$4$ current can be calculated using the current algebra in the numerator 
$SU(N)$ currents of spin-$1$. 
The spin-$5$  Casimir current can be extracted explicitly 
from the second-order pole of 
this OPE \footnote{This will continue to the generation of 
general higher spin-$s$ ($s \geq 3$) currents. In other words, 
once the structure of the higher spin-$s$ current is known, 
the OPE between the spin-$3$ current and this spin-$s$ current
will provide the structure of the higher spin-$(s+1)$ current 
at the second order pole.}. 

The explicit form for the spin-$5$ current is needed   
because it was necessary in reference \cite{GG1}
to obtain the commutator between the 
spin-$3$ Laurent mode and the spin-$4$ Laurent mode and determine 
the minimal representation \cite{GG1,GG2} 
of the asymptotic quantum symmetry algebra 
${\cal W}_{\infty} [\mu]$ of the higher spin theory on the 
$AdS_3$ space, where $\mu$ is a parameter of the algebra \footnote{   
The eigenvalues of the zero mode of 
the spin-$3$ and spin-$4$ fields were expressed as
the conformal spin $h$ of the ground state and the central charge $c$,
which is also a parameter of the algebra.
Furthermore, the eigenvalue of the zero mode of the
spin-$5$ field was described as 
the eigenvalues of the zero mode of the spin-$3$ and spin-$4$ fields,  
the conformal spin $h$ of the ground state, the central charge $c$ and 
$\mu$-dependent 
structure constant appearing in front of the spin-$3$ field Laurent mode 
in the right hand side of the above commutator.
The cubic equation for the conformal spin $h$ contains 
two solutions in terms of the two parameters $(\mu, c)$.
Under these conditions, $\mu =N$ and $c=c_{N,k}$, where $c_{N,k}$ 
is equal to (\ref{ctilde}),
the above two conformal spins are exactly those of the physical 
representations, coset primaries denoted by $(f;0)$ and $(0;f)$ 
\cite{GG,GG2}.}.   
The zero mode eigenvalue equations 
of the spin-$5$ Casimir current in the 2-dimensional coset model 
(\ref{cosetexp}) behave like those of the spin-$5$ field 
in  asymptotic quantum symmetry algebra 
${\cal W}_{\infty} [\mu]$ of the higher spin theory on the 
$AdS_3$ space. 

In this paper, 
the spin-5 Casimir current was obtained and  
consists of quintic, quartic (with one derivative) 
and cubic (with two derivatives) WZW currents in the coset (\ref{cosetexp}) 
contracted with 
$SU(N)$ invariant tensors. The relative $52$ coefficients depend on 
$N$ and $k$ explicitly.
To arrive at this spin-$5$ current,
one should realize that 
the second order pole of the OPE between the spin-$3$ and
spin-$4$ currents contains the other part as well as the 
new spin-$5$ current \footnote{  How does one subtract this extra piece 
from the second
order pole to obtain the correct spin-$5$ primary current? 
Any OPE between the two quasi-primary fields with given spins
produce other quasi-primary fields with fixed spins 
in the right hand side of this OPE \cite{Bowcock,BFKNRV,Nahm1,Nahm2,Ahn1211}.
Of course, the structure constant is generally unknown.
On the other hand, the coefficients arising in the descendant fields of this OPE
are determined completely by the spins of these three quasi-primary fields (two
of them are the operators in the left hand side and one is the operator 
in the right hand side)
and 
the number of derivatives in the quasi-primary field in the right hand side.}.

The structure constant mentioned above can be determined explicitly because 
this study was interested in the specific WZW model and the basic current 
algebra is known explicitly.
Furthermore, in general, the second order pole after calculating the various
OPEs between the spin-$3$ current with four terms and 
spin-$4$ current with $19$ terms contains 
many unwanted normal ordered products between the fields.
They should be rearranged in terms of the fully normal ordered products 
\cite{BBSS1}.
This step is necessary because the zero mode for any 
given multiple product of the fields should be taken very carefully, and  
the zero mode can be obtained from this fully normal ordered product systematically.  
The rearrangement lemmas in \cite{BBSS1} are useful for obtaining the desired 
normal ordered product.

The zero modes of the spin-$1$ currents that play the role of
$SU(N)$ generators act on the coset primaries.
The singlet condition suggests that the zero mode of
the spin-$1$ current with level $k$ in the numerator of (\ref{cosetexp}) 
acting on the state $|(0;f) >$, where the element 
$0$ means a trivial representation 
of the coset,  vanishes while 
the zero mode of
the diagonal spin-$1$ current in the denominator of (\ref{cosetexp}) 
acting on the state $|(f;0) >$ vanishes \cite{GH}.
These are the eigenvalue equations, and 
the rightmost zero mode should be taken first 
when this singlet condition is applied
to the multiple product (quintic, quartic and cubic terms) 
of zero modes.  
After that, the $SU(N)$ generators appear.
Then the next rightmost zero mode can be repeated with the singlet condition
(each zero mode disappears and the generators with $SU(N)$ 
group indices do appear) \footnote{
Note that the state  $|(0;f) >$ transforms as a fundamental representation 
with respect to the zero mode of the second numerator current with level $1$
and  the state  $|(f;0) >$ transforms as a fundamental representation 
with respect to the zero mode of the first numerator current with level $k$.}.    
At the final stage after calculating this procedure, we are left with 
the eigenvalue, which has the form of a trace of the 
$SU(N)$ generators contracted group indices with a range of $SU(N)$ invariant
tensors, times the state.
The eigenvalue depends on $N$.

The relative $52$ coefficients appearing in the 
spin-$5$ current are very complicated expressions in terms of 
$N$ and $k$. This is because  they arise from  
each coefficient function in the spin-$3$ current and in the spin-$4$ 
current. The latter consists of  rather complicated fractional functions, whereas
the former has a simple factorized form. 
Furthermore, the final expression for the coefficient functions is rather involved
due to ordering in the composite fields.
Although each $52$ coefficient function is complicated, 
the eigenvalue equations for the spin-$5$ current have
very simple factorized
forms \footnote{
To reveal the duality between the higher spin theory on the $AdS_3$
and $W_N$ minimal model coset 
CFT,  the eigenvalues for the higher
spin currents in the coset model can be compared with those of the higher spin fields
in the asymptotic quantum symmetry algebra 
${\cal W}_{\infty} [\mu]$ of the higher spin theory on the 
$AdS_3$.
As described before, the CFT computations for the eigenvalue equations 
of the higher spin-$5$ current coincide with those in ${\cal W}_{\infty} [\mu]$.
The corresponding eigenvalue equations for the 
spin-$3$ and spin-$4$ currents (including the spin-$2$ current) 
can be calculated at finite $N$ and $k$ (the large 
$N$ limit result can be obtained by taking the limit appropriately).}.
   
In section $2$, 
after reviewing the GKO coset construction for the spin-$2$ current,
the generalization to the spin-$3$ and spin-$4$ currents is described. 

In section $3$, from the explicit expressions for the spin-$3$ and spin-$4$
Casimir currents, the second order pole of the OPE between them
is obtained. The new spin-$5$ Casimir current is found by subtracting both the quasi primary field of spin-$5$
and the descendant field (of spin-$5$) of spin-$3$ Casimir current with correct
coefficient functions.  

Section $4$ analyzes
the zero mode eigenvalue equations of spin-$5$ Casimir current   
in the large $N$ limit. The three-point functions are obtained. 
The exact eigenvalue equations and the exact 
three-point functions are also described.

Section $5$ presents a summary of this work as well as a
discussion on the future directions.

Appendices $A$-$F$ provide  
detailed descriptions of sections $3$ and $4$.
 
\section{Generalization of the GKO coset construction: review}

The diagonal coset WZW model is given by the following coset $\frac{G}{H}$, which is expressed as
\bea
\frac{G}{H} =
\frac{\widehat{SU}(N)_k \oplus \widehat{SU}(N)_1}{\widehat{SU}(N)_{k+1}}.
\label{coset}
\eea
The numerator spin-$1$ currents are denoted by $K^a(z)$ of level 
$k$ and $J^a(z)$ of level $1$, whereas the denominator spin-$1$ current is  
denoted by $J'^a(z)$ of level $(k+1)$.
The standard OPE between
$J^a(z)$ and $J^b(w)$ is
\bea
J^a (z) \, J^b (w) = -\frac{1}{(z-w)^2} \, k_1 \delta^{ab} +\frac{1}{(z-w)}
\, f^{abc} J^c(w) + \cdots,
\label{JJcoset}
\eea
where the level is characterized by the positive integer $k_1$, which is fixed 
as $1$ in the coset model (\ref{coset}).
The other spin-$1$ current with level $k$ living in the 
other $SU(N)$ factor of the above coset model (\ref{coset}) 
satisfies the following OPE between $K^a(z)$ and $K^b(w)$,
\bea
K^a (z) \, K^b (w) = -\frac{1}{(z-w)^2} \, k_2 \delta^{ab} +\frac{1}{(z-w)}
\, f^{abc} K^c(w) + \cdots,
\label{KKcoset}
\eea
where the level is given by $k_2$, which is rewritten as $k$ in the coset 
(\ref{coset}). 
These two spin-$1$ currents are independent of each other
because the OPE 
satisfies $J^a(z) K^b(w)=0$ and there are no singular terms between them.
The diagonal spin-$1$ current living in the denominator of
the coset (\ref{coset}) is the sum of the above two numerator currents 
\bea
J'^a (z) = J^a (z) +K^a (z).
\label{diag}
\eea
By adding the OPEs (\ref{JJcoset}) 
and (\ref{KKcoset}), the following OPE between the currents (\ref{diag}) 
satisfy
\bea
J'^a (z) \, J'^b (w) = -\frac{1}{(z-w)^2} \, k' \delta^{ab} +\frac{1}{(z-w)}
\, f^{abc} J'^c(w) + \cdots, \qquad k' \equiv k_1 +k_2 \equiv k+1.
\label{opej'j'}
\eea
By construction, the level of the denominator current in (\ref{opej'j'}) 
is the sum of each level
of the numerator current.

The stress energy tensor with arbitrary levels $(k_1, k_2)$
via a Sugawara construction 
is expressed as 
\bea
T(z) =  -\frac{1}{2(k_1+N)} \, J^a J^a (z)
 -\frac{1}{2(k_2+N)} \, K^a K^a (z)
  +\frac{1}{2(k_1 + k_2 +N)} \, J'^a J'^a (z).
\label{stressT}
\eea
The corresponding stress energy tensor 
in the coset model (\ref{coset}) can be read off from (\ref{stressT}).
The OPE of the stress energy tensor (\ref{stressT}) 
with itself from the OPEs (\ref{JJcoset}), 
(\ref{KKcoset}) and (\ref{opej'j'})
is given by
\bea 
T(z) \, T(w)
& = & \frac{1}{(z-w)^4} \, \frac{c}{2} +\frac{1}{(z-w)^2} \, 2 T(w) +
\frac{1}{(z-w)} \, \pa T(w) +\cdots,
\label{cosettt}
\eea 
where $c$ in the fourth-order pole of (\ref{cosettt}) 
denotes the central charge in the coset (\ref{coset}) that depends on
$N$ and $k$ and is given by
\bea
c & = & 
(N-1) \left[ 1-\frac{N(N+1)}{(N+k)(N+k+1)} \right].
\label{ctilde}
\eea

The higher spin-$3$ current consists of 
cubic WZW currents contracted with the $SU(N)$ completely symmetric traceless 
invariant $d^{abc}$ tensor.
The relative coefficient functions were fixed by two conditions:
1) 
this spin-$3$ current transforms as a primary field under the
above stress energy tensor (\ref{stressT}) with levels $(1,k)$; and 
2)  
the OPE between this spin-$3$ current and the diagonal current (\ref{diag})
does not have any singular terms.  
Finally, the overall coefficient can be fixed by calculating the highest-order 
singular term in the spin-$3$ spin-$3$ OPE, 
which is equal to $\frac{c}{3}$ with (\ref{ctilde}).  

A higher spin-$4$ current can be constructed similarly, and consists of 
quartic, cubic (with one derivative) and quadratic (with two derivative) 
WZW currents.
This spin-$4$ current can be obtained in two ways. 
One is to consider the most general spin-$4$ current and determine 
the various unknown coefficient functions using the above two requirements.
The other is to calculate the OPE between the above spin-$3$ current 
and itself and read off the second-order pole structure. 
The normalization of the spin-$4$ current is related to the normalization of the 
spin-$3$ current and the structure constant appearing in the spin-$4$
current from the OPE between the spin-$3$ current and itself. 
In this case, the structure of the spin-$4$ current is determined completely
except for the normalization constant because the full expression of 
the spin-$3$ current is known.

\section{ Spin-$5$ Casimir operator}

Subsection 3.1 reviews the relevant facts about the 
spin-$3$ and spin-$4$ currents first. 
This subsection will focus on the second-order pole in the OPE between the spin-$3$ current 
and the spin-$4$ current.
Subsection 3.2 then constructs 
the next higher spin-$5$ current from the known spin-$3$ and 
spin-$4$ currents
explicitly.

\subsection{Review on the spin-$3$ and spin-$4$ currents}

For later convenience, let us introduce spin-$3$ operators 
\bea
Q(z) \equiv d^{abc} J^a J^b J^c(z), \qquad R(z) \equiv d^{abc} K^a K^b K^c(z),
\label{spin3Q}
\eea
and spin-$2$ operators
\bea
Q^a(z) \equiv d^{abc} J^b J^c(z), \qquad R^a(z) \equiv d^{abc} K^b K^c(z).
\label{spin2Q}
\eea
As described before, 
the $d$ symbol appearing in (\ref{spin3Q}) and (\ref{spin2Q})
is a completely symmetrical traceless $SU(N)$ invariant tensor of rank $3$. 
The operators 
$Q(z)$ and $Q^a(z)$ are the primary fields of spin-$2$ and spin-$3$, respectively, 
under the stress energy tensor
defined in the second numerator $SU(N)_1$ factor in coset (\ref{coset}),
corresponding to the first term of (\ref{stressT}).
Similarly, 
the operators, 
$R(z)$ and $R^a(z)$, are the primary fields of spin-$2$ and spin-$3$, respectively, 
under the stress energy tensor
defined in the first numerator $SU(N)_k$ factor, 
corresponding to the second term of (\ref{stressT}).
This can also be generalized to the spin-$4$ operators but 
the $d$ symbol of rank $4$ is reduced to the product of the 
$d$ symbol of rank $3$ 
and the $\delta$ symbol of rank $2$ with the appropriate contractions.
See equation (\ref{dabcd}).

Calculating the following OPEs is straightforward \cite{BBSS1,BBSS2}
\bea
J^a(z) \, Q^b(w)  & = &  -\frac{1}{(z-w)^2} \, (2k_1+N) d^{abc} J^c(w)
+\frac{1}{(z-w)} \, f^{abc} Q^c(w) +\cdots,
\label{JaQb}
\\
J^a (z) \, Q(w) & = & -\frac{1}{(z-w)^2} \, 3(k_1+N) Q^a(w) +\cdots,
\label{JaQ}
\\
K^a(z) \, R^b(w)  & = & 
 -\frac{1}{(z-w)^2} \, (2k_2+N) d^{abc} K^c(w)
+\frac{1}{(z-w)} \, f^{abc} R^c(w) +\cdots,
\label{KaRb}
\\
K^a (z) \, R(w) & = & -\frac{1}{(z-w)^2} \, 3(k_2+N) R^a(w) 
+\cdots.
\label{KaR}
\eea
Occasionally, some of the OPEs can be obtained 
from the known OPEs by taking the symmetries between the currents and 
levels.
For example, 
the OPE (\ref{KaRb}) can be obtained from the OPE (\ref{JaQb}) 
using $J^a \rightarrow K^a$ (and $Q^a \rightarrow R^a$) 
and $k_1 \rightarrow k_2$ without calculating (\ref{KaRb}) independently 
and vice versa.
Similarly, the OPE (\ref{KaR}) is related to the OPE 
(\ref{JaQ}) with $J^a \leftrightarrow K^a$ and $k_1 \leftrightarrow k_2$. 
Therefore, some OPEs and some operators are expressed
with arbitrary levels $(k_1, k_2)$ rather than fixing them. 

The coset spin-$3$ primary field can be expressed as
\bea
W^{(3)}(z) &=&  d^{abc} \left( A_1 J^a J^b J^c(z) + A_2 J^a J^b K^c(z) + A_3
  J^a K^b K^c(z) + A_4 K^a K^b K^c(z) \right) \nonu\\
  & \equiv &   A_1 \, Q(z) + A_2 \, K^a Q^a(z) + A_3 \, J^a R^a(z)
  + A_4 \, R(z) ,
\label{primaryspin3}
\eea
where the coefficient functions that depend on $k_1, k_2$ or $N$ 
were determined elsewhere \cite{BBSS2}
\bea
A_1  & \equiv&  k_2 (N+k_2)(N+2k_2), \qquad
A_2   \equiv  -3 (N+k_1)(N+k_2)(N+2k_2), \nonu \\
A_3  & \equiv&  3(N+k_1)(N+k_2)(N+2k_1), \qquad
A_4   \equiv  -k_1 (N+k_1)(N+2k_1).
\label{BA}
\eea
Actually, there is an overall normalization factor 
$B(N,k_1,k_2)$ 
in (\ref{primaryspin3}), but  $B(N,k_1,k_2)$ is not considered.
Compared to the stress energy tensor (\ref{stressT}), 
the spin-$3$ current is not the difference between the spin-$3$ current in the 
numerator and the spin-$3$ current in the denominator in the coset model
(\ref{coset}). As described previously, 
by assuming the possible terms of spin-$3$, 
all the relative coefficient functions are fixed by the above two conditions.
The normalization factor $B(N, k_1, k_2)$ is determined by 
calculating the OPE of the spin-$3$ current and itself.
See the equation (\ref{Vtilde}) for the explicit form.

By substituting (\ref{dabcdFormula}) and   (\ref{dabcdFormula2})
into (\ref{primaryspin4}),
the following form of the spin-$4$ current was obtained 
\bea
&& W^{(4)}(z)  =  
\nonu \\
&& d^{abe} d^{cde} \left( h_6 J^a J^b J^c J^d+  
h_7   J^a J^b J^c K^d  +  h_8 J^a J^b K^c K^d
+ h_9  J^a K^b K^c K^d 
\right.
\nonu \\
&& + \left. h_{10} K^a K^b K^c K^d \right)(z) 
+ h_{11} J^a J^a J^b J^b(z) +
h_{12} J^a J^a J^b K^b(z)
 + 
h_{13} J^a J^a K^b K^b(z)  
\nonu \\
&&+ 
h_{14}  J^a K^a K^b K^b(z) 
+h_{15} K^a K^a K^b K^b(z) 
+
h_{17} \pa^2 J^a K^a(z) 
+  h_{20} \pa J^a \pa K^a(z)   
 +  
h_{22} J^a \pa^2 K^a(z) \nonu \\
&&+
h_{23} J^a J^b K^a K^b(z) 
+
h_{24} d^{ace} d^{bde} J^a J^b K^c K^d(z) 
+
h_{25}  f^{abc} \pa J^a J^b K^c(z)+
h_{26}  f^{abc} \pa K^a K^b J^c(z)  \nonu \\
&&+
h_{27} \pa^2 J^a J^a(z) 
+
h_{28} \pa^2 K^a K^a(z)+
h_{29} \pa J^a \pa J^a(z)  +
h_{30} \pa K^a \pa K^a(z),   
\label{modifiedspin4}
\eea
where the $21$ coefficient functions are
\bea
&&h_6 \equiv  c_6+3 c_1 ,\qquad
h_7 \equiv c_7+3 c_2  ,\qquad
h_8 \equiv c_8+c_3  ,\qquad
h_9 \equiv  c_9+3 c_4 ,\qquad
h_{10} \equiv c_{10}+3 c_5  ,\nonu \\
&&h_{11} \equiv c_{11}-
\frac{12 \left(N^2-4\right)}{N\left(N^2+1\right)} c_1,\qquad
h_{12} \equiv c_{12}-\frac{12 \left(N^2-4\right)}{N\left(N^2+1\right)} c_2,\qquad
h_{13} \equiv c_{13}-\frac{4 \left(N^2-4\right)}{N\left(N^2+1\right)} c_3,\nonu \\
&&h_{14} \equiv 
c_{14}-\frac{12 \left(N^2-4\right)}{N\left(N^2+1\right)} c_4,\qquad
h_{15} \equiv 
c_{15}-\frac{12 \left(N^2-4\right)}{N\left(N^2+1\right)} c_5,\nonu \\
&&h_{17} \equiv c_{17}+\frac{2(N^2-4)(N^2-3)}{N^2+1}c_2,\qquad
h_{20} \equiv c_{20}-\frac{(N^2-4)(N^2-3)}{N^2+1}c_3,\nonu \\
&&h_{22} \equiv c_{22}-\frac{(N^2-4)(N^2-3)}{N^2+1}c_4,\qquad
h_{23}\equiv c_{23}-\frac{8(N^2-4)}{N(N^2+1)} c_3,\qquad
h_{24}\equiv 2 c_3,\nonu \\
&&h_{25}\equiv -\frac{3(N^2-4)(N^2-3)}{N(N^2+1)} c_2,\qquad
h_{26}\equiv \frac{3(N^2-4)(N^2-3)}{N(N^2+1)} c_4,\nonu \\
&&h_{27}\equiv \frac{2(N^2-4)(N^2-3)}{N^2+1} c_1,\qquad
h_{28}\equiv \frac{2(N^2-4)(N^2-3)}{N^2+1} c_5,\nonu \\
&&h_{29}\equiv -\frac{3(N^2-4)(N^2-3)}{N^2+1} c_1,\qquad
h_{30}\equiv -\frac{3(N^2-4)(N^2-3)}{N^2+1} c_5.
\label{hcoeff}
\eea
The last seven-terms of (\ref{modifiedspin4}) newly appear
and the first five-terms of (\ref{primaryspin4}) disappear.
From (\ref{hcoeff}), the precise relationships between 
the spin-$4$ current (\ref{primaryspin4}) and its 
different versions (\ref{modifiedspin4}) and the subscripts of the 
$h$ coefficients are kept unchanged. 

\subsection{The spin-$5$ current}

The next step is to construct the spin-$5$ current 
from the spin-$3$ current (\ref{primaryspin3}) and  
spin-$4$ current (\ref{modifiedspin4}). 
Although in principle the complete OPE can be calculated, 
this subsection focuses on the second order pole only to extract 
the spin-$5$ current.
The OPE between the spin-$3$ and spin-$4$ currents  is given by \cite{GG1} 
\bea
W^{(3)}(z) \, 
W^{(4)}(w)&=& \frac{1}{(z-w)^4} \, \eta_{W} W^{(3)}(w) + 
\frac{1}{(z-w)^3} \, \frac{1}{3}\eta_{W} \pa W^{(3)}(w)
\nonu \\
&+& \frac{1}{(z-w)^2} \left[ \eta_{\Lambda} \left( T W^{(3)}-\frac{3}{14}\pa^2 W^{(3)} \right) +\frac{\eta_{W}}{14}\pa^2 W^{(3)} +W^{(5)} \right](w)
\nonu \\
& + & ({\cal O}(z-w)^{-1}),
\nonu 
\eea
where the $(N, c)$-dependent coefficients are 
\bea 
\eta_{\Lambda}=\frac{39}{(114+7c)} 
\left(C_{33}^{4}\right)^{2},\qquad 
\eta_{W}=\frac{3}{4} \left(C_{33}^{4}\right)^{2}. 
\label{pole2coefficient}
\eea 
 
Because the primary currents are not normalized properly, 
there is little difference from \cite{GG1} in an overall factor
\bea 
W^{(3)}(z) \,
W^{(4)}(w) |_{\frac{1}{(z-w)^2}}  = \eta_{\Lambda} \left( T W^{(3)}(w)-\frac{3}{14}\pa^2 W^{(3)}(w) \right) +\frac{\eta_{W}}{14}\pa^2 W^{(3)}(w) +W^{(5)}(w).
\label{cosetspin3spin4}
\eea 
The combination 
$ (T W^{(3)} -\frac{3}{14}\pa^2 W^{(3)})(w) \equiv \Lambda(w)$ 
in (\ref{cosetspin3spin4}) 
is a quasi-primary field \footnote{
The coefficient $\frac{3}{14}$ comes from  
general formula $\frac{3}{2(2s+1)}$, where $s$ is the spin 
for the field $\Phi$ \cite{BFKNRV,Ahn1211}.
In other words, the OPE $T(z) \, \Lambda(w)$
has no third order pole and the nonvanishing fourth order pole 
of this OPE is given by $\frac{1}{14}(114+7c)W^{(3)}$.
The numerical factor $\frac{1}{14}$ can be obtained from 
the formula
$\frac{1}{n!} \prod_{x=0}^{n-1} \frac{(h_i-h_j+h_k+x)}{(2h_k +x)}$ with $h_i=3, 
h_j=4$, $h_k=3$ and $n=2$ \cite{BFKNRV,Ahn1211}.  
Note that the equation, $C_{34}^3 =\frac{3}{4} \, C_{33}^4$ holds for 
general $N$.
Furthermore, for $N=4$, expression (\ref{cosetspin3spin4}) without 
the spin-$5$ current 
is already known in \cite{KW,BFKNRV,Ozer}.}.
The explicit $(N, c)$-dependence of 
the structure constant $C_{33}^{4}$ in (\ref{pole2coefficient}) 
was obtained from the literature 
\cite{Hornfeck92,GG1}
\bea 
\left(C_{33}^{4}\right)^{2}=
\frac{64 (c+2) (N-3) \left[ c (N+3)
+2 (4 N+3) (N-1) \right]}{( 5 c+22) (N-2) \left[ c
   (N+2) + (3 N+2) (N-1) \right]} .
\label{C334}
\eea 
Note that due to the $(N-3)$ factor in (\ref{C334}), 
the coset model (\ref{coset}) for $N=3$ does not produce the spin-$4$ current.
The spin-$5$ current appearing in the OPE 
between the spin-$3$ and the spin-$4$ currents 
should contain the factor $(N-4)$ in its
expression. 
The $N$ should be greater than $4$.
This issue will be addressed later. See equation (\ref{Vtilde}).

The spin-$5$ current can be obtained from 
\bea 
W^{(5)}(w)= W^{(3)}(z)\, 
W^{(4)}(w) |_{\frac{1}{(z-w)^2}}-\eta_{\Lambda} 
\left( T W^{(3)}(w)-\frac{3}{14}\pa^2 W^{(3)}(w) \right) -\frac{\eta_{W}}{14}\pa^2 W^{(3)}(w). 
\label{spin5construction}
\eea 
The stress energy tensor and the spin-$3$ current are given in (\ref{stressT})
and (\ref{primaryspin3}). Furthermore, 
the relative coefficients are given in 
(\ref{pole2coefficient}) together with (\ref{C334}). 
To ensure the right relative coefficients in the right hand side 
of (\ref{spin5construction}), two arbitrary unknown 
constants can be introduced in front of $\eta_{\Lambda}$ and $\eta_{W}$.
For $N=3$, they are vanishing by calculating the OPE between $T(z)$ and 
$W^{(5)}(w)$ and requiring $W^{(5)}(w)$ is a primary field of spin-$5$. 
This is consistent with the fact that 
the structure constant $C_{33}^4$ in (\ref{C334}) vanishes at $N=3$.
For $N=4$ and $N=5$, these extra parameters are equal to $1$ by solving the 
primary condition.    
The next step is to calculate the second order pole for the OPE 
$W^{(3)}(z) \, W^{(4)}(w)$ explicitly.
Appendix $B$ presents all the detailed calculations. 
Appendix $C$ describes the fully normal ordering procedure. 
For $N=3$ or $N=4$, all the calculations in Appendices 
$B$ and $C$ were checked explicitly using Thielemans package 
\cite{Thielemans}.

After the complicated calculations, 
the spin-$5$ current is summarized by the following $52$ terms:
\bea 
&& W^{(5)}(z)=
\nonu \\
&& a_1 \,d^{abf} d^{fcg} d^{gde} J^a J^b J^c J^d J^e(z) + 
\left(a_2 \,d^{abf} d^{fcg} d^{gde}+a_3 \,d^{abf} d^{feg} d^{gcd} \right) 
J^a J^b J^c J^d K^e(z) \nonu\\
&&+ 
\left(a_4 \,d^{abf} d^{fcg} d^{gde}+a_5 \,d^{bdf} d^{fag} d^{gce}+
a_6 \,d^{abf} d^{fdg} d^{gce} \right) J^a J^b J^c K^d K^e(z) \nonu\\
&&+\left( a_7 \,d^{abf} d^{fcg} d^{gde} + a_8  \,d^{bcf} d^{fag} d^{gde}
+ a_9  \,d^{acf} d^{fdg} d^{gbe} \right) J^a J^b K^c K^d K^e(z) \nonu\\
&&+ \left(  a_{10} \,d^{abf} d^{fcg} d^{gde} 
+ a_{11}  \,d^{bcf} d^{fag} d^{gde} \right)  J^a K^b K^c K^d K^e(z)
+ a_{12}  \,d^{abf} d^{fcg} d^{gde}  K^a K^b K^c K^d K^e(z) \nonu\\
&&+ a_{13} \,d^{abc} \delta^{de}  J^a J^b J^c J^d J^e(z)
+ \left( a_{14} \,d^{abc} \delta^{de} + a_{15} \,d^{abe} \delta^{cd}  \right)
J^a J^b J^c J^d K^e(z)
\nonu\\ 
&&+\left( a_{16} \,d^{abc} \delta^{de} 
+  a_{17} \,d^{abd} \delta^{ce} 
+  a_{18} \,\delta^{ab} d^{cde}  \right) J^a J^b J^c K^d K^e(z)
\nonu\\
&&+ \left( a_{19} \,d^{abc} \delta^{de} + a_{20} \,\delta^{ab} d^{cde} 
+a_{21} \,\delta^{ac} d^{bde}  \right) J^a J^b K^c K^d K^e(z)
\nonu\\
&& + \left( a_{22} \,d^{abc} \delta^{de} + a_{23} \, \delta^{ab} d^{cde} 
\right) J^a K^b K^c K^d K^e(z)
+ a_{24} \, d^{abc} \delta^{de}  K^a K^b K^c K^d K^e(z)
\nonu\\
&&+ a_{25} \, f^{ade} d^{bce} J^a J^b J^c \pa J^d(z)
+ a_{26} \, d^{abe} f^{cde}  J^a J^b J^c \pa K^d(z)
+\left( a_{27} \, d^{abe} f^{cde} +a_{28} \, d^{ace} f^{bde}  \right.
\nonu\\
&&+ \left. a_{29} \, d^{ade} f^{bce}  \right) J^a J^b \pa J^c K^d(z)
+\left( a_{30} \, d^{abe} f^{cde} +a_{31} \, d^{ace} f^{bde}
+a_{32} \, d^{ade} f^{bce} \right) J^a J^b K^c \pa K^d(z)
\nonu\\
&&+ \left( a_{33} \, d^{ace} f^{bde} +a_{34} \, f^{ade} d^{bce} 
+a_{35} \, f^{abe} d^{cde}  \right) J^a \pa J^b K^c  K^d(z)
+\left( a_{36} \, d^{abe} f^{cde} +a_{37} \, f^{ade} d^{bce} \right.
\nonu\\
&&+ \left. a_{38} \, f^{ace} d^{bde} \right)  J^a K^b K^c \pa K^d(z)
+ a_{39} \, f^{ade} d^{bce} \pa J^a K^b K^c K^d(z) 
+ a_{40} \, f^{ade} d^{bce}  K^a K^b K^c \pa K^d(z) 
\nonu\\
&&+  a_{41} \, d^{abc} J^a J^b \pa^2 J^c(z)
+a_{42} d^{abc} \, J^a \pa J^b \pa J^c(z)
+a_{43} d^{abc} \, J^a \pa^2 J^b  K^c(z)
+a_{44} \, d^{abc} J^a  J^b \pa^2 K^c(z)
\nonu\\
&&+ a_{45} \, d^{abc} \pa J^a  \pa J^b  K^c(z)
+a_{46} \, d^{abc}  J^a  \pa J^b  \pa K^c(z)
+a_{47} \, d^{abc}  \pa^2 J^a  K^b  K^c(z)
\nonu \\
&& +  a_{48} \, d^{abc}   J^a  K^b  \pa^2 K^c(z)
+a_{49} \, d^{abc}  \pa J^a  K^b  \pa K^c(z)
+a_{50} \, d^{abc}   J^a \pa K^b  \pa K^c(z)
\nonu \\
&& +  a_{51} \, d^{abc}   K^a K^b  \pa^2 K^c(z)
+ a_{52} \, d^{abc}   K^a \pa K^b  \pa K^c(z),
\label{spin5detail}
\eea 
where the $52$ 
coefficient functions $a_i$ are given in Appendix $D$ (\ref{spin5coeff}).
The spin-$5$ current (\ref{spin5detail}) at $N=5$ was checked to confirm that 
there are no singular terms (in the OPE between the 
stress energy tensor and the spin-$5$ current) with an order greater than $2$.
For consistency check,  
the OPE between the spin-$5$ current and the diagonal current 
was also calculated, which showed that are no singular terms when $N=4$.
The primary condition with $N=4$ was checked until the third order 
singular terms. Furthermore,
when $N=3$, the above regularity and primary conditions (up to the second 
order pole) 
were checked explicitly.  
The spin-$5$ current consists of the first $24$ quintic terms, 
the middle sixteen quartic terms with the $f$ symbol (with one derivative), 
and the remaining 
twelve cubic terms with the two derivatives. 
The last twelve-terms can be observed from the second derivative of 
spin-$3$ current $W^{(3)}(z)$.
The twelve quintic terms containing the $\delta$ symbol 
also can be seen from the composite operator, $T W^{(3)}(z)$. 
The vacuum character $(7.18)$ of \cite{BS}, which is equal to 
$\frac{1}{\prod_{s=2}^N \prod_{n=s}^{\infty} (1-q^n)} =
1+ q^2 + 2q^3 + 4q^4 + 6 q^5 + 12 q^6 + {\cal O}(q^7)$, 
contains the numerical
factor, $6$, in front of the $q^5$ term.   
This suggests that there are six spin-$5$ fields,
$ T W^{(3)}(z)$, $\pa^2 W^{(3)}(z)$, $\pa W^{(4)}(z)$, $\pa^3 T(z)$, $\pa T^2(z)$, and $W^{(5)}(z)$.
Among these fields, the only primary spin-$5$ field is given by
$W^{(5)}(z)$. For the next higher spin-$6$ fields, 
there are eleven nonprimary fields, which can be obtained from 
the higher spin currents of spin $s=3,4,5$, the stress energy tensor, and 
derivatives,
and a single spin-$6$ primary current.  

\section{Three-point functions }

\subsection{Eigenvalue equation of the zero mode of spin-$5$ current acting on the state $| (f;0) >$ in the large $N$ limit}


This section describes the three-point functions with scalars 
for the spin-$5$ current found in the previous 
section. 
The large $N$ 't Hooft limit \cite{GG} is defined as 
\bea
N, k \rightarrow \infty, \qquad \lambda \equiv \frac{N}{N+k}
\qquad \mbox{fixed}.
\label{limit}
\eea
The product of the $SU(N)$ generators 
has the following decompositions 
with $\delta, d$ and $f$ symbols (for example, see reference \cite{Ahn1111})
\bea
T^a T^b = -\frac{1}{N} \delta^{ab} -\frac{i}{2} d^{abc} T^c
+\frac{1}{2} f^{abc} T^c.
\label{TT1}
\eea
A range of quintic products with three $d$ symbols can be obtained 
from equation (\ref{TT1}) and the traceless condition for the anti-hermitian basis. 

The zero modes of the current satisfy the commutation relations 
of the underlying finite dimensional Lie algebra $SU(N)$.
For the state $|(f;0)>$, $T^a$ corresponds to $K_0^a$ 
and for the state
$|(0;f)>$, $T^a$ corresponds to $J_0^a$ as follows
\bea
|(f;0)>:  \quad T^a \leftrightarrow  K_0^a, 
\qquad
|(0;f)>: \quad T^a \leftrightarrow J_0^a.
\nonu
\eea

\subsubsection{Eigenvalue equation}

Now the zero mode eigenvalue equation   of the spin $5$ primary 
field $W^{(5)}(z)$ acting on the primary state $(f;0)$ in the 
large $N$ 't Hooft limit can be calculated using the  
relations in  Appendix (\ref{limitspin5coeff}).
The ground state transforms as a fundamental representation with respect to
$K_0^a$ (and as an antifundamental representation with respect to 
$J_0^a$). 
For example, calculate 
the following eigenvalue equation corresponding to the first term of
the spin-$5$ current
\bea
d^{abf} d^{fcg} d^{gde} (J^a J^b J^c J^d J^e )_0 |(f;0)>.
\label{eigenexp}
\eea
How does one obtain the zero mode of the composite operator?
Using the standard definition for the zero mode, $(A.2)$ of \cite{BBSS1},
\bea
d^{abf} d^{fcg} d^{gde} J_0^e J_0^d J_0^c J_0^b J_0^a  |(f;0)>. 
\label{interinter}
\eea
That is, the zero mode is the product of each zero mode but the ordering is
reversed (the indices of $a,b,c,d$, and $e$ go to the indices of
$e,d,c,b$ and $a$).
The next step is to use the eigenvalue equation.
Because the singlet condition for 
the state  $(f;0)$ is expressed as
\bea
\left( J_0^a+K_0^a \right)|(f;0)>=0,
\label{singletcondition}
\eea
The expression (\ref{interinter}) is equivalent to
\bea
-d^{abf} d^{fcg} d^{gde} J_0^e J_0^d J_0^c J_0^b K_0^a  |(f;0)>=
-d^{abf} d^{fcg} d^{gde} K_0^a J_0^e J_0^d J_0^c J_0^b   |(f;0)> 
\label{toto}
\eea
where the zero mode $K_0^a$ commutes with the zero mode $J_0^b$
and $K_0^a$ is moved to the left.
The singlet condition (\ref{singletcondition}) is applied to the rightmost
zero mode $J_0^b$ (\ref{toto})
and the following result can be obtained
\bea
d^{abf} d^{fcg} d^{gde} K_0^a J_0^e J_0^d J_0^c K_0^b   |(f;0)> 
=d^{abf} d^{fcg} d^{gde} K_0^a K_0^b J_0^e J_0^d J_0^c    |(f;0)>. 
\label{middstep}
\eea
This procedure can be repeated until all the $J_0^a$'s are exchanged
with the $K_0^a$'s with the appropriate ordering. 
Therefore, the above eigenvalue equation (\ref{eigenexp})
becomes
\bea
-d^{abf} d^{fcg} d^{gde} K_0^a K_0^b K_0^c K_0^d K_0^e  |(f;0)> 
& = & -\frac{1}{N} d^{abf} d^{fcg} d^{gde}  \mbox{Tr} ( T^a T^b T^c T^d T^e) 
|(f;0)>
\nonu\\
& \rightarrow &i N^4  |(f;0)>.
\label{aboveabove1}
\eea
In the first line of (\ref{aboveabove1}), $\frac{1}{N}$ is multiplied
because the eigenvalue is needed (not the trace).
Of course, the zero mode eigenvalue equation can be expressed using the 
zero mode $J_0^a$ rather than $K_0^a$ with the corresponding $SU(N)$ generator
rather than $T^a$ \footnote{
Similarly, the following eigenvalue equation corresponding to 
the ninth-term of the spin-$5$ current  
can be obtained
\bea
d^{acf} d^{fdg} d^{gbe} (J^a J^b K^c K^d K^e )_0 |(f;0)>&=&
\frac{1}{N} d^{abf} d^{fcg} d^{gde}  \mbox{Tr} ( T^a T^b T^c T^d T^e) 
|(f;0)>
\nonu\\
& \rightarrow &-i N^4 |(f;0)>.
\label{expres}
\eea
In the first line of (\ref{expres}),
the zero mode is obtained by changing the ordering of the current reversely.
The singlet condition (\ref{singletcondition}) 
is used in the second line. There is no change in sign because of the even number of $J_0^a$'s.
In the third line, $\frac{1}{N}$ is multiplied and the cyclic 
property of the trace is used.
Finally, 
the previous result of (\ref{ddd1}) is used in the last line of 
(\ref{expres}).}.
Here the following quantity (related to the first term of spin-$5$ current) 
can be used:
\bea
d^{abf} d^{fcg} d^{gde} \mbox{Tr} (T^a T^b T^c T^d T^e) & = & 
-\frac{i}{8} d^{abf} d^{fcg} d^{gde}  d^{abh} d^{hci} d^{ide} 
\nonu \\
& = & 
-\frac{i}{8} \frac{2}{N} (N^2-4) \delta^{fh} \frac{2}{N} (N^2-4) \delta^{gi}
d^{fcg} d^{hci}
\nonu \\
& = &  -i \frac{1}{N^3} (N^2-4)^3 (N^2-1)
\rightarrow - i N^5,
\label{ddd1}
\eea
where the identities of Appendix $A$ of \cite{Ahn1111}, $(A.3)$ and 
$(A.4)$,
are used. Note $\delta^{aa} =N^2-1$. For $N=3$, 
the above identity is checked explicitly. 
At the final stage, a large $N$ limit (\ref{limit}) is taken.
Similarly, the second-, third-, fourth-, sixth-, seventh-, ninth-, tenth-, 
eleventh- and twelfth-terms of the spin-$5$ current can be analyzed.

One can proceed to calculate the remaining trace identities.
For the fifth-term (and eighth-term) of the spin-$5$ current,
the following calculation can be performed:  
\bea
d^{acf} d^{fbg} d^{gde} \mbox{Tr} (T^a T^b T^c T^d T^e) 
& = & 
d^{acf} d^{fbg} d^{gde} \left( -\frac{i}{2N}\delta^{ab}d^{cde} 
-\frac{i}{8}d^{abh}d^{hci}d^{ide}
+\frac{i}{8}f^{abh}f^{hci}d^{ide}
\right)
\nonu\\
&=&\frac{4}{N^3}i(N^2-4)^2(N^2-1)
\rightarrow 4i N^3.
\label{ddd2}
\eea
The $d d d $ product can be reduced to a single $d$ and
the $d d f$ product can be written in terms of a single $f$ 
using the identities involving the $f$- and $d$-tensors 
of $SU(N)$.  For the $N=3$, 
the above identity has been checked explicitly. 
The large $N$ behavior of (\ref{ddd2}) is different from that of 
(\ref{ddd1}).
The large $N$ limit for the coefficient functions was not considered.
Once the $N$ behavior of these coefficient functions is included, 
(\ref{ddd1}) and (\ref{ddd2}) 
do to the final eigenvalue equations. 

For the thirteenth-term (fourteenth-, $\cdots$, $24$th-terms) 
of the spin-$5$ current,
the quintic product with $d$ symbol and $\delta$ symbol is obtained
as follows:
\bea
d^{abc} \delta^{de}  \mbox{Tr} (T^a T^b T^c T^d T^e) & = & 
d^{abc} \delta^{de} \left( - \frac{i}{2N}d^{abc} \delta^{de}\right)
=- \frac{i}{N^2}(N^2-4)(N^2-1)^2
\nonu \\
& \rightarrow & - i N^4.
\label{ddelta}
\eea
Note that the $d d$ product becomes a single $\delta$ symbol. 
The large $N$ behavior of (\ref{ddelta}) is different from (\ref{ddd1}) or
(\ref{ddd2}).
 For $N=3$, 
the above identity has been checked explicitly. 

One has the quartic products with $d$ symbol and $f$ symbol
for the $25$th-term ($26$th-, 
$27$th-, $29$th-, $30$th-, $31$st-,
$33$rd-, $35$th-, $36$th-, $37$th-, $39$th-,
$40$th-terms) of the spin-$5$ current
\bea
d^{abe} f^{cde} \mbox{Tr} (T^a T^b T^c T^d) & = & \frac{i}{4} d^{abe} f^{cde}
d^{abf} f^{cdf} 
 =  i (N^2-4)(N^2-1)
\nonu \\
& \rightarrow & i N^4.
\label{dfidentity1}
\eea
The large $N$ behavior of (\ref{dfidentity1})
is the same as that of (\ref{ddelta}).

For the $28$th-term of the spin-$5$ current,
\bea
d^{ace} f^{bde} \mbox{Tr} (T^a T^b T^c T^d) 
& = & d^{ace} f^{bde} \left( \frac{i}{4} d^{abf}f^{cdf}+\frac{i}{4} f^{abf}d^{cdf}  \right) \nonu\\
&=&  \frac{i}{4} d^{ace} \left( f^{bde}d^{abf}f^{cdf}+ f^{bde} f^{abf}d^{cdf} \right) \nonu\\
&=&  \frac{i}{4} d^{ace} \left( -N d^{ace} + N d^{ace} \right) =0.
\label{dfidentity2}
\eea
The $d f f$ product reduces to a single $d$ symbol. 
Furthermore, the contributions from 
the $32$nd-, $34$th-, and $38$th-terms vanish.
For $N=3, 4$, 
the above identities  (\ref{dfidentity1}) and 
(\ref{dfidentity2}) have been checked explicitly. 

Finally,  the cubic product with the $d$ symbol is calculated
for the $41$st-term of the spin-$5$ current (and $42$nd-term, 
$\cdots$, $52$nd-term),
\bea
d^{abc} \mbox{Tr} (T^a T^b T^c ) & = & 
\frac{i}{2} d^{abc} d^{abc} 
=  \frac{i}{N}(N^2-4)(N^2-1) 
\rightarrow i N^3.
\label{dtrace}
\eea

Moreover, the large $N$ behaviors of spin-$5$ coefficient functions $a_i$ 
in the large $N$ limit are given in Appendix (\ref{limitspin5coeff}). 
 
Let us move on the 
zero mode eigenvalue with one derivative corresponding to the 
$25$th-term of the spin-$5$ current
\bea
f^{ade} d^{bce} (J^a J^b J^c \pa J^d )_0 |(f;0)>&=&
f^{ade} d^{bce} (\pa J^d)_0 J_0^c J_0^b J_0^a |(f;0)> 
=-f^{ade} d^{bce}  J_0^d J_0^c J_0^b J_0^a |(f;0)> \nonu\\
&=&-f^{ade} d^{bce}  K_0^a K_0^b K_0^c K_0^d |(f;0)> 
\nonu \\
& = & 
\frac{1}{N} d^{abe} f^{cde}   \mbox{Tr} ( T^a T^b T^c T^d)  |(f;0)> 
 \rightarrow i N^3  |(f;0)>.
\label{zero1round}
\eea
In the first line of (\ref{zero1round}),
the zero mode is taken by reversing the ordering of the current as before.
One then uses the following property,
$\pa J^a(z)=\pa \left( \sum_{m} \frac{J^a_m}{z^{m+1}} \right) 
=\sum_{m} \frac{(-m-1)J^a_m}{z^{m+2}}
\equiv \sum_{m} \frac{(\pa J^a)_m}{z^{m+2}}$,
which leads to the zero mode, $(\pa J^a)_0=-J^a_0$.
In the second line, the singlet condition (\ref{singletcondition}) 
is used, and there is no sign change due to the even number of the $J^a_0$'s.
Furthermore, the previous result (\ref{dfidentity1}) is used \footnote{
Consider the contribution from the $41$st-term of spin-$5$ current
with two derivative terms
\bea
d^{abc} (J^a J^b \pa^2 J^c )_0 |(f;0)>
&=&d^{abc}  (\pa^2 J^c )_0 J_0^b J_0^a |(f;0)> 
=2d^{abc}   J_0^c J_0^b J_0^a |(f;0)> \nonu\\
&=&-2d^{abc}   K_0^a K_0^b K_0^c |(f;0)> 
=-\frac{2}{N}d^{abc}     \mbox{Tr} ( T^a T^b T^c) |(f;0)> \nonu\\
&\rightarrow &-2i N^2  |(f;0)>.
\label{zero2round}
\eea 
In the first line of (\ref{zero2round}), the zero mode is taken.
The zero mode of the second derivative for the current can be obtained from the
fact that 
$\pa^2 J^a(z)=\pa^2 \left( \sum_{m} \frac{J^a_m}{z^{m+1}} \right) 
=\sum_{m} \frac{(m+1)(m+2)J^a_m}{z^{m+3}}
\equiv \sum_{m} \frac{(\pa^2 J^a)_m}{z^{m+3}}$,
which leads to $(\pa^2 J^a)_0=2J^a_0$.
In the second line, the singlet condition (\ref{singletcondition})
is used. Note the extra minus sign because of the odd number of the current.
As before, $\frac{1}{N}$ is multiplied.
At the final stage, the result (\ref{dtrace}) is used.
Appendix $E$ describes the complete results  for the  eigenvalue equations in the spin-$5$ zero modes.}. 

In this way, the nonzero zero mode contributions 
of all the terms in $W^{(5)}(z)$ (\ref{spin5detail}) can be obtained. 
Let us describe the nonzero contributions.
From the analysis in (\ref{1})-(\ref{4}), (\ref{6}), 
(\ref{7}), (\ref{9})-(\ref{11}),
these nine-terms behave as $N^4$ and each coefficient function in 
(\ref{Nbehavior}) has a factor, $N$.
These terms then contribute to the final result in the large $N$ limit. 
Furthermore, the non-zero contributions from (\ref{25})-(\ref{27}), 
(\ref{29})-(\ref{31}), (\ref{33}), (\ref{35})-(\ref{37}) and (\ref{39}) have a factor, $N^3$, and the coefficient
functions for these terms behave as $N^2$ from (\ref{Nbehavior}). 
Finally, from (\ref{41})-(\ref{50}), these 
contributions are given by $N^2$ and the corresponding coefficients
behave  as $N^3$  from (\ref{Nbehavior}).
By summing these contributions, the following 
eigenvalue equation can be obtained for the zero mode of the 
spin-$5$ current $W^{(5)}_0$ acting on the state $|(f;0)>$ in the large $N$ limit
\bea
W^{(5)}_0 |(f;0) >&=&
-\left[\frac{24 i (1+\lambda ) (3+\lambda ) 
(4+\lambda ) }{7\lambda ^3 (2-\lambda ) } \right] N^5 | (f;0) >.
\label{main1}
\eea
The eigenvalue has a very simple factorized form. 
The same eigenvalue equation at  finite $N$ and $k$ will be evaluated.

\subsection{Eigenvalue equation of the zero mode of the spin-$5$ current acting on the state $| (0;f) >$ in the large $N$ limit}

Next the zero mode eigenvalue of the 
spin-$5$ primary field $W^{(5)}(z)$ acting 
on the primary state $|(0;f)>$ is described  
in a large $N$ 't Hooft limit using the 
above (\ref{ddd1})-(\ref{dtrace}) 
relations and Appendix (\ref{limitspin5coeff}).
The ground state transforms as a fundamental representation with respect to
$J_0^a$ and the zero mode
$K_0^a$ 
has a vanishing eigenvalue equation
 (i.e. a fundamental representation 
with respect to the zero mode of the diagonal current)
\bea
K_0^a  |(0;f)> =0.
\label{sing}
\eea 
For example,
calculate 
the following eigenvalue equation corresponding to the first-term of the
spin-$5$ current
\bea
d^{abf} d^{fcg} d^{gde} (J^a J^b J^c J^d J^e )_0 |(0;f)>&=&
d^{abf} d^{fcg} d^{gde} J_0^e J_0^d J_0^c J_0^b J_0^a  |(0;f)> \nonu\\
&=&\frac{1}{N} d^{abf} d^{fcg} d^{gde}  \mbox{Tr} ( T^e T^d T^c T^b T^a) 
|(0;f)>,
\label{other}
\eea
where the previous result
(\ref{interinter}) is used.
Equation (\ref{other}) becomes the following 
expression using  the property of the $d$ symbol
\bea
d^{abf} d^{fcg} d^{gde} (J^a J^b J^c J^d J^e )_0 |(0;f)>&=&
\frac{1}{N} d^{abf} d^{fcg} d^{gde}    \mbox{Tr} ( T^a T^b T^c T^d T^e) 
|(0;f)> \nonu \\
& 
 \rightarrow & -i N^4  |(0;f)>.
\label{first}
\eea

The next non-zero contribution appears in the thirteenth-term.
The following nontrivial contribution can be obtained
from the thirteenth-term
\bea
d^{abc} \delta^{de} (J^a J^b J^c J^d J^e )_0 |(0;f)>&=&
d^{abc} \delta^{de} J_0^e J_0^d J_0^c J_0^b J_0^a |(0;f)> \nonu\\
&=&\frac{1}{N} d^{abc} \delta^{de}  \mbox{Tr} (  T^a T^b T^c T^d T^e) 
|(0;f)>
\nonu\\
& \rightarrow &-i N^3  |(0;f)>,
\label{refref}
\eea
where, as before, the trace property is used and the $\frac{1}{N}$
is multiplied to obtain the eigenvalue.
The large $N$ limit is taken at the final stage in (\ref{refref}).
On the other hand, due to the $N$ factor from (\ref{Nbehavior}),
the final contribution becomes zero in the large $N$ limit because
$N \times N^3 =N^4$ (the leading large $N$ behavior has $N^5$).

The next nonzero contribution appears in the $25$th-term 
as follows:
\bea
f^{ade} d^{bce} (J^a J^b J^c \pa J^d )_0 |(0;f)>&=&
-f^{ade} d^{bce}  J_0^d J_0^c J_0^b J_0^a |(0;f)> \nonu\\
&=&-\frac{1}{N} d^{abe} f^{cde}   \mbox{Tr} ( T^a T^b T^c T^d)  |(0;f)> \nonu\\
& \rightarrow &-i N^3 |(0;f)>.
\label{twentyfive}
\eea
In the first line, the previous relation (\ref{zero1round})
is used.
The nonzero contribution in the $41$st-term 
can be obtained 
\bea
d^{abc} (J^a J^b \pa^2 J^c )_0 |(0;f)>
&=&2d^{abc}   J_0^a J_0^b J_0^c |(0;f)> 
=\frac{2}{N}d^{abc}     \mbox{Tr} ( T^a T^b T^c) |(0;f)> \nonu\\
& \rightarrow &2i N^2  |(0;f)>.
\label{fortyfirst}
\eea
The first line comes from (\ref{zero2round}).
The final nonzero contribution comes from the next $42$nd-term
and the following is obtained:
\bea
d^{abc} (J^a \pa J^b \pa J^c )_0 |(0;f)>
& = & d^{abc}   J_0^a J_0^b J_0^c |(0;f)> 
=\frac{1}{N}d^{abc}     \mbox{Tr} ( T^a T^b T^c)|(0;f)> \nonu\\
& \rightarrow &i N^2  |(0;f)>.
\label{fortytwo}
\eea

By combining 
(\ref{first}), (\ref{twentyfive}), (\ref{fortyfirst}) 
and (\ref{fortytwo}) with the corresponding coefficients in the 
large $N$ limit  (\ref{Nbehavior}), the final zero mode eigenvalue 
equation can be obtained
as follows:
\bea
&&W^{(5)}_0|  (0;f) > =
\nonu\\
&& \left[ -i N^4 \times N \left(\frac{6 (\lambda -2)^2 
(\lambda -1) }{\lambda ^3 (\lambda +2)} \right)
-i N^3 \times N^2 
\left(-\frac{6 (\lambda -2) (\lambda -1)}{\lambda ^3} \right)
\right. \nonu \\
&& \left. +i N^2 \times N^3 
\left( \frac{24 (\lambda -1) \left(3 \lambda ^2-20\right)}{7 
\lambda ^3 (\lambda +2)}  
-\frac{48 (\lambda -1) \left(2 \lambda ^2-11\right)}{7 \lambda ^3 (\lambda +2)}\right) \right] |  (0;f) >
\nonu \\
&& =
\left[ 
\frac{24 i (1-\lambda )(3-\lambda )(4-\lambda)   }{7 \lambda ^3 (2+\lambda )}
\right] N^5   |(0;f)>.
\label{main2}
\eea
The same eigenvalue equation at  finite $N$ and $k$ can be 
determined later.

\subsection{Three-point functions in the large $N$ limit}

Consider the diagonal modular invariant, by pairing 
up identical representations on the left (holomorphic) and 
right (antiholomorphic) sectors. One of the primaries 
is given by $(f ; 0) \otimes (f ; 0)$ \cite{CY}. 
From the previous result (\ref{main1}), the eigenvalue of 
the spin-$5$ zero mode  for $( f ; 0) \otimes ( f ; 0)$ is 
\bea
W^{(5)}_0 | {\cal{O}}_{+} > & = &   
- \left[\frac{24i(1+\lambda)(3+\lambda)(4+\lambda)}{7 \lambda^3(2-\lambda) }
\right] N^5 | {\cal{O}}_{+} >, \qquad {\cal{O}}_{+} \equiv 
( f ; 0) \otimes ( f ; 0).
\label{primary1}
\eea
In addition,  
the eigenvalue of  the spin-$5$ zero mode for 
$( 0; f ) \otimes ( 0; f )$, which is another primary, can be obtained. 
In this case, only four terms with no $K^a$ can 
survive because  
$K_0^a |( 0; f )>=0$. Therefore, the result can be calculated from (\ref{main2}) as follows:
\bea
W^{(5)}_0 | {\cal{O}}_{-} > & = &   
\left[ \frac{24i(1-\lambda)(3-\lambda)(4-\lambda)}{7 \lambda^3(2+\lambda) }
 \right] N^5| {\cal{O}}_{-} >, \qquad  {\cal{O}}_{-} \equiv 
( 0; f ) \otimes ( 0; f ).
\label{primary2}
\eea
Symmetry exists between 
(\ref{primary1}) and (\ref{primary2}). One can be obtained from the other 
by taking $\lambda \rightarrow -\lambda$ and vice versa.
Similar relations can be obtained for the other primaries  $(\overline{f} ; 0) \otimes (\overline{f} ; 0)$ and 
$( 0; \overline{f} ) \otimes ( 0; \overline{f} )$
after taking the generators of $SU(N)$ in the antifundamental representation
carefully, 
\bea
W^{(5)}_0 | {\overline{\cal{O}}}_{+} > & = &   
\left[ \frac{24i(1+\lambda)(3+\lambda)(4+\lambda)}{7 \lambda^3(2-\lambda) }
\right] N^5| \overline{\cal{O}}_{+} >, \qquad {\overline{\cal{O}}}_{+} \equiv 
(\overline{f} ; 0) \otimes (\overline{f} ; 0), 
\nonu\\
W^{(5)}_0 | \overline{\cal{O}}_{-} > & = &   
-\left[ \frac{24i(1-\lambda)(3-\lambda)(4-\lambda)}{7 \lambda^3(2+\lambda) }
\right]
N^5| \overline{\cal{O}}_{-} >, \qquad  \overline{\cal{O}}_{-} \equiv 
( 0; \overline{f} ) \otimes ( 0; \overline{f} ).
\label{primary34}
\eea
Compared to (\ref{primary1}) and (\ref{primary2}), there is an overall 
sign change.

The three-point functions of spin-$5$ current 
with two scalars from (\ref{primary1}) 
and (\ref{primary2}) (or (\ref{primary34})) 
can then be expressed as
\bea
<\overline{\cal{O}}_{+} {\cal{O}}_{+} W^{(5)}>  & = &
-\left[ \frac{24i(1+\lambda)(3+\lambda)(4+\lambda)}{7 \lambda^3(2-\lambda) }
\right] N^5,
\label{threepoint1}
\\
<\overline{\cal{O}}_{-} {\cal{O}}_{-} W^{(5)}> & = &
\left[ \frac{24i(1-\lambda)(3-\lambda)(4-\lambda)}{7 \lambda^3(2+\lambda) }
\right] N^5.
\label{threepoint2}
\eea
The following result can be obtained by dividing these 
two relations (\ref{threepoint1}) and (\ref{threepoint2}):
\bea
\frac{<\overline{\cal{O}}_{+} {\cal{O}}_{+} W^{(5)}>}{
<\overline{\cal{O}}_{-} {\cal{O}}_{-} W^{(5)}>}
&=&-\frac{(1+\lambda)(2+\lambda)
(3+\lambda)(4+\lambda)}{(1-\lambda)(2-\lambda)(3-\lambda)(4-\lambda)}.
\label{largeNratio}
\eea
When $\lambda=\frac{1}{2}$, this becomes 
$-9$, which is precisely the same as the bulk computation for spin-$5$ 
in \cite{CY}.
For a general $\lambda$, the bulk result in \cite{AKP} for spin-$5$
current agrees with (\ref{largeNratio}).
Recall that the ratio (\ref{largeNratio}) containing constant $1$ without a minus
sign is precisely the ratio for the spin-$2$ current,
 the ratio (\ref{largeNratio}) containing constants $1$ and $2$ 
is exactly the ratio for the spin-$3$ current and 
 the ratio (\ref{largeNratio}) containing constants $1, 2, 3$ without a minus
sign is exactly the ratio for the spin-$4$ current.
For the spin-$5$ current, the ratio of the three-point function  is obtained 
from that of the spin-$4$ current by multiplying 
$-\frac{(4+\lambda)}{(4-\lambda)}$.  See footnote \ref{largeNbe}.
The ratio of three-point function for the spin-$s$ current 
can be expressed as $(-1)^s \prod_{n=1}^{s-1} \frac{(n+\lambda)}{(n-\lambda)}$
\footnote{More explicitly, one can use the equations $(4.51)$ 
and $(4.53)$ of \cite{AKP} to check the bulk result corresponding to 
(\ref{largeNratio}). What does this imply in the bulk theory side?
The three-point function in the bulk theory 
can be determined from the asymptotic behavior of
a scalar field in $AdS_3$ as the radial coordinate goes to infinity. On the 
other hand, the CFT 
result of this paper (i.e. the equation (\ref{largeNratio})) 
indicates that for fixed spin $s=5$, the $\lambda$-dependent 
part of this scalar field asymptotics has very simple factorized forms
$ \pm \prod_{n=1}^{4} (n \pm \lambda)$.
This provides the nontrivial information on the change in the $AdS_3$ scalar
under the gauge transformation for the higher spin deformation in the bulk
theory. Of course, this feature arises in section $4$ of \cite{AKP} via 
bulk theory computation. Therefore, one might think that the physical 
implication for the three-point function (\ref{largeNratio}) is to 
indicate a hidden 
functional dependence of the change in the 
$AdS_3$ scalar (under the gauge transformation) on the deformation parameter 
$\lambda$ where the $\lambda =\frac{1}{2}$ in undeformed bulk theory.       }.
This issue for the finite $N$ and $k$ will be addressed later.

Thus far, the normalization for the spin-$5$ current is not considered.
The normalized spin 5 primary field $\widetilde{W}^{(5)}(z)$ can be obtained
using the unnormalized spin-$5$ current $W^{(5)}(z)$ (\ref{spin5detail})
as follows:
\bea
\widetilde{W}^{(5)}(z) &\equiv& \left[\frac{B}{C_{33}^4 C_{34}^5} \right] W^{(5)}(z), 
\nonu \\
B^2 & \equiv & -\frac{ N}{18 (N+k)^2 (1+N)^2 (1+k+ N)^2 
(N+2)(2 k+N) (2+2 k+3 N)(-4+N^2)},
\nonu \\
(C_{34}^5)^2 & = &
\frac{25 (5c+22) (N-4) \left[ c (N+4)
+3 (5 N+4) (N-1) \right]}{( 7 c+114) 
(N-2) \left[ c
   (N+2) + (3 N+2) (N-1) \right]}, 
\label{Vtilde}
\eea
where $B$ is the overall coefficient function of 
the spin-$3$ current \cite{BBSS2},  $C_{34}^{5}$  is 
the structure constant and is given in \cite{BEHHH,GG1}. 
Note the factor $(N-4)$ in this expression.
This suggests that the $SU(4)$ coset model does not produce 
a spin-$5$ current, as mentioned before.
Moreover the structure constant 
$C_{33}^4$ was given in (\ref{C334}). 
The structure constant $C_{34}^5$ can be re-expressed as follows \cite{GG1}:
$(C_{34}^5)^2=25\left[ \frac{(C_{33}^4)^2}{(\hat{C}_{33}^4)^2}-1\right]$, where 
$\hat{C}_{33}^4$ is the structure constant for $N=4$.
Therefore, for $N=4$, the structure constant vanishes.
Accordingly, how does one obtain 
the relative coefficient factor in $\widetilde{W}^{(5)}(z)$ in (\ref{Vtilde})? 
Originally, the spin-$3$ current has an overall factor $B$.
One can multiply $B$ at both sides in (\ref{cosetspin3spin4}). 
For the spin-$4$ current, this study did not consider the structure constant 
$C_{33}^4$ in \cite{Ahn1111} in front of $W^{(4)}(w)$ of the second order pole
in the OPE $W^{(3)}(z) W^{(3)}(w)$.
This analysis checked that for $N=4$, the spin-$4$ current 
$\frac{W^{(4)}(z)}{C_{33}^4}$ provides the highest singular term, $\frac{c}{4}$,
correctly. 
This suggests that one should divide out the 
structure constant $C_{33}^4$ in (\ref{cosetspin3spin4})
\footnote{
\label{largeNbe}
The eigenvalue equations can be calculated for the other currents.
For the spin-$4$ current, the large $N$ limit was given in $(3.47)$ 
of \cite{Ahn1111}. By dividing $C_{33}^4$ with a large $N$ limit, 
\bea
\widetilde{W}^{(4)}_0 | {\cal{O}}_{+} >  =     
\frac{\sqrt{5}}{20} (1+\lambda) \sqrt{\frac{(2+\lambda)(3+\lambda)}{(2-\lambda)(3-\lambda) }}| {\cal{O}}_{+} >,
\widetilde{W}^{(4)}_0 | {\cal{O}}_{-} >  =    
\frac{\sqrt{5}}{20} (1-\lambda) \sqrt{\frac{(2-\lambda)(3-\lambda)}{(2+\lambda)(3+\lambda) }}| {\cal{O}}_{-} >.
\label{foureigen}
\eea
For the spin-$3$ current, by multiplying $(4.22)$ or $(4.20)$ of \cite{GH}
by the constant $A_1$ (\ref{BA}) 
and constant $B$ (\ref{Vtilde}) with a large $N$
limit, respectively, results in similar equations as follows:
\bea
\widetilde{W}^{(3)}_0 | {\cal{O}}_{+} >  =    
\frac{\sqrt{2}}{6} (1+\lambda) \sqrt{\frac{(2+\lambda)}{(2-\lambda)}}| {\cal{O}}_{+} >, \qquad
\widetilde{W}^{(3)}_0 | {\cal{O}}_{-} >  =    
-\frac{\sqrt{2}}{6} (1-\lambda) \sqrt{\frac{(2-\lambda)}{(2+\lambda)}}| {\cal{O}}_{-} >.
\label{threeeigen}
\eea
The $\lambda$-dependent part of the eigenvalue equation of
spin-$4$ current in (\ref{foureigen}) can be 
obtained from that of the spin-$3$ current in (\ref{threeeigen}) 
by multiplying $
\sqrt{\frac{(3+\lambda)}{(3-\lambda)}}$ and  $
\sqrt{\frac{(3-\lambda)}{(3+\lambda)}}$, respectively, up to the sign.
This feature holds for the spin-$3$ and spin-$2$ currents. 
That is, by  multiplying $
\sqrt{\frac{(2+\lambda)}{(2-\lambda)}}$ and  $
\sqrt{\frac{(2-\lambda)}{(2+\lambda)}}$
with  the following spin-$2$ eigenvalue equations,
\bea
T_0 | {\cal{O}}_{+} >  =   
\frac{1}{2} (1+\lambda) | {\cal{O}}_{+} >, \qquad
T_0 | {\cal{O}}_{-} >  =    
\frac{1}{2} (1-\lambda) | {\cal{O}}_{-} >,
\label{twoeigen}
\eea
one obtains the above eigenvalue equations (\ref{threeeigen}) up to the sign. }.
$\frac{B}{C_{33}^4} \, W^{(5)}(w)$ is on the right hand side.
Once again, this quantity should be multiplied by the structure constant 
$C_{34}^5$ and divide it out. This leaves 
the quantity
$\frac{B}{C_{33}^4 C_{34}^5} \, W^{(5)}(w)$, which is denoted by $\widetilde{W}^{(5)}(w)$.
The central charge term of the OPE 
between $\widetilde{W}^{(5)}$ and itself will have the form
\bea
\widetilde{W}^{(5)} (z) \, \widetilde{W}^{(5)} (w) &=& \frac{1}{(z-w)^{10}} \, \frac{c}{5} 
+ \cdots.
\label{nor}
\eea 
In this normalization (\ref{nor}), 
the previous eigenvalue equations (\ref{primary1}) and (\ref{primary2}) become 
\bea
\widetilde{W}^{(5)}_0 | {\cal{O}}_{+} > & = &   
\frac{\sqrt{14}}{70} (1+\lambda) \sqrt{\frac{(2+\lambda)(3+\lambda)(4+\lambda)}{(2-\lambda)(3-\lambda)(4-\lambda) }}| {\cal{O}}_{+} >,
\nonu \\
\widetilde{W}^{(5)}_0 | {\cal{O}}_{-} > & = &   
-\frac{\sqrt{14}}{70} (1-\lambda) \sqrt{\frac{(2-\lambda)(3-\lambda)(4-\lambda)}{(2+\lambda)(3+\lambda)(4+\lambda) }}| {\cal{O}}_{-} >.
\label{normalresult}
\eea
From the description of footnote \ref{largeNbe}, 
the general $\lambda$ dependent behavior 
for the eigenvalue equations of a higher spin-$s$ current can be read off and 
\bea
\widetilde{W}_0^{(s)} | {\cal{O}}_{+} >  & = &    
\left[(1+\lambda) 
\sqrt{\frac{(2+\lambda)(3+\lambda) \cdots (s-1+\lambda)}
{(2-\lambda)(3-\lambda) \cdots (s-1-\lambda) }} \right]
| {\cal{O}}_{+} >, \nonu \\
\widetilde{W}_0^{(s)} 
| {\cal{O}}_{-} >   & = &     
(-1)^s \left[ (1-\lambda) 
\sqrt{\frac{(2-\lambda)(3-\lambda) \cdots (s-1-\lambda)}
{(2+\lambda)(3+\lambda) \cdots (s-1 +\lambda) }} \right]
| {\cal{O}}_{-} >.
\label{general}
\eea
In (\ref{general}), the second equation can be obtained from 
the first equation 
by taking $\lambda \rightarrow -\lambda$ and vice versa 
up to an overall sign.

Note that 
the ratio of unnormalized spin-$5$ and 
normalized one in the large $N$ limit
is
\bea
\frac{W^{(5)}(z)}{\widetilde{W}^{(5)}(z)} 
= \frac{C_{33}^4 C_{34}^5}{B} \rightarrow 
- i N^5 \frac{120 \sqrt{14}}{7 \lambda^3} \sqrt 
{\frac{(3-\lambda)(3+\lambda)(4-\lambda)(4+\lambda)}{(2-\lambda)(2+\lambda)}}.
\eea

\subsection{Eigenvalue equations and three-point functions at a finite $(N,k)$}

At a finite $N$ and $k$, the $52$ coefficient functions are given in terms of
$(N,k)$ in 
(\ref{spin5coeff})  and
the identities (\ref{ddd1})-(\ref{dtrace}) hold for any $N$.
The exact zero mode eigenvalue equations can be obtained. 
Surprisingly, the eigenvalue equations  are simple factorized forms.
For the primary state, $| (f;0)>$, the $48$ terms contribute to the
final result. The traces corresponding to the $28$th, $32$nd, 
$34$th, and $38$th terms of the spin-$5$ current are identically 
zero.
The eigenvalue equation can be expressed as follows:
\bea
&&W^{(5)}_0 |(f;0)> =
\nonu\\
&&
\left[ -24 i (k+1) (N-4) (N-3) (N-1) (N+1)^2 (N+2) (k+N+1) (k+2 N) \right.
\nonu\\
&&  (k+2 N+1)(3 k+4 N+3)(4 k+5 N+4)
/
\left( N^3 (2 k+N)  (7 k^2 N+107 k^2+14 k N^2  \right.
\nonu\\
&& + \left. \left.  
221 k N+107 k+114 N^2+114 N) \right) \right] |(f;0)>.
\label{exact1}
\eea
Note the factor $(N-4)(N-3)$ in the numerator of (\ref{exact1}).
$(N-s+1) (N-s+2) \cdots (N-4)(N-3)$
is expected in the general spin-$s$ eigenvalue equation.
Of course, the large $N$ limit result (\ref{main1})
can be reproduced from the more general result (\ref{exact1}).

For the other primary state $| (0;f)>$, 
the eigenvalue equation becomes 
\bea
&&W^{(5)}_0 |(0;f)> =
\nonu\\
&& 
\left[ 24 i k (k+1) (N-4) (N-3) (N-1) (N+1)^2 (N+2) (k+N) (k+2 N) (3 k+2 N) \right.
\nonu\\
&&  (4 k+3 N) 
/
\left( N^3 (2 k+3 N+2) (7 k^2 N+107 k^2+14 k N^2+221 k N+107 k+114 N^2 
\right. 
\nonu\\
&&+ \left. \left. 114 N )\right) \right]
|(0;f)>.
\label{exact2}
\eea
The factor, $(N-4)(N-3)$, in (\ref{exact2}) appears in this case and
 the large $N$ limit (\ref{main2})
can be obtained from the more general result (\ref{exact2}).
The three-point functions from (\ref{exact1}) and (\ref{exact2}) 
can be summarized as
\bea
&& <\overline{\cal{O}}_{+} {\cal{O}}_{+} W^{(5)}>
 = 
\nonu \\ 
&& \left[ -24 i (k+1) (N-4) (N-3) (N-1) (N+1)^2 (N+2) (k+N+1) (k+2 N) \right.
\nonu\\
&&  (k+2 N+1)(3 k+4 N+3)(4 k+5 N+4)
/
\left( N^3 (2 k+N)  (7 k^2 N+107 k^2+14 k N^2  \right.
\nonu\\
&& + \left. \left.  
221 k N+107 k+114 N^2+114 N) \right) \right],
%
\nonu \\ 
&& <\overline{\cal{O}}_{-} {\cal{O}}_{-} W^{(5)}> =
\nonu \\
&& 
\left[ 24 i k (k+1) (N-4) (N-3) (N-1) (N+1)^2 (N+2) (k+N) (k+2 N) (3 k+2 N) \right.
\nonu\\
&&  (4 k+3 N) 
/
\left( N^3 (2 k+3 N+2) (7 k^2 N+107 k^2+14 k N^2+221 k N+107 k+114 N^2 
\right. 
\nonu\\
&&+ \left. \left. 114 N )\right) \right].
\label{threepointnk}
\eea
Obviously, the previous results (\ref{threepoint1}) and (\ref{threepoint2})
can be obtained from (\ref{threepointnk}) with the appropriate limit.

From (\ref{exact1}) and (\ref{exact2}) or (\ref{threepointnk}),
the ratio between the three-point functions at a finite $N$ and $k$,
can be expressed as
\bea
&& \frac{<\overline{\cal{O}}_{+} {\cal{O}}_{+} W^{(5)}>}{
<\overline{\cal{O}}_{-} {\cal{O}}_{-} W^{(5)}>}
= \nonu \\
&& 
-\frac{(k+N+1) (k+2 N+1) (2 k+3 N+2) (3 k+4 N+3) (4 k+5 N+4)}{k (k+N) (2 k+N) (3 k+2 N) (4 k+3 N)}.
\label{ratio}
\eea
Of course, the large $N$ behavior in (\ref{largeNratio})
can be observed from this general expression by taking the appropriate limit 
as before. In other words, (\ref{largeNratio}) is obtained once  
the numerical constants, $1, 1, 2, 3, 4$ in the numerator
of (\ref{ratio}) are ignored. 
This result comes from  the 
$W_N$ coset CFT at a finite $N$ and $k$, and corresponds to
the ratio of three-point functions in the deformed $AdS_3$ bulk theory
\footnote{For the lower spin currents, the eigenvalues and three-point functions can also be analyzed.
For the spin-$4$ current (\ref{primaryspin4}) or (\ref{modifiedspin4}),  
the following result is described in Appendix $F$ (\ref{spin4result})
\bea
W^{(4)}_0 |(f;0)>&=& \left[
\frac{2 (k+1) (N-3) (N^2-1) (k+2 N) (k+2 N+1) (3 k+4 N+3)}{N^2 
(k+N) (2 k+N) \, d(N,k)}\right] 
|(f;0)>,
\nonu\\
W^{(4)}_0 |(0;f)>&=& \left[ \frac{2 k (k+1) (N-3) (N^2-1) (k+2 N) (3 k+2 N)}{
N^2 (k+N+1)  (2 k+3 N+2) \, d(N,k)} \right] |(0;f)>,
\label{fourspin}
\eea
where $d(N,k )  \equiv  
17 k+17 k^2+22 N+39 k N+5 k^2 N+22 N^2+10 k N^2$ is introduced.
Note the presence of the factor $(N-3)$.
All the terms in (\ref{primaryspin4}) or (\ref{modifiedspin4})
contribute to the first eigenvalue, whereas 
three terms in (\ref{primaryspin4}) (or four terms in (\ref{modifiedspin4})) 
can survive in the 
second eigenvalue in (\ref{fourspin}).
For the spin-$3$ current (\ref{primaryspin3}), 
the two eigenvalue equations from  Appendix $F$ (\ref{spin3result}) are 
\bea
W^{(3)}_0 |(f;0)>&=&- \left[\frac{i}{N^2} 
(N^2-4) (N^2-1) (k+N+1) (k+2 N+1) (2 k+3 N+2) 
\right]|(f;0)>,
\nonu\\
W^{(3)}_0 |(0;f)>&=& \left[ \frac{i}{N^2} 
k (N^2-4) (N^2-1) (k+N) (2 k+N) \right] |(0;f)>.
\label{threespin}
\eea
The first term of (\ref{primaryspin3}) can contribute to 
the second eigenvalue of (\ref{threespin}). Note the factor $(N-2)$
in (\ref{threespin}). 
Finally, the spin-$2$ current (\ref{stressT}) 
has the following eigenvalue equations with Appendix $F$ (\ref{spin2result})
\bea
T_0 |(f;0)>= \left[ \frac{(N-1) (k+2 N+1)}{2 N (k+N)} \right] |(f;0)>,
\qquad
T_0 |(0;f)>= \left[ \frac{k (N-1)}{2 N (k+N+1)} \right] |(0;f)>.
\label{twospin}
\eea
The eigenvalue equations for the general spin-$s$ are difficult to obtain from
(\ref{exact1}), (\ref{exact2}), 
(\ref{fourspin}), (\ref{threespin}), and (\ref{twospin}).}.

In \cite{GG1,GG2}, the minimal representation of ${\cal W}_{\infty}[\mu]$, where 
the fewest number of low lying states exists, was studied.
Indirectly, the eigenvalue of the zero mode of the spin-$5$ current, $x$,
can be expressed in terms of the spin (or conformal dimension), 
$h$, the eigenvalue of spin-$3$ current, 
$w$, the eigenvalue of spin-$4$ current, $u$ (note that four times
their spin-$4$ field is
equal to the spin-$4$ current in the present study and see $(A.22)$ of \cite{GG1}), 
the central charge $c$, the 
constant, $N_3$, and 
the structure constant, $N_4=\frac{75}{16 \cdot 56} N_3^2 (C_{33}^4)^2$,
which depends on $(c, \mu)$.
Explicitly, equations $(B.8), (B.15), (B.16)$ and $(B.17)$ of \cite{GG1}
(where the version $2$  of arXiv 
corrected some typographical errors that appeared in the version $1$ of arXiv)
contain the eigenvalue of the zero mode of the spin-$5$ current, $x$. 
By substituting the following quantities
\bea
N_4  & = &  
\frac{6 (1+k) (-3+N) (1+N) (k+2 N) (3 k+2 N) (3+3 k+4 N)}{7 (-2+N) (2 k+N) (2+2 k+3 N) \, d(N,k)}, \nonu \\
d(N,k ) & \equiv & 
(17 k+17 k^2+22 N+39 k N+5 k^2 N+22 N^2+10 k N^2), 
\nonu \\
 N_3  & = & \frac{2}{5},
\nonu \\
h(f;0)  & = &  \frac{(N-1) (k+2 N+1)}{2 N (k+N)},
\nonu \\
w(f;0) & =  &
-\frac{i}{N^2} (N^2-4) (N^2-1) 
(k+N+1) (k+2 N+1) (2 k+3 N+2),
\nonu \\
u(f;0) & = & \frac{2 (k+1) (N-3) (N^2-1) (k+2 N) (k+2 N+1) (3 k+4 N+3)}
{N^2 (k+N) (2 k+N) \, d(N,k)
},
\nonu \\
c & = & 
\frac{k (-1+N) (1+k+2 N)}{(k+N) (1+k+N)},
\label{relation}
\eea
into the formula $(B.15)$ of \cite{GG1}
\bea
20 \times 
w \left[ \frac{4N_4}{25N_3} -\frac{208N_4}{25N_3(c+\frac{114}{7})} 
(h+\frac{3}{7}) +\frac{1}{4} \times \frac{6u}{5h} \right],
\label{xexp}
\eea
the eigenvalue given in (\ref{exact1}) can be derived.
Note that $B$ times 
a spin-$5$ current is equal to $4 \times 5 (=20)$ times
their spin-$5$ field. This is why  
the extra numerical factor $20$ is placed in (\ref{xexp}).
In addition, the eigenvalues in the expressions 
(\ref{fourspin}), (\ref{threespin}) and (\ref{twospin}) 
are substituted in (\ref{relation}).
Similarly, using the following eigenvalues with the other quantities in 
(\ref{relation}) 
\bea
h(0;f)  & = & \frac{k (N-1)}{2 N (k+N+1)},
\nonu \\
w(0;f) & =  & \frac{i}{N^2} k (N^2-4) (N^2-1) 
(k+N) (2 k+N),
\nonu \\
u(0;f) & = &  \frac{2 k (k+1) (N-3) (N^2-1) (k+2 N) (3 k+2 N)}{
N^2 (k+N+1)(2 k+3 N+2) \, d(N,k)},
\label{relation1}
\eea
the expression (\ref{xexp}) leads to the other eigenvalue in (\ref{exact2}).
Checking that the eigenvalues for the spin-$2,3,4$ fields
written in terms of the conformal dimension, $h$, and the central charge $c$ 
in \cite{GG1} coincide with those expressed as $N$ and $k$
in (\ref{relation}) and (\ref{relation1}) is quite simple.

The following ratios between the three-point functions 
for spin $s=2, 3, 4$ can be obtained from (\ref{twospin}), (\ref{threespin}), and (\ref{fourspin})
\bea
\frac{<\overline{\cal{O}}_{+} {\cal{O}}_{+} T>}{
<\overline{\cal{O}}_{-} {\cal{O}}_{-} T>}
& = & \frac{(k+N+1) (k+2 N+1)}{k (k+N)},
\nonu \\
\frac{<\overline{\cal{O}}_{+} {\cal{O}}_{+} W^{(3)}>}{
<\overline{\cal{O}}_{-} {\cal{O}}_{-} W^{(3)}>}
& = & -\frac{(k+N+1) (k+2 N+1) (2 k+3 N+2)}{k (k+N) (2 k+N)},
\nonu \\
\frac{<\overline{\cal{O}}_{+} {\cal{O}}_{+} W^{(4)}>}{
<\overline{\cal{O}}_{-} {\cal{O}}_{-} W^{(4)}>}
& = & 
\frac{(k+N+1) (k+2 N+1) (2 k+3 N+2) (3 k+4 N+3)}{k (k+N) (2 k+N) 
(3 k+2 N)}.
\label{Expp}
\eea
At $s=5$, equation (\ref{ratio}) can be derived.
As a spin $s$ increases, the extra factors
in the numerator and denominator appear up to the sign.  
This can be generalized to the arbitrary spin-$s$ current, $W^{(s)}$, 
whose ratio is expressed as
\bea
\frac{<\overline{\cal{O}}_{+} {\cal{O}}_{+} W^{(s)}>}{
<\overline{\cal{O}}_{-} {\cal{O}}_{-} W^{(s)}>}
= (-1)^s \frac{(k+N+1)}{(k+N)} \prod_{n=1}^{s-1} \left[
\frac{n k +(n+1) N +n}{n k +
(n-1) N} \right].
\label{arbitrays}
\eea
Of course, this generalizes the fixed $\lambda=\frac{1}{2}$ case in  
bulk theory \cite{CY}  and the arbitrary $\lambda$ in the boundary/bulk theory
\cite{AKP} to the  finite 
$N$ and $k$. In other words, for fixed spin $s$,
each factor in the numerator of (\ref{arbitrays}) has the extra numerical 
constants, $1, 1, 2, 3, 4, \cdots, (s-1)$. The previous results in \cite{AKP}, where they assumed 
${\cal W}_{\infty}[\lambda]$ symmetry in the CFT computations, can be obtained if these constants are ignored.

\subsection{More general coset model}

The more general coset model, $\frac{G}{H}$, 
can be described by two arbitrary levels $(k_1,k_2)$
\bea
\frac{G}{H}=
\frac{\widehat{SU}(N)_{k_2} \oplus \widehat{SU}(N)_{k_1}}
{\widehat{SU}(N)_{k_2+k_1}}.
\label{generalcoset}
\eea
Many additional extra currents can be expected in the general coset model (\ref{generalcoset}). 
The corresponding algebra should be larger than the conventional $W_N$ algebra.
For example,
the explicit form of the structure constant, 
$C_{33}^{4}$, is not known so far 
in the general coset model. 
Recall that the spin-$3$ and spin-$4$ currents with arbitrary 
levels $(k_1,k_2)$
are known 
in the general coset model. If the central charge term 
of the OPE between the spin-$4$ current with itself (which is equal to 
$\frac{c}{4}$ with the central charge in the general coset model) is calculated,  
the explicit form for the structure constant, $C_{33}^{4}$, 
in the general coset model can be determined completely.

The explicit forms of $C_{33}^{4}$ for some fixed $N$ 
from the Thielemans package \cite{Thielemans} are presented.
The left hand side of (\ref{cosetspin3spin4}) is known 
for an arbitrary $(k_1, k_2)$ with a fixed $N=4, 5$, and 
two arbitrary parameters in front of $\eta_{\Lambda}$ and $\eta_{W}$ are introduced.
The spin-$5$ current can be written as in (\ref{spin5construction})
with two parameters.  Now the OPE between $T(z)$
and the spin-$5$ current can be calculated by requiring the primary condition.
The above two parameters can be fixed completely and the structure
constants for a fixed $N=4,5$ with arbitrary levels are
as follows:
\bea
\left[ C_{33}^4 (N=5,k_1,k_2) \right]^2&=&
\left[ 64 (k_1^4 (2771 k_2^2+10530 k_2+2575)+k_1^3 (5542 k_2^3+76480 k_2^2
 \right.
\nonu\\
&+&  215750 k_2+51500)+k_1^2 (2771
   k_2^4+76480 k_2^3+600725 k_2^2+1399250 k_2 
   \nonu\\
   &+&   406250)+10 k_1 (1053 k_2^4+21575 k_2^3+139925 k_2^2+309625
   k_2+148750)
   \nonu\\
   &+& \left.  25 (k_2+5)^2 (103 k_2^2+1030 k_2+3375))
\right] 
/ 
\left[ 21 (2 k_1+5) (2 k_2+5) \right.
\nonu\\
&  \times &
(2 k_1^3 (71 k_2+55)+k_1^2 
(284 k_2^2+2705 k_2+1925)+k_1 (142 k_2^3+2705
   k_2^2 
   \nonu\\
   &+& \left. 
13675 k_2+11000)+55 (k_2+5)^2 (2 k_2+15)) \right],
\nonu \\
\left[ C_{33}^4(N=4,k_1,k_2) \right]^2&=&
\left[ 8 (k_1^4 (409 k_2^2+1296 k_2
+320)+k_1^3 (818 k_2^3+9136 k_2^2+21376 k_2  \right.
\nonu\\
&+&   5120)+k_1^2 (409 k_2^4+9136
   k_2^3+58240 k_2^2+112976 k_2+33664)
   \nonu\\
   &+&  16 k_1 (81 k_2^4+1336 k_2^3+7061 k_2^2+13280 k_2+6592)
   \nonu\\
   &+&\left.  64 (k_2+4)^2 (5
   k_2^2+40 k_2+126)) \right]
   /
   \left[ 3 (k_1+2) (k_2+2) (k_1^3 (97 k_2+88) 
   \right. \nonu\\
   &+& 2 k_1^2 (97 k_2^2+767 k_2+616)+k_1 (97 k_2^3+1534 k_2^2+6592
   k_2+5632)
   \nonu\\
   &+&\left.  88 (k_2+4)^2 (k_2+6)) \right].
\label{aboveexpr}
\eea
These structure constants are reduced to the 
structure constants of the minimal model when $k_1$ is fixed to $1$ 
(Of course, $k_2=k$).
Furthermore,  
the highest singular term of the spin-$4$ current 
(when $N=4$) was checked to ensure that it behaves correctly with (\ref{aboveexpr}).

\section{Conclusions and outlook }
 
The spin-5 Casimir operator (\ref{spin5detail}) was obtained by calculating the second-order pole in the OPE 
between 
the spin-$3$ Casimir operator (\ref{primaryspin3}) 
and the spin-$4$ Casimir operator (\ref{modifiedspin4}).   
The three-point functions 
with two scalars  (\ref{threepoint1}) and (\ref{threepoint2}) 
for all values of the 't Hooft
coupling in the large $N$ limit (and without any limit) were obtained by analyzing the zero-mode eigenvalue equations carefully.
These three-point functions (\ref{threepointnk}) were dual to those 
in the $AdS_3$ higher spin gravity theory with matter. 
A future study should calculate the corresponding three-point functions 
in the bulk theory.

$\bullet$ Although the spin-$5$ current contains 
the multiple product between the $d$, $f$ symbol of rank $3$, or the $\delta$ symbol
of rank $2$, one does not observe  
a completely symmetric traceless $d$ symbol of rank $5$.
A future study should examine the generalization of (\ref{dabcd}) to a 
higher rank. From the second order pole of 
the OPE between the spin-$3$ current and the spin-$s$ 
current, where $s \geq 4$, one can imagine that a higher $d$ symbol of
rank $s$
can be written in terms of the multiple product of $d$ symbol of rank $3$, where 
the number of $d$ symbols is $(s-2)$. 
The generalization of \cite{macfar} to the multiple product of 
$d$ symbols is needed.

$\bullet$ Because the spin-$4$ current is known for any arbitrary 
level $(k_1,k_2)$,
it would be interesting to 
calculate the eighth-order pole of the OPE between the spin-$4$ current 
and itself, as mentioned before.
Once this is done,
the structure constant $C_{33}^4$ in the general coset model 
can then be obtained.
Of course, this structure constant will reduce to the previous result 
(\ref{C334}) 
in the 
coset minimal model (\ref{cosetexp}) when $k_1=1$ and $k_2=k$.
Furthermore, for $N=4$ or $N=5$, one should obtain the relations 
(\ref{aboveexpr}). 

$\bullet$ What of the spin-$6$ current?
From the vacuum character computation described before, 
there are eleven nonprimary
spin-$6$ currents that can be constructed from 
the spin-$2,3, 4,5$ currents (with some derivatives) 
and a single primary spin-$6$ current. 
This current might be constructed by calculating the OPE between the 
spin-$3$ current (\ref{primaryspin3}) and spin-$5$ current 
(\ref{spin5detail}).
Definitely,
the independent terms will be greater than $52$ terms for the spin-$5$ 
current.
At least, relation (\ref{arbitrays}) for $s=6$ can be observed.
Furthermore, because of this relation, one also expects that 
the eigenvalue equations for two coset primary states has factorized forms,
as in the other lower higher spin cases.
In a bulk theory point of view, the asymptotic quantum symmetry algebra
$\cal{W}_{\infty}[\mu]$ in \cite{GG1} 
should be extended to the algebra containing 
the commutator between the 
spin-$3$ Laurent mode and the spin-$5$ Laurent mode. 
The previous work given in \cite{GHJ} (classical algebra) 
is expected to be helpful in obtaining this 
commutator. When $N=6$, the above construction should be related to 
the work reported elsewhere \cite{Hornfeck92-1}, where $W_6$ algebra was studied.

$\bullet$ Because the spin-$3,4,5$ Casimir 
currents are known for the coset model 
(\ref{cosetexp}), one can also check whether the other structure constants,
$C_{44}^4, C_{45}^5$ in \cite{BEHHH}, can be expressed in terms of 
$C_{33}^4$ (\ref{C334}). 
The former can be obtained from the highest singular
term in the OPE of the spin-$4$ current and itself and the latter can be obtained 
from the OPE of the spin-$4$ current and the spin-$5$ current as well as
the OPE between the spin-$5$ current and itself. 
In principle, the other OPEs, spin-$3$ spin-$5$, spin-$4$ spin-$4$, 
spin-$4$ spin-$5$, and spin-$5$ spin-$5$, can be  obtained 
and the algebra should be 
related to the $W_5$ algebra \cite{Hornfeck92-1}, where $N$ is fixed to $N=5$.
Extracting several quasi-primary fields in these OPEs is nontrivial.
For example, 
the OPE between the spin-$5$ current and itself provides the second order pole,
where the possible quasi-primary fields of spin-$8$ arise.
Even in the well-known quantum Miura transformations \cite{BS}, 
the formula appears quite complicated.
As mentioned in the introduction, the general procedure to obtain 
the correct quasi-primaries is based on previous studies 
 \cite{Bowcock,BFKNRV,Nahm1,Nahm2,Ahn1211}.

$\bullet$ The ${\cal N}=1$ supersymmetric coset model can be obtained by taking 
one of the levels to be $N$ in the context of that reported elsewhere \cite{Ahn1211,Ahn1305,ASS}. 
The immediate question in this direction is what are the spin contents of
this coset model?
A future study should examine this particular coset model
because this might be the possible dual theory for the $AdS_3$ string theory. 
Determining the $N$-generalization for the character 
technique is an open problem.

$\bullet$ The duality in \cite{GG,GG1} has been generalized
to the duality with the other group, which is analogous to the $O(N)$ vector model
in one dimension higher \cite{Ahn1106,GV}. 
In \cite{Ahn1202,AP}, the explicit Casimir currents are constructed.
In contrast to the present work, the Casimir construction with orthogonal group
has the symmetric $SO(N)$ invariant tensor of rank $2$. 
In other words, the general $N$-dependence in the OPEs  
can be read off from the fixed several $N$ case results.
No complicated contractions were observed between the $d$ symbols  
in the $SU(N)$
coset model. 

$\bullet$ The large ${\cal N}=4$ holography was found recently in \cite{GG1305}.
The 2-dimensional CFT has many supersymmetries.   
Sometimes it is useful for describing the explicit construction of
Casimir operators in ${\cal N}=2$ superspace. For example, 
in the context of the ${\cal N}=2$ Kazama-Suzuki model. Some relevant descriptions 
can be found in \cite{Ahn1206,Ahn1208}.
Thus far, it is unknown how to construct the ${\cal N}=4$ extension of
$W_3$ algebra. Among the sixteen currents (in the spin contents, they 
are given by four spin-$\frac{1}{2}$ currents, 
seven spin-$1$ currents, four spin-$\frac{3}{2}$
currents and a spin-$2$ current) 
in this large ${\cal N}=4$ superconformal
algebra, only eleven currents, after projecting out four 
spin-$\frac{1}{2}$ currents and a spin-$1$ current, play an important role   
in higher spin theory.
The possible lowest higher spin current contains 
the following spin contents
$(1,\frac{3}{2},\frac{3}{2},2)$, $(\frac{3}{2},2,2,\frac{5}{2})$, 
$(\frac{3}{2},2,2,\frac{5}{2})$ and $(2,\frac{5}{2},\frac{5}{2},3)$ in 
${\cal N}=2$ superspace notation.
In a future study, it would be very interesting to construct the above eleven currents 
living in 
large ${\cal N}=4$ nonlinear algebra and the above sixteen currents
that act as extra higher spin currents  
in the specific coset model suggested in \cite{GG1305}.
Therefore, it would be interesting to find 
$(11+16)$ currents explicitly in terms of the coset WZW primaries. 

\vspace{.7cm}

\centerline{\bf Acknowledgments}

CA would like to thank M.R. Gaberdiel
for discussions. 
This work was supported by the Mid-career Researcher Program through
the National Research Foundation of Korea (NRF) grant 
funded by the Korean government (MEST) (No. 2012-045385).
This research was supported by Kyungpook National University 
Research Fund 2012.
CA acknowledges warm hospitality from 
the School of  Liberal Arts (and Institute of Convergence Fundamental
Studies), Seoul National University of Science and Technology.

\newpage

\appendix

\renewcommand{\thesection}{\large \bf \mbox{Appendix~}\Alph{section}}
\renewcommand{\theequation}{\Alph{section}\mbox{.}\arabic{equation}}

\section{ Coefficient functions appearing in the spin-$4$ current }

The spin-$4$ current can be obtained by calculating the second-order pole in the 
OPE between the spin-$3$ current and itself.
In this calculation,
the quasi-primary field of spin-$4$ and
the descendant term of stress energy tensor of spin-$4$ with the correct 
coefficients should be subtracted from the above 
second-order pole.  
The coset spin-$4$ primary field is given by \cite{Ahn1111}
\bea
&& W^{(4)}(z) = 
\nonu \\
&& d^{abcd} \left( c_1  J^a J^b J^c J^d +
c_2  J^a J^b J^c K^d +
c_3  J^a J^b K^c K^d
+
c_4 J^a K^b K^c K^d +c_5  K^a K^b K^c K^d \right)(z) 
\nonu \\
&& +  
d^{abe} d^{cde} \left( c_6 J^a J^b J^c J^d+  
c_7   J^a J^b J^c K^d 
+
c_8   J^a J^b K^c K^d
+   c_9  J^a K^b K^c K^d + 
c_{10}  K^a K^b K^c K^d \right)(z)
\nonu \\
&& +  c_{11} J^a J^a J^b J^b(z) +
c_{12} J^a J^a J^b K^b(z) 
 + 
c_{13} J^a J^a K^b K^b(z)  
+
c_{14}  J^a K^a K^b K^b(z) 
\nonu \\
&& + 
c_{15} K^a K^a K^b K^b(z) 
+
c_{17} \pa^2 J^a K^a(z) 
+ c_{20} \pa J^a \pa K^a(z)   
+ c_{22}  J^a \pa^2 K^a(z) \nonu \\
&& +  
c_{23} J^a J^b K^a K^b(z),
\label{primaryspin4}
\eea
where the coefficient functions $c_i$ are given in \cite{Ahn1111} or 
in Appendix 
$A$ (\ref{a-1})-(\ref{a-23}).  
As the OPE between the spin-$3$ current (\ref{primaryspin3}) 
and spin-$4$ current (\ref{primaryspin4}) is calculated, 
the identity between the $d$ symbols should be used. 
Therefore, it is better to rewrite the above spin-$4$ current 
using the $d$ symbol of rank $3$. 
The various identities of these $d$ symbols of rank $3$
can be used.

Using the identity \cite{Ahn1111}
\bea
&& d^{abcd} J^a J^b J^c J^d(z)   =   
  3 d^{abe} d^{cde} J^a J^b J^c J^d(z)
-\frac{12(N^2-4)}{N(N^2+1)} J^a J^a J^b J^b(z)  
\nonu\\
&& -  \frac{3(N^2-4)(N^2-3)}{N^2+1}  
\pa J^a \pa J^a(z)
+\frac{2(N^2-4)(N^2-3)}{N^2+1} \pa^2 J^a J^a(z),
\label{dabcdFormula}
\eea
the first term of (\ref{primaryspin4}) can be 
decomposed into the right hand side of (\ref{dabcdFormula}), where
the following identity is used \cite{Schoutens}
\bea
d^{abcd} = d^{abe} d^{ecd} +d^{ace} d^{ebd} +d^{ade} d^{ebc}
-\frac{4(N^2-4)}{N(N^2+1)} \left( \delta^{ab} \delta^{cd} +\delta^{ac}
\delta^{bd} +\delta^{ad} \delta^{bc} \right).
\label{dabcd}
\eea
The first term of (\ref{primaryspin4})
can be absorbed into the sixth- and eleventh-terms of (\ref{primaryspin4})
besides the derivative terms.

One can calculate similar identities as follows \cite{Ahn1111}:
\bea
 d^{abcd} J^a J^b J^c K^d(z)   &=& 
  3 d^{abe} d^{cde} J^a J^b J^c K^d(z)
-\frac{12(N^2-4)}{N(N^2+1)} J^a J^a J^b K^b(z) \nonu \\
 &-&\frac{3(N^2-4)(N^2-3)}{N(N^2+1)} f^{abc} \pa J^a J^b K^c(z)  
+\frac{2(N^2-4)(N^2-3)}{N^2+1} \pa^2 J^a K^a(z),  
\nonu \\
 d^{abcd} J^a J^b K^c K^d(z)   &=& 
  d^{abe} d^{cde} J^a J^b K^c K^d(z)+
2d^{ace} d^{bde} J^a J^b K^c K^d(z) \nonu \\
&-&\frac{4(N^2-4)}{N(N^2+1)}J^a J^a K^b K^b(z) 
-\frac{8(N^2-4)}{N(N^2+1)}J^a J^b K^a K^b(z) 
\nonu\\
&-&\frac{(N^2-4)(N^2-3)}{N^2+1} \pa J^a \pa K^a(z), 
\nonu\\
d^{abcd} J^a K^b K^c K^d(z)   &=& 
  3 d^{abe} d^{cde} J^a K^b K^c K^d(z)
-\frac{12(N^2-4)}{N(N^2+1)} J^a K^a K^b K^b(z) \nonu \\
 &+&\frac{3(N^2-4)(N^2-3)}{N(N^2+1)} f^{abc} \pa K^a K^b J^c(z)  
-\frac{(N^2-4)(N^2-3)}{N^2+1} J^a  \pa^2 K^a(z), 
\nonu\\
d^{abcd} K^a K^b K^c K^d(z)   &=&  d^{abcd} J^a J^b J^c J^d(z) |_{J^a \leftrightarrow K^a}.
\label{dabcdFormula2}
\eea
Checking whether the second term of (\ref{primaryspin4})
can be absorbed into the seventh- and twelfth-terms of (\ref{primaryspin4}) is relatively simple.
Furthermore, the $f$ symbol term (due to the normal ordering of the currents
$J^a(z)$ and $K^a(z)$) 
in the first equation of (\ref{dabcdFormula2}) 
newly arises. 
The third term of (\ref{primaryspin4}) can contribute to 
the eighth- and thirteenth-terms and the two new independent terms
appear.  
Similarly, the fourth-term of (\ref{primaryspin4})
can be absorbed into the ninth- and fourteenth-terms.
In this case, the $f$ symbol term also arises. 
The symmetry 
between the spin-$1$ currents is used at the final equation of (\ref{dabcdFormula2}). That is, 
the left hand side of that equation can be obtained from 
(\ref{dabcdFormula}) by changing $J^a(z)$ to $K^a(z)$ and vice versa.

The following presents the various coefficient functions in (\ref{primaryspin4})
that appeared in \cite{Ahn1111}:
\bea
c_1&=&
\left[ -k_2 (N^2+1) (k_1^2 k_2 N^2-3 k_1^2 k_2-2
   k_1^2 N+k_1 k_2^2 N^2-3 k_1 k_2^2+2 k_1 k_2
   N^3-8 k_1 k_2 N  \right.
   \nonu\\
   &-&  \left. 4 k_1 N^2-2 k_2^2 N-4 k_2 N^2-2
   N^3) \right]/
   \left[ 2 (N-2) (N+2) (N^2-3) (k_1+N) (k_1+k_2+N) \right.
   \nonu\\
   &\times&\left.  D(k_1, k_2, N) \right],
\label{a-1}
\\
c_2&=& \left[ -2 N (N^2+1) (10 k_1^3 k_2 N^2-30 k_1^3 k_2-20
   k_1^3 N+30 k_1^2 k_2^2 N^2-90 k_1^2 k_2^2+35 k_1^2
   k_2 N^3 \right. 
   \nonu\\
   &-& 133 k_1^2 k_2 N-38 k_1^2 N^2+20 k_1 k_2^3
   N^2-60 k_1 k_2^3+55 k_1 k_2^2 N^3-161 k_1 k_2^2
   N+30 k_1 k_2 N^4
   \nonu\\
   &-&\left.  96 k_1 k_2 N^2+24 k_2^3 N+50 k_2^2
   N^2+44 k_2 N^3+18 N^4) \right]
   /
   \left[ 3 (N-2) (N+2) (N^2-3) \right.
   \nonu\\
   &\times&\left.  (2 k_1+N) (k_1+k_2+N) (2 k_1+2
   k_2+3 N) D(k_1, k_2, N) \right] ,
\label{a-2} \\
c_3&=&
\frac{N}{(N-2) (N+2) (k_1+k_2+N) (2 k_1+2 k_2+3 N)},
\label{a-3}
\\
c_4&=&
\left[ 2 N (N^2+1) (5 k_1^2 k_2 N^2-15 k_1^2 k_2-10
   k_1^2 N+5 k_1 k_2^2 N^2-15 k_1 k_2^2+10 k_1
   k_2 N^3  \right.
   \nonu\\
   &-&\left.  56 k_1 k_2 N-36 k_1 N^2-10 k_2^2 N-36 k_2
   N^2-26 N^3) \right] 
/
\left[ 3 (N-2) (N+2) (N^2-3)  \right.
\nonu\\
&\times&\left.  (k_1+k_2+N) (2 k_1+2 k_2+3 N) D(k_1, k_2, N) \right],
\label{a-4}\\
c_5&=&
\left[ -k_1 (N^2+1) (k_1^2 k_2 N^2-3 k_1^2 k_2-2
   k_1^2 N+k_1 k_2^2 N^2-3 k_1 k_2^2+2 k_1 k_2
   N^3-8 k_1 k_2 N  \right.
   \nonu\\
   &-&\left.  4 k_1 N^2-2 k_2^2 N-4 k_2 N^2-2
   N^3) \right]
/
\left[ 2 (N-2) (N+2) (N^2-3) (k_2+N) (k_1+k_2+N)  \right.
\nonu\\
&\times&\left. D(k_1, k_2, N) \right],
\label{a-5} \\
c_6&=&
\left[ k_2 (6 k_1^4 k_2 N^4-12 k_1^4 k_2 N^2-18 k_1^4
   k_2-12 k_1^4 N^3-12 k_1^4 N+12 k_1^3 k_2^2 N^4-24
   k_1^3 k_2^2 N^2  \right.
   \nonu\\
   &-& 36 k_1^3 k_2^2+24 k_1^3 k_2 N^5-72
   k_1^3 k_2 N^3-96 k_1^3 k_2 N-48 k_1^3 N^4-48 k_1^3
   N^2+6 k_1^2 k_2^3 N^4
   \nonu\\
   &-& 12 k_1^2 k_2^3 N^2-18 k_1^2
   k_2^3+32 k_1^2 k_2^2 N^5-76 k_1^2 k_2^2 N^3-156
   k_1^2 k_2^2 N+31 k_1^2 k_2 N^6-112 k_1^2 k_2
   N^4
   \nonu\\
   &-& 255 k_1^2 k_2 N^2-58 k_1^2 N^5-102 k_1^2 N^3+8 k_1
   k_2^3 N^5-16 k_1 k_2^3 N^3-72 k_1 k_2^3 N+23 k_1
   k_2^2 N^6
   \nonu\\
   &-& 48 k_1 k_2^2 N^4-279 k_1 k_2^2 N^2+14 k_1
   k_2 N^7-42 k_1 k_2 N^5-324 k_1 k_2 N^3-20 k_1
   N^6-108 k_1 N^4
   \nonu\\
   &+&\left.  16 k_2^3 N^4-72 k_2^3 N^2+34 k_2^2 N^5-186
   k_2^2 N^3+20 k_2 N^6-156 k_2 N^4+2 N^7-42 N^5) \right]
/
\nonu\\
&\times& \left[ 
(N-2) (N+2) (N^2-3) (k_1+N) (2 k_1+N) (k_1+k_2+N) (2
   k_1+2 k_2+3 N) \right. 
   \nonu\\
   &\times&\left. D(k_1, k_2, N) \right],
\label{a-6} \\
c_7&=&
\left[ 4 N (5 k_1^3 k_2 N^4-10 k_1^3 k_2 N^2-15 k_1^3
   k_2-10 k_1^3 N^3-10 k_1^3 N+10 k_1^2 k_2^2 N^4-32
   k_1^2 k_2^2 N^2  \right.
   \nonu\\
   &+& 6 k_1^2 k_2^2+15 k_1^2 k_2 N^5-72
   k_1^2 k_2 N^3+25 k_1^2 k_2 N-30 k_1^2 N^4+14 k_1^2
   N^2+5 k_1 k_2^3 N^4-22 k_1 k_2^3 N^2
   \nonu\\
   &+& 21 k_1
   k_2^3+15 k_1 k_2^2 N^5-80 k_1 k_2^2 N^3+113 k_1
   k_2^2 N+10 k_1 k_2 N^6-90 k_1 k_2 N^4+168 k_1
   k_2 N^2
   \nonu\\
   &-& 22 k_1 N^5+66 k_1 N^3-10 k_2^3 N^3+78 k_2^3 N-30
   k_2^2 N^4+190 k_2^2 N^2-22 k_2 N^5+154 k_2 N^3
   \nonu\\
   &-&\left.  2 N^6+42
   N^4) \right]
/
\left[ (N-2) (N+2) (N^2-3) (2 k_1+N) (k_1+k_2+N) (2 k_1+2
   k_2+3 N)  \right.
   \nonu\\
   &\times&\left. D(k_1, k_2, N) \right],
\label{a-7} \\
c_8&=&\frac{2 N^2}{(N-2) (N+2) (2 k_1+N) (2 k_2+N) (2 k_1+2 k_2+3 N)},
\label{a-8}
\\
c_9&=&
\left[ -4 N (5 k_1^3 k_2 N^4+2 k_1^3 k_2 N^2-51 k_1^3
   k_2+22 k_1^3 N^3-66 k_1^3 N+10 k_1^2 k_2^2 N^4-8
   k_1^2 k_2^2 N^2  \right.
   \nonu\\
   &-& 66 k_1^2 k_2^2+15 k_1^2 k_2 N^5+12
   k_1^2 k_2 N^3-211 k_1^2 k_2 N+50 k_1^2 N^4-170
   k_1^2 N^2+5 k_1 k_2^3 N^4
   \nonu\\
   &-& 10 k_1 k_2^3 N^2-15 k_1
   k_2^3+15 k_1 k_2^2 N^5-28 k_1 k_2^2 N^3-155 k_1
   k_2^2 N+10 k_1 k_2 N^6-2 k_1 k_2 N^4
   \nonu\\
   &-& 280 k_1
   k_2 N^2+26 k_1 N^5-150 k_1 N^3-10 k_2^3 N^3-10 k_2^3 N-30
   k_2^2 N^4-74 k_2^2 N^2-22 k_2 N^5
   \nonu\\
   &-& \left. 110 k_2 N^3-2 N^6-46
   N^4) \right]
/
\left[ (N-2) (N+2) (N^2-3) (2 k_2+N) (k_1+k_2+N)  \right.
\nonu\\
&\times&\left. (2 k_1+2 k_2+3 N) D(k_1, k_2, N) \right],
\label{a-9} \\
c_{10}&=&
\left[ k_1 (6 k_1^3 k_2^2 N^4-12 k_1^3 k_2^2 N^2-18
   k_1^3 k_2^2+8 k_1^3 k_2 N^5-16 k_1^3 k_2 N^3-72
   k_1^3 k_2 N+16 k_1^3 N^4  \right.
   \nonu\\
   &-& 72 k_1^3 N^2+12 k_1^2
   k_2^3 N^4-24 k_1^2 k_2^3 N^2-36 k_1^2 k_2^3+32
   k_1^2 k_2^2 N^5-76 k_1^2 k_2^2 N^3-156 k_1^2 k_2^2
   N
   \nonu\\
   &+& 23 k_1^2 k_2 N^6-48 k_1^2 k_2 N^4-279 k_1^2 k_2
   N^2+34 k_1^2 N^5-186 k_1^2 N^3+6 k_1 k_2^4 N^4-12 k_1
   k_2^4 N^2
   \nonu\\
   &-& 18 k_1 k_2^4+24 k_1 k_2^3 N^5-72 k_1
   k_2^3 N^3-96 k_1 k_2^3 N+31 k_1 k_2^2 N^6-112 k_1
   k_2^2 N^4-255 k_1 k_2^2 N^2
   \nonu\\
   &+& 14 k_1 k_2 N^7-42 k_1
   k_2 N^5-324 k_1 k_2 N^3+20 k_1 N^6-156 k_1 N^4-12
   k_2^4 N^3-12 k_2^4 N
   \nonu\\
   &-&\left.  48 k_2^3 N^4-48 k_2^3 N^2-58 k_2^2
   N^5-102 k_2^2 N^3-20 k_2 N^6-108 k_2 N^4+2 N^7-42 N^5) \right]
/
\nonu\\
&\times& \left[ (N-2) (N+2) (N^2-3) (k_2+N) (2 k_2+N) (k_1+k_2+N) (2
   k_1+2 k_2+3 N)  \right.
   \nonu\\
   &\times&\left. D(k_1, k_2, N) \right],
\label{a-10}
\\
c_{11}&=&
\left[ -2 k_2 (3 k_1^2 k_2 N^2-9 k_1^2 k_2-6 k_1^2 N+3
   k_1 k_2^2 N^2-9 k_1 k_2^2+6 k_1 k_2 N^3-24
   k_1 k_2 N-12 k_1 N^2  \right.
   \nonu\\
   &+&\left.  4 k_2^2 N^3-18 k_2^2 N+4 k_2
   N^4-24 k_2 N^2-6 N^3) \right]
/
\left[ N (N^2-3) (k_1+N) (k_1+k_2+N)  \right.
\nonu\\
&\times&\left. D(k_1, k_2, N) \right],
\label{a-11} \\
c_{12}&=&
\left[ 8 (-10 k_1^3 k_2 N^2+30 k_1^3 k_2+20 k_1^3 N-14
   k_1^2 k_2^2 N^2+42 k_1^2 k_2^2-19 k_1^2 k_2 N^3+85
   k_1^2 k_2 N  \right.
   \nonu\\
   &+& 38 k_1^2 N^2-4 k_1 k_2^3 N^2+12 k_1
   k_2^3-7 k_1 k_2^2 N^3+17 k_1 k_2^2 N+2 k_1
   k_2 N^4+8 k_2^3 N^3-48 k_2^3 N
   \nonu\\
   &+&\left.  20 k_2^2 N^4-110 k_2^2
   N^2+12 k_2 N^5-80 k_2 N^3-18 N^4) \right]
/
\left[ (N^2-3) (2 k_1+N) \right.
\nonu\\
&\times&\left. (k_1+k_2+N) (2 k_1+2 k_2+3 N) D(k_1, k_2, N) \right],
\label{a-12} \\
c_{13}&=&
\left[ -4 (3 k_1^2 k_2 N^2-9 k_1^2 k_2-22 k_1^2 N+3 k_1
   k_2^2 N^2-9 k_1 k_2^2+2 k_1 k_2 N^3-44 k_1
   k_2 N-44 k_1 N^2  \right.
   \nonu\\
   &-&\left.  22 k_2^2 N-44 k_2 N^2-22 N^3) \right]
/
\left[ (N^2+1) (k_1+k_2+N) (2 k_1+2 k_2+3 N) \right.
\nonu\\
&\times&\left. D(k_1, k_2, N) \right],
\label{a-13} \\
c_{14}&=&
\left[ 8 (8 k_1^3 N^2-24 k_1^3+13 k_1^2 k_2 N^2-39 k_1^2
   k_2+20 k_1^2 N^3-70 k_1^2 N+5 k_1 k_2^2 N^2-15 k_1
   k_2^2  \right.
   \nonu\\
   &+& \left. 18 k_1 k_2 N^3-80 k_1 k_2 N+12 k_1 N^4-72
   k_1 N^2-10 k_2^2 N-36 k_2 N^2-26 N^3) \right]
/
\left[ (N^2-3) \right.
\nonu\\
&\times&\left. (k_1+k_2+N) (2 k_1+2 k_2+3 N) D(k_1, k_2, N) \right],
\label{a-14} \\
c_{15}&=&
\left[ -2 k_1 (3 k_1^2 k_2 N^2-9 k_1^2 k_2+4 k_1^2
   N^3-18 k_1^2 N+3 k_1 k_2^2 N^2-9 k_1 k_2^2+6 k_1
   k_2 N^3-24 k_1 k_2 N  \right.
   \nonu\\
   &+& \left. 4 k_1 N^4-24 k_1 N^2-6 k_2^2
   N-12 k_2 N^2-6 N^3) \right]
/
\left[ N (N^2-3) (k_2+N) (k_1+k_2+N) \right.
\nonu\\
&\times&\left. D(k_1, k_2, N) \right],
\label{a-15} \\
c_{17}&=&
\left[ 2 (6 k_1^4 k_2 N^2-18 k_1^4 k_2-12 k_1^4 N+12 k_1^3
   k_2^2 N^2-36 k_1^3 k_2^2+29 k_1^3 k_2 N^3-111 k_1^3
   k_2 N  \right.
   \nonu\\
   &-& 58 k_1^3 N^2+6 k_1^2 k_2^3 N^2-18 k_1^2 k_2^3+42
   k_1^2 k_2^2 N^3-150 k_1^2 k_2^2 N+46 k_1^2 k_2 N^4-230
   k_1^2 k_2 N^2
   \nonu\\
   &-& 88 k_1^2 N^3+13 k_1 k_2^3 N^3-51 k_1
   k_2^3 N+38 k_1 k_2^2 N^4-166 k_1 k_2^2 N^2+24 k_1 k_2
   N^5-156 k_1 k_2 N^3
   \nonu\\
   &-& \left. 42 k_1 N^4+6 k_2^3 N^2+4 k_2^2 N^3-2
   k_2 N^4) \right]
/
\left[ 3 (2 k_1+N) (k_1+k_2+N) (2 k_1+2 k_2+3 N) \right.
\nonu\\
&\times&\left. D(k_1, k_2, N) \right],
\label{a-17} \\
c_{20}&=&
\left[ -2 (3 k_1^3 k_2 N^4-6 k_1^3 k_2 N^2-9 k_1^3 k_2-6
   k_1^3 N^3-6 k_1^3 N+6 k_1^2 k_2^2 N^4-12 k_1^2 k_2^2 N^2-18
   k_1^2 k_2^2  \right.
   \nonu\\
   &+& 8 k_1^2 k_2 N^5-34 k_1^2 k_2 N^3-18 k_1^2
   k_2 N-32 k_1^2 N^4+12 k_1^2 N^2+3 k_1 k_2^3 N^4-6 k_1
   k_2^3 N^2-9 k_1 k_2^3
   \nonu\\
   &+& 8 k_1 k_2^2 N^5-34 k_1 k_2^2
   N^3-18 k_1 k_2^2 N+4 k_1 k_2 N^6-64 k_1 k_2 N^4+24
   k_1 k_2 N^2-46 k_1 N^5
   \nonu\\
   &+&\left.  42 k_1 N^3-6 k_2^3 N^3-6 k_2^3 N-32
   k_2^2 N^4+12 k_2^2 N^2-46 k_2 N^5+42 k_2 N^3-20 N^6+24 N^4) \right]
/
\nonu\\
&\times&\left[ (N^2+1) (k_1+k_2+N) (2 k_1+2 k_2+3 N) D(k_1, k_2, N) \right],
\label{a-20} \\
c_{22}&=&
\left[ 2 (k_1+k_2+2 N) (3 k_1^2 k_2 N^2-9 k_1^2 k_2-6
   k_1^2 N+3 k_1 k_2^2 N^2-9 k_1 k_2^2+2 k_1 k_2 N^3-12
   k_1 k_2 N  \right.
   \nonu\\
   &-&\left.  4 k_1 N^2-6 k_2^2 N-4 k_2 N^2+2 N^3) \right]
/
\left[ 3 (k_1+k_2+N) (2 k_1+2 k_2+3 N) \right.
\nonu \\
&\times&\left. D(k_1, k_2, N) \right],
\label{a-22} \\
c_{23}&=&
\left[ -8 (3 k_1^2 k_2 N^2-9 k_1^2 k_2+8 k_1^2
   N^3-14 k_1^2 N+3 k_1 k_2^2 N^2-9 k_1
   k_2^2+18 k_1 k_2 N^3-28 k_1 k_2 N  \right.
   \nonu\\
   &+&\left.  20
   k_1 N^4-24 k_1 N^2+8 k_2^2 N^3-14 k_2^2 N+20
   k_2 N^4-24 k_2 N^2+12 N^5-10 N^3) \right]
/
\nonu\\
&\times& \left[ (N^2+1) (k_1+k_2+N) (2 k_1+2 k_2+3 N) D(k_1, k_2, N) \right],
\label{a-23}
\eea
where 
\bea
D(k_1, k_2, N) &\equiv&  
5 k_1^2 k_2 N^2+17 k_1^2 k_2+22 k_1^2 N+5 k_1 k_2^2 N^2
+17 k_1 k_2^2+10 k_1 k_2 N^3+56 k_1 k_2 N
\nonu\\
&+& 44 k_1 N^2 + 22 k_2^2 N+44 k_2 N^2+22 N^3.
\eea
Note that  $(c_i +b_i)$ in \cite{Ahn1111} is replaced with $c_i$ here.
Of course, in the coset model (\ref{cosetexp}), the levels 
are fixed to $k_1=1$ and $k_2=k$. 

\section{ Intermediate spin-$5$ field contents in
the second-order pole of the OPE $W^{(3)}(z) W^{(4)}(w)$ }

The second-order pole in $W^{(3)}(z) W^{(4)}(w)$  (\ref{spin5construction}) 
can be calculated explicitly
using the spin-$3$ current in 
(\ref{primaryspin3}) and the spin-$4$  current in (\ref{modifiedspin4}). 
Consider the first term $Q(z) \equiv 
d^{abc}J^a J^b J^c(z)$ of (\ref{primaryspin3}) and make an OPE between $Q(z)$ and 
$21$ terms in the spin-$4$ current.
For the spin-$4$ current, only the fields that did not have a 
$d$ symbol were considered.
Twelve quartic terms can be obtained 
from the following five terms with the free indices.
\bea
J^a J^b J^c J^d(w), \quad
J^a J^b J^c K^d(w), \quad
J^a J^b K^c K^d(w), \quad
J^a K^b K^c K^d(w), \quad
K^a K^b K^c K^d(w).
\label{five}
\eea
After calculating the OPE between $Q(z)$ with 
the five fields in (\ref{five}),
the nine OPEs between $Q(z)$ with 
the remaining nine terms in the spin-$4$ currents are given. 
The following fourteen OPEs are then obtained:
\bea 
&&Q(z) \, J^a J^b J^c J^d(w)|_\frac{1}{(z-w)^2}=
-3(k_1+N) \left(Q^a J^b J^c J^d+J^a Q^b J^c J^d+J^a J^b Q^c J^d
+J^a J^b J^c Q^d\right)(w),\nonu\\
&&Q(z) \, J^a J^b J^c K^d(w)|_\frac{1}{(z-w)^2}=
-3(k_1+N) \left(Q^a J^b J^c K^d+J^a Q^b J^c K^d  
+ J^a J^b Q^c K^d \right)(w), 
\nonu\\
&&Q(z) \, J^a J^b K^c K^d(w)|_\frac{1}{(z-w)^2}=
-3(k_1+N) \left(Q^a J^b K^c K^d+J^a Q^b K^c K^d\right)(w),\nonu\\
&&Q(z)   \, J^a K^b K^c K^d(w)|_\frac{1}{(z-w)^2}=
-3(k_1+N) Q^a K^b K^c K^d(w), \nonu\\
&&Q(z) \, K^a K^b K^c K^d(w)|_\frac{1}{(z-w)^2}=0,
\nonu\\
&&Q(z) \, \pa^2 J^a K^a(w)|_\frac{1}{(z-w)^2}=-9(k_1+N) \pa^2 Q^a K^a(w),\nonu\\
&&Q(z) \, \pa J^a \pa K^a(w)|_\frac{1}{(z-w)^2}=-6(k_1+N) \pa Q^a 
\pa K^a(w),\nonu\\
&&Q(z) \, J^a \pa^2 K^a(w)|_\frac{1}{(z-w)^2}=-3(k_1+N) Q^a \pa^2 K^a(w),
\nonu\\
&&Q(z) \, f^{abc} \pa J^a J^b K^c(w)|_\frac{1}{(z-w)^2}=
-3(k_1+N) \left( 2f^{abc} \pa Q^a J^b K^c
+f^{abc} \pa J^a Q^b K^c \right)(w), 
\nonu\\
&&Q(z) \, f^{abc} \pa K^a K^b J^c(w)|_\frac{1}{(z-w)^2}=
-3(k_1+N) \pa K^a K^b Q^c(w),
\nonu\\
&&Q(z) \, \pa^2 J^a J^a(w)|_\frac{1}{(z-w)^2}=
-3(k_1+N) \left( 3 J^a \pa^2 Q^a + \pa^2 J^a Q^a \right)(w),
\nonu\\
&&Q(z) \, \pa^2 K^a K^a(w)|_\frac{1}{(z-w)^2}=0,
\nonu\\
&&Q(z) \, \pa J^a \pa J^a(w)|_\frac{1}{(z-w)^2}=
-12(k_1+N) \pa J^a \pa Q^a(w), 
\nonu\\
&&Q(z) \, \pa K^a \pa K^a(w)|_\frac{1}{(z-w)^2}=0.
\label{QUpole2}
\eea  
For the quartic terms in the left hand side, the extra 
$d$ or $\delta$ symbols are multiplied to obtain the 
final result for the second order pole.
Further simplifications for the right hand side of (\ref{QUpole2}) is needed
to analyze the zero mode eigenvalue equations.
In other words, field $Q^a(z)$ should be located at the right hand side of 
field $J^a(z)$ by moving $Q^a(z)$ to the right.
Field $K^a(z)$ can be placed to the right.
This issue will be addressed in Appendix $C$.

Let us introduce spin-$2$ operator for convenience.
\bea
L^a(z) \equiv d^{abc} J^b K^c(z).
\label{Ldef}
\eea
Consider the second term  
$d^{abc}J^a J^b K^c(z) \equiv K^e Q^e(z)$ of (\ref{primaryspin3}) and 
calculate the fourteen OPEs as follows:
\bea
&&K^e  Q^e(z)  \, J^a J^b J^c J^d(w)|_\frac{1}{(z-w)^2}
=
-(2 k_1+N) \left[ L^a J^b J^c J^d + J^a L^b J^c J^d  
+ J^a J^b L^c J^d \right.  
\nonu\\
&&+ J^a J^b J^c L^d - f^{afg}d^{beg} \pa (J^e K^f) J^c J^d 
-f^{afg}d^{ceg} J^b \pa(J^e K^f) J^d  
- f^{afg}d^{deg} J^b J^c \pa(J^e K^f) 
\nonu\\
&&-f^{bfg}d^{ceg} J^a \pa(J^e K^f) J^d
-f^{bfg}d^{deg} J^a J^c \pa(J^e K^f)  
\left.-f^{cfg}d^{deg} J^a J^b \pa(J^e K^f) \right](w)
\nonu\\
&&-f^{afg}f^{beg} (Q^e K^f) J^c J^d(w) -f^{afg}f^{ceg} J^b (Q^e K^f) J^d(w) 
-  f^{afg}f^{deg} J^b J^c (Q^e K^f)(w) 
\nonu\\
&&-f^{bfg}f^{ceg} J^a (Q^e K^f) J^d(w) 
-f^{bfg}f^{deg} J^a J^c (Q^e K^f)(w) 
-f^{cfg}f^{deg} J^a J^b (Q^e K^f)(w),
\nonu\\
&&K^e Q^e(z)  \, J^a J^b J^c K^d(w)|_\frac{1}{(z-w)^2}
=- k_2 Q^d J^a J^b J^c(w) -(2 k_1+N) \left[ K^d L^a J^b J^c 
+K^d J^a L^b J^c \right. 
\nonu\\
&&+ K^d J^a J^b L^c - f^{afg}d^{beg}K^d \pa (J^e K^f)J^c
-f^{afg}d^{ceg}K^d J^b \pa (J^e K^f) -f^{bfg}d^{ceg}K^d J^a \pa (J^e K^f) 
\nonu\\
&&\left. +f^{dfg}d^{aeg} \pa (J^e K^f) J^b J^c
+f^{dfg}d^{beg} J^a \pa (J^e K^f) J^c 
+f^{dfg}d^{ceg}  J^a J^b \pa (J^e K^f)  \right](w)
\nonu\\
&&-f^{afg}f^{beg}K^d (Q^e K^f) J^c(w)
-f^{afg}f^{ceg}K^d J^b (Q^e K^f)(w)
-f^{bfg}f^{ceg}K^d J^a (Q^e K^f)(w) 
\nonu\\
&&+f^{dfg}f^{aeg} (Q^e K^f) J^b J^c(w)
    +f^{dfg}f^{beg} J^a (Q^e K^f)  J^c(w)
    +f^{dfg}f^{ceg} J^a J^b (Q^e K^f)(w),
\nonu\\
&&K^e Q^e(z)  \, J^a J^b K^c K^d(w)|_\frac{1}{(z-w)^2}
=-k_2 J^a J^b Q^c K^d(w)
-k_2 J^a J^b K^c Q^d(w) 
-(2 k_1 + N) \left[ L^a J^b K^c K^d  \right.
\nonu\\
&&+\left. J^a L^b K^c K^d - f^{afg}d^{beg} \pa (J^e K^f) K^c K^d \right](w)
+k_2 \left( f^{ace} J^b \pa Q^e K^d 
+ f^{ade} J^b K^c \pa Q^e \right.
\nonu\\
&&+ f^{bce} J^a \pa Q^e K^d 
+ \left. f^{bde} J^a K^c \pa Q^e -  f^{cde} J^a J^b \pa Q^e  \right)(w)
-f^{afg}f^{beg} ( Q^e K^f ) K^c K^d(w)
\nonu\\
&&+f^{cfg}f^{aeg} J^b ( Q^e K^f ) K^d(w)
+f^{dfg}f^{aeg} J^b K^c ( Q^e K^f )(w)
+f^{beg}f^{cfg} J^a ( Q^e K^f ) K^d(w)
\nonu\\
&&+f^{beg}f^{dfg} J^a K^c ( Q^e K^f )(w)
-f^{ceg}f^{dfg} J^a J^b ( Q^e K^f )(w),
\nonu\\
&&K^e Q^e(z)  \, J^a K^b K^c K^d(w)|_\frac{1}{(z-w)^2}
=
-(2 k_1+N) L^a K^b K^c K^d(w)
-k_2 \left( J^a Q^b K^c K^d + J^a K^b Q^c K^d \right.
\nonu\\
&&+ J^a K^b K^c Q^d - f^{abe} \pa Q^e K^c K^d
-f^{ace} K^b \pa Q^e K^d - f^{ade} K^b K^c \pa Q^e
+f^{bce} J^a \pa Q^e K^d 
\nonu\\
&& + \left. f^{bde} J^a K^c \pa Q^e + f^{cde} J^a K^b \pa Q^e  \right)(w)
+f^{aeg}f^{bfg} (Q^e K^f) K^c K^d(w)
+f^{aeg}f^{cfg} K^b (Q^e K^f) K^d(w) \nonu\\
&&+ f^{aeg}f^{dfg} K^b K^c (Q^e K^f)(w) 
-f^{beg}f^{cfg} J^a (Q^e K^f)  K^d(w)
-f^{beg}f^{dfg} J^a K^c (Q^e K^f)(w) \nonu\\
&&-f^{ceg}f^{dfg} J^a K^b (Q^e K^f)(w),
\nonu\\
&&K^e Q^e(z)  \, K^a K^b K^c K^d(w)|_\frac{1}{(z-w)^2}
=
-k_2 \left( Q^a K^b K^c K^d + K^a Q^b K^c K^d + K^a K^b Q^c K^d \right.
\nonu\\
&&+ K^a K^b K^c Q^d 
+f^{abe} \pa Q^e K^c K^d
+f^{ace} K^b \pa Q^e  K^d
+f^{ade} K^b K^c \pa Q^e 
+f^{bce} K^a \pa Q^e  K^d \nonu\\
&&+f^{bde} K^a K^c \pa Q^e 
+\left. f^{cde} K^a K^b \pa Q^e \right)(w)
-f^{aeg}f^{bfg} (Q^e K^f) K^c K^d(w) -f^{aeg}f^{cfg} K^b (Q^e K^f) K^d(w) \nonu\\
&&-f^{aeg}f^{dfg}  K^b K^c (Q^e K^f)(w) 
-f^{beg}f^{cfg}  K^a  (Q^e K^f) K^d(w)
-f^{beg}f^{dfg}  K^a K^c (Q^e K^f)(w) \nonu\\
&&-f^{ceg}f^{dfg}  K^a K^b (Q^e K^f)(w),
\nonu\\
&&K^e Q^e(z) \, \pa^2 J^a K^a(w)|_\frac{1}{(z-w)^2}
=
-3(2k_1+N) \pa^2 L^a K^a(w) -k_2 \pa^2 J^a Q^a(w) 
+2f^{abc} \pa (Q^a K^b) K^c(w),
\nonu\\
&&K^e Q^e(z)  \, \pa J^a \pa K^a(w)|_\frac{1}{(z-w)^2}
=
-2(2k_1+N) \pa L^a \pa K^a(w) -2k_2 \pa J^a \pa Q^a(w)
+f^{abc} (Q^a K^b) \pa K^c(w) \nonu\\
&&-f^{abc} \pa J^a (Q^b K^c)(w),
\nonu\\
&&K^e Q^e(z)   \, J^a \pa^2 K^a(w)|_\frac{1}{(z-w)^2}
=
-(2k_1+N)  L^a \pa^2 K^a(w)
-3k_2 J^a \pa^2 Q^a(w)
+2N \pa^2 (Q^a K^a)(w) \nonu\\
&&-2f^{abc} J^a \pa (Q^b K^c)(w),
\nonu\\
&&K^e Q^e(z) \,  f^{abc} \pa J^a J^b K^c(w)|_\frac{1}{(z-w)^2}
=k_2 \left( -f^{abc} \pa J^a J^b Q^c + 2N \pa J^a \pa Q^a \right)(w)
\nonu\\
&&+(2 k_1+N) \left( -2 d^{abe}f^{cde} \pa (J^a K^b) J^c K^d
+f^{ade}d^{bce} \pa J^a (J^b K^c) K^d \right)(w)
+f^{abe}f^{cde}(Q^a K^b) J^c K^d(w)
\nonu\\
&&+N f^{abc} \pa J^a Q^b K^c(w),
\nonu\\
&&K^e Q^e(z) \,  f^{abc} \pa K^a K^b J^c(w)|_\frac{1}{(z-w)^2}
=k_2 \left( 2 f^{abc} \pa Q^a J^b K^c +f^{abc} \pa K^a J^b Q^c
-2N \pa K^a \pa Q^a \right)(w) \nonu\\
&&-(2 k_1+N) f^{ade}d^{bce} \pa K^a (J^b K^c) K^d(w)
+f^{abe}f^{cde} (Q^a K^b) J^c K^d(w) + N f^{abc} Q^a \pa K^b K^c(w),
\nonu\\
&&K^e Q^e(z) \,   \pa^2 J^a J^a(w)|_\frac{1}{(z-w)^2}
=-(2k_1+N) \left( 3 \pa^2 L^a J^a + \pa^2 J^a L^a \right)(w)
+2f^{abc} \pa(Q^a K^b) J^c(w),
\nonu\\
&&K^e Q^e(z) \,  \pa^2 K^a K^a(w)|_\frac{1}{(z-w)^2}
=-k_2 \left( 3 \pa^2 Q^a K^a + Q^a \pa^2 K^a  \right)(w)
-2 f^{abc} \pa( Q^a K^b) K^c(w),
\nonu\\
&&K^e Q^e(z) \,  \pa J^a \pa J^a(w)|_\frac{1}{(z-w)^2}
=-2(2k_1+N) \left( \pa L^a \pa J^a+\pa J^a \pa L^a \right)(w)
+f^{abc} \left( (Q^a K^b) \pa J^c \right.
\nonu\\
&&+ \left. \pa J^a (Q^b K^c) \right)(w),
\nonu\\
&&K^e Q^e(z)  \, \pa K^a \pa K^a(w)|_\frac{1}{(z-w)^2}
=-4 k_2 \pa Q^a \pa K^a(w)
-f^{abc} \left( (Q^a K^b) \pa K^c + \pa K^a (Q^b K^c)  \right)(w).
\label{KQUpole2}
\eea
Further simplifications for the right hand side of (\ref{KQUpole2})
with the appropriate normal ordering procedure 
are needed.

The second order pole for the third term  
$d^{abc}J^a K^b K^c(z) \equiv J^e R^e(z)$ 
of (\ref{primaryspin3}) and the spin-$4$ current 
without using an explicit calculation can be obtained from the previous result 
(\ref{KQUpole2})
\bea
&&J^e R^e(z) \, J^a J^b J^c J^d(w)|_\frac{1}{(z-w)^2}
=
\left( K^e Q^e(z) \,  K^a K^b K^c K^d(w)|_\frac{1}{(z-w)^2} \right)  |_{k_1 \leftrightarrow k_2, J^a \leftrightarrow K^a , Q^a \leftrightarrow R^a},  
\nonu\\
&&J^e R^e(z) \, J^a J^b J^c K^d(w)|_\frac{1}{(z-w)^2}
=
\left( K^e Q^e(z)  \, J^d K^a K^b K^c(w)|_\frac{1}{(z-w)^2} \right)  |_{k_1 \leftrightarrow k_2, J^a \leftrightarrow K^a , Q^a \leftrightarrow R^a},  
\nonu\\
&&J^e R^e(z) \, J^a J^b K^c K^d(w)|_\frac{1}{(z-w)^2}
=
\left( K^e Q^e(z) \, J^c J^d K^a K^b(w)|_\frac{1}{(z-w)^2} \right)  |_{k_1 \leftrightarrow k_2, J^a \leftrightarrow K^a , Q^a \leftrightarrow R^a},  
\nonu\\
&&J^e R^e(z) \, J^a K^b K^c K^d(w)|_\frac{1}{(z-w)^2}
=
\left( K^e Q^e(z)  \, J^b J^c J^d K^a(w)|_\frac{1}{(z-w)^2} \right)  |_{k_1 \leftrightarrow k_2, J^a \leftrightarrow K^a , Q^a \leftrightarrow R^a},  
\nonu\\
&&J^e R^e(z) \, K^a K^b K^c K^d(w)|_\frac{1}{(z-w)^2}
=
\left( K^e Q^e(z) \, J^a J^b J^c J^d(w)|_\frac{1}{(z-w)^2} \right)  |_{k_1 \leftrightarrow k_2, J^a \leftrightarrow K^a , Q^a \leftrightarrow R^a},  
\nonu\\
&&J^e R^e(z) \, \pa^2 J^a K^a(w)|_\frac{1}{(z-w)^2}
=
\left( K^e Q^e(z) \,   J^a \pa^2 K^a(w)|_\frac{1}{(z-w)^2} \right)  |_{k_1 \leftrightarrow k_2, J^a \leftrightarrow K^a , Q^a \leftrightarrow R^a},  
\nonu\\
&&J^e R^e(z) \, \pa J^a \pa K^a(w)|_\frac{1}{(z-w)^2}
=
\left( K^e Q^e(z) \,  \pa J^a \pa K^a(w)|_\frac{1}{(z-w)^2} \right)  |_{k_1 \leftrightarrow k_2, J^a \leftrightarrow K^a , Q^a \leftrightarrow R^a},  
\nonu\\
&&J^e R^e(z) \, J^a \pa^2 K^a(w)|_\frac{1}{(z-w)^2}
=
\left( K^e Q^e(z) \,  \pa^2 J^a  K^a(w)|_\frac{1}{(z-w)^2} \right)  |_{k_1 \leftrightarrow k_2, J^a \leftrightarrow K^a , Q^a \leftrightarrow R^a},
\nonu\\
&&J^e R^e(z)  \, f^{abc}\pa  J^a J^b K^c(w)|_\frac{1}{(z-w)^2}
=
\left( K^e Q^e(z) \, f^{abc}\pa  K^a K^b J^c(w)|_\frac{1}{(z-w)^2} \right)  |_{k_1 \leftrightarrow k_2, J^a \leftrightarrow K^a , Q^a \leftrightarrow R^a},
\nonu\\
&&J^e R^e(z)  \, f^{abc}\pa  K^a K^b J^c(w)|_\frac{1}{(z-w)^2}
=
\left( K^e Q^e(z) \, f^{abc}\pa  J^a J^b K^c(w)|_\frac{1}{(z-w)^2} \right)  |_{k_1 \leftrightarrow k_2, J^a \leftrightarrow K^a , Q^a \leftrightarrow R^a},
\nonu\\
&&J^e R^e(z) \, \pa^2 J^a J^a(w)|_\frac{1}{(z-w)^2}
=
\left( K^e Q^e(z)  \, \pa^2 K^a K^a(w)|_\frac{1}{(z-w)^2} \right)  |_{k_1 \leftrightarrow k_2, J^a \leftrightarrow K^a , Q^a \leftrightarrow R^a},  
\nonu\\
&&J^e R^e(z) \, \pa^2 K^a K^a(w)|_\frac{1}{(z-w)^2}
=
\left( K^e Q^e(z) \, \pa^2 J^a J^a(w)|_\frac{1}{(z-w)^2} \right)  |_{k_1 \leftrightarrow k_2, J^a \leftrightarrow K^a , Q^a \leftrightarrow R^a},  
\nonu\\
&&J^e R^e(z) \, \pa J^a \pa J^a(w)|_\frac{1}{(z-w)^2}
=
\left( K^e Q^e(z) \,  \pa K^a \pa K^a(w)|_\frac{1}{(z-w)^2} \right)  |_{k_1 \leftrightarrow k_2, J^a \leftrightarrow K^a , Q^a \leftrightarrow R^a},  
\nonu\\
&&J^e R^e(z) \, \pa K^a \pa K^a(w)|_\frac{1}{(z-w)^2}
=
\left( K^e Q^e(z)  \, \pa J^a \pa J^a(w)|_\frac{1}{(z-w)^2} \right)  |_{k_1 \leftrightarrow k_2, J^a \leftrightarrow K^a , Q^a \leftrightarrow R^a}.
\label{JRUpole2}
\eea

Finally, the second order pole in the OPE between  
the fourth term  $d^{abc}K^a K^b K^c(z) \equiv R(z)$ 
of (\ref{primaryspin3}) and the spin-$4$ current 
can be obtained from the previous result (\ref{QUpole2}).
\bea
&&R(z)  \, J^a J^b J^c J^d(w)|_\frac{1}{(z-w)^2}=0, 
\nonu\\
&&R(z)  \, J^a J^b J^c K^d(w)|_\frac{1}{(z-w)^2}
=  \left(Q(z)  \, J^d K^a K^b K^c(w)|_\frac{1}{(z-w)^2}\right)|_{k_1 \leftrightarrow k_2,J^a \leftrightarrow K^a, Q^a \leftrightarrow R^a}, 
\nonu\\
&&R(z)  \, J^a J^b K^c K^d(w)|_\frac{1}{(z-w)^2}=
\left( Q(z)  \, J^c J^d K^a K^b(w)|_\frac{1}{(z-w)^2} \right) |_{k_1 \leftrightarrow k_2, J^a \leftrightarrow K^a , Q^a \leftrightarrow R^a} ,
\nonu\\
&&R(z)  \, J^a K^b K^c K^d(w)|_\frac{1}{(z-w)^2}=
\left( Q(z)  \, J^b J^c J^d K^a(w)|_\frac{1}{(z-w)^2} \right) |_{k_1 \leftrightarrow k_2, J^a \leftrightarrow K^a , Q^a \leftrightarrow R^a} , 
\nonu\\
&&R(z) \, K^a K^b K^c K^d(w)|_\frac{1}{(z-w)^2}
=
\left( Q(z) \, J^a J^b J^c J^d(w)|_\frac{1}{(z-w)^2} \right) |_{k_1 \leftrightarrow k_2, J^a \leftrightarrow K^a , Q^a \leftrightarrow R^a}, 
\nonu\\
&&R(z) \, \pa^2 J^a K^a(w)|_\frac{1}{(z-w)^2}=
\left( Q(z)  \, J^a \pa^2 K^a(w)|_\frac{1}{(z-w)^2} \right) |_{k_1 \leftrightarrow k_2, J^a \leftrightarrow K^a , Q^a \leftrightarrow R^a}, 
\nonu\\
&&R(z) \, \pa J^a \pa K^a(w)|_\frac{1}{(z-w)^2}=
\left( Q(z)  \, \pa J^a \pa K^a(w)|_\frac{1}{(z-w)^2} \right) |_{k_1 \leftrightarrow k_2, J^a \leftrightarrow K^a , Q^a \leftrightarrow R^a}, 
\nonu\\
&&R(z) \, J^a \pa^2 K^a(w)|_\frac{1}{(z-w)^2}=
\left( Q(z) \, \pa^2 J^a K^a(w)|_\frac{1}{(z-w)^2} \right) |_{k_1 \leftrightarrow k_2, J^a \leftrightarrow K^a , Q^a \leftrightarrow R^a},
\nonu\\
&&R(z) \, f^{abc} \pa J^a J^b K^c(w)|_\frac{1}{(z-w)^2}=
\left( Q(z) \, f^{abc} \pa K^a K^b J^c(w)|_\frac{1}{(z-w)^2} \right)|_{k_1 \leftrightarrow k_2, J^a \leftrightarrow K^a , Q^a \leftrightarrow R^a},
\nonu\\
&&R(z) \, f^{abc} \pa K^a K^b J^c(w)|_\frac{1}{(z-w)^2}=
\left( Q(z) \, f^{abc} \pa J^a J^b K^c(w)|_\frac{1}{(z-w)^2} \right)|_{k_1 \leftrightarrow k_2, J^a \leftrightarrow K^a , Q^a \leftrightarrow R^a},
\nonu\\
&&R(z) \, \pa^2 J^a J^a(w)|_\frac{1}{(z-w)^2}=
0,
\nonu\\
&&R(z) \, \pa^2 K^a K^a(w)|_\frac{1}{(z-w)^2}=
\left( Q(z) \, \pa^2 J^a J^a(w)|_\frac{1}{(z-w)^2}\right) |_{k_1 \leftrightarrow k_2, J^a \leftrightarrow K^a , Q^a \leftrightarrow R^a} ,
\nonu\\
&&R(z) \, \pa J^a \pa J^a(w)|_\frac{1}{(z-w)^2}=
0, 
\nonu\\
&&R(z) \, \pa K^a \pa K^a(w)|_\frac{1}{(z-w)^2}=
\left( Q(z) \, \pa J^a \pa J^a(w)|_\frac{1}{(z-w)^2}\right) |_{k_1 \leftrightarrow k_2, J^a \leftrightarrow K^a , Q^a \leftrightarrow R^a}.
\label{RUpole2}
\eea
Therefore, the second order pole structures are exhausted.  
Further simplifications are given in Appendix $C$.

\section{Details of the fully normal ordered spin-$5$ operator }

Thus far, the results from the previous Appendices are 
not fully ordered in the sense of \cite{BBSS1}. 
Further simplifications are needed with the help of the rearrangement lemma.

The following commutators with the Appendices in \cite{BBSS1} 
from OPEs (\ref{JJcoset}) and (\ref{JaQb}). 
\bea
\left[J^a, J^b \right](z) & = & f^{abc} \pa J^c(z),
\qquad
\left[J^a, \pa J^b \right](z)=\frac{1}{2} f^{abc} \pa^2 J^c(z),
\nonu\\
\left[J^a, Q^b \right](z) & = & f^{abc} \pa Q^c(z) 
+ \frac{1}{2} (2 k_1+N) d^{abc} \pa^2 J^c(z),
\nonu\\
\left[J^a, \pa Q^b \right](z) & = & \frac{1}{2} 
f^{abc} \pa^2 Q^c(z) + \frac{1}{6} (2 k_1+N) d^{abc} \pa^3 J^c(z).
\label{JJJQcommutator}
\eea
For example, take 
$Q(z) d^{abe}d^{cde} J^a J^b K^c K^d(w)|_\frac{1}{(z-w)^2}$ 
from the third equation of (\ref{QUpole2}) in Appendix $B$ by multiplying
two $d$ symbols
\bea
Q(z) \, d^{abe}d^{cde} J^a J^b K^c K^d(w)|_\frac{1}{(z-w)^2}
=-3(k_1+N) d^{abe}d^{cde} \left(Q^a J^b K^c K^d+J^a Q^b K^c K^d\right)(w).
\label{aboveabove}
\eea
As stated before, the field $Q^a(z)$ in (\ref{aboveabove}) 
should be moved to the right.
The right hand side of (\ref{aboveabove})
can be expressed as
\bea
-3(k_1+N) d^{abe}d^{cde} \left(J^b Q^a  K^c K^d - [J^b,Q^a]  K^c K^d 
+J^a Q^b K^c K^d\right)(w)
\label{above1}
\eea
and using the third equation of (\ref{JJJQcommutator})
this (\ref{above1}) can be expressed as
\bea
-3(k_1+N) d^{abe}d^{cde} 
\left( - \left(f^{baf} \pa Q^f + \frac{1}{2} (2 k_1+N) d^{baf} \pa^2 J^f \right)  K^c K^d 
+2J^a Q^b K^c K^d\right)(w).
\label{above2}
\eea
Owing to the antisymmetric property of the $f$ symbol and the symmetric property of
the $d$ symbol, the first term of (\ref{above2}) vanishes 
and the remaining terms are
\bea
\frac{3}{2}(k_1+N)(2 k_1+N)  d^{abe}d^{abf}d^{cde}  \pa^2 J^f K^c K^d(w)
-6 (k_1+N) d^{abe}d^{cde} J^a Q^b K^c K^d(w).
\label{exp}
\eea
The first term in (\ref{exp}) 
can be simplified using the identity between two $d$ symbols
\bea
\frac{3}{N}(k_1+N)(2 k_1+N)(N^2-4)  d^{abc}  \pa^2 J^a K^b K^c(w)
 -6 (k_1+N) d^{abe}d^{cde} J^a Q^b K^c K^d(w).
\label{fullordex}
\eea
The last term of (\ref{fullordex}) can be simplified further 
as follows:
\bea
&& d^{abe}d^{cde} J^a Q^b K^c K^d(z)
 =  d^{abe}d^{cde} d^{bfg}J^a (J^f J^g) K^c K^d(z)
\nonu\\
& & =  
d^{abe}d^{cde} d^{bfg} \left( J^f   J^g J^a K^c K^d -[J^f ,J^a]  J^g K^c K^d 
-J^f   [J^g, J^a] K^c K^d\right)(z)
\label{expexp}
\eea
where the definition of $Q^b(z)$ is used.
The last two terms of (\ref{expexp}) can be expressed as
\bea
-d^{abe}d^{cde} d^{bfg} f^{fah} \pa J^h J^g K^c K^d(z)
-d^{abe}d^{cde} d^{bfg} f^{gah} J^f \pa J^h K^c K^d(z)
\label{this}
\eea
after substituting the commutators in (\ref{JJJQcommutator}).
Using the identity of $(A.7)$ of \cite{Ahn1111} (the triple product 
$d d f$ is reduced to a single $f$), 
this expression (\ref{this}) can be simplified
as
\bea
-\frac{N^2-4}{N}f^{abe}d^{cde} \pa J^a J^b K^c K^d(z)
+ \frac{N^2-4}{N} f^{abe}d^{cde}  J^a \pa J^b K^c K^d(z).
\label{expp}
\eea
The field $\pa J^a$ in the first term of (\ref{expp}) 
should be moved to the right
to yield the following: 
\bea
-\frac{N^2-4}{N}f^{abe}d^{cde} \left(  J^b \pa J^a K^c K^d 
- [J^b, \pa J^a] K^c K^d \right)(z)
+ \frac{N^2-4}{N} f^{abe}d^{cde}  J^a \pa J^b K^c K^d(z).
\label{inter}
\eea
Again, the commutator can be simplified, and after a small calculation, 
(\ref{inter})  leads to
\bea
\frac{N^2-4}{2N}f^{abe}d^{cde} f^{baf} \pa^2 J^f K^c K^d(z)
+ \frac{2(N^2-4)}{N} f^{abe}d^{cde}  J^a \pa J^b K^c K^d(z).
\label{ffterm}
\eea
The $f f$ term in (\ref{ffterm}) 
can be simplified further and the final fully normal ordered product
becomes
\bea
 d^{abe}d^{cde} J^a Q^b K^c K^d(z) & = &
 d^{abf}d^{fcg} d^{gde} J^a  J^b J^c K^d K^e(z)
+(4-N^2) d^{abc}  \pa^2 J^a K^b K^c(z)
\nonu \\
& + & \frac{2(N^2-4)}{N} f^{abe}d^{cde}  J^a \pa J^b K^c K^d(z).
\label{indepex}
\eea 
where the right hand side of (\ref{indepex}) appears in the fourth,
$47$th, and $35$th terms in (\ref{spin5detail}). 
The final form for  
$Q(z) \, d^{abe}d^{cde} J^a J^b K^c K^d(w)|_\frac{1}{(z-w)^2}$ can be obtained from 
(\ref{fullordex}) and (\ref{indepex}) 
as follows:
\bea
&&Q(z) \, d^{abe}d^{cde} J^a J^b K^c K^d(w)|_\frac{1}{(z-w)^2}
\nonu\\
&& =-6(k_1+N)d^{abf}d^{fcg} d^{gde} J^a  J^b J^c K^d K^e(w)
-\frac{12}{N}(k_1+N)(N^2-4) f^{abe}d^{cde}  J^a \pa J^b K^c K^d(w)
\nonu\\
&& +\frac{3}{N}(k_1+N)(N^2-4)(2k_1+3N) d^{abc}  \pa^2 J^a K^b K^c(w).
\label{formfinal}
\eea
This identity (\ref{formfinal}) can be used continually. 

The other commutators are needed to rearrange all the 
normal ordered products in 
(\ref{QUpole2}), (\ref{KQUpole2}),  (\ref{JRUpole2}) and (\ref{RUpole2}),
which are not fully normal ordered product. 
First, calculate the OPE $J^a(z) \, J^b K^c(w)$.
Recall the mixed spin-$2$ field (\ref{Ldef}).  
\bea 
J^a(z) \, J^b K^c(w)&=&-\frac{1}{(z-w)^2} \, k_1 \delta^{ab} K^c(w)
+\frac{1}{(z-w)} \, f^{abd} J^d K^c(w) + \cdots .
\label{OPEJJK}
\eea
From the OPE (\ref{OPEJJK}), 
the following various commutators, via $(A.10)$ of \cite{BBSS1},  
are obtained 
\bea
\left[J^a, L^b \right](z) & = & \frac{1}{2} k_1 d^{abc} \pa^2  K^c(z)
+ f^{acd}d^{bce} \pa (J^d K^e)(z),
\nonu\\
\left[J^a, \pa (J^b K^c)  \right](z) & = & 
\frac{1}{6} k_1 \delta^{ab} \pa^3  K^c(z)
+ \frac{1}{2}f^{abd} \pa^2 (J^d K^c)(z),
\nonu\\
\left[J^a, \pa^2 (J^b K^c)  \right](z) & = & 
\frac{1}{12} k_1 \delta^{ab} \pa^4  K^c(z)
+ \frac{1}{3}f^{abd} \pa^3 (J^d K^c)(z),
\nonu\\
\left[K^a, L^b \right](z) & = & 
\left( \left[J^a, L^b \right](z) \right)|_{k_1 \leftrightarrow k_2, J^a \leftrightarrow K^a },
\nonu\\
\left[K^a, \pa (J^b K^c)  \right](z) & = & 
\left( \left[J^a, \pa (J^c K^b)  \right](z)
\right) |_{k_1 \leftrightarrow k_2, J^a \leftrightarrow K^a },
\nonu\\
\left[K^a, \pa^2 (J^b K^c)  \right](z) & = & 
\left( \left[J^a, \pa^2 (J^c K^b)  \right](z)
\right) |_{k_1 \leftrightarrow k_2, J^a \leftrightarrow K^a }.
\label{comm1}
\eea
Similarly, the OPEs 
$J^a(z)\, Q^b K^c(w)$ and $K^a(z) \, Q^b K^c(w)$
can be derived
\bea
J^a(z) \, Q^b K^c(w)&=&-\frac{1}{(z-w)^2}\, (2 k_1+N) d^{abd} J^d K^c(w)
+\frac{1}{(z-w)} \, f^{abd} Q^d K^c(w) + \cdots,
\nonu\\
K^a(z) \, Q^b K^c(w)&=&-\frac{1}{(z-w)^2} \, k_2 \delta^{ac} Q^b(w)
+\frac{1}{(z-w)} \, f^{acd} Q^b K^d(w) +\cdots .
\label{that}
\eea
The relevant commutators, which can be obtained from (\ref{that}), 
are as follows:
\bea
\left[J^a, Q^b K^c \right](z) & = & 
\frac{1}{2}(2 k_1+N) d^{abd} \pa^2 ( J^d K^c)(z)
+ f^{abd} \pa (Q^d K^c)(z),
\nonu\\
\left[J^a, \pa(Q^b K^c) \right](z) & = & 
\frac{1}{6}(2 k_1+N) d^{abd} \pa^3 ( J^d K^c)(z) +\frac{1}{2} f^{abd} \pa^2 (Q^d K^c)(z),
\nonu\\
\left[K^a, Q^b K^c \right](z) & = & \frac{1}{2}k_2 \delta^{ac} \pa^2 Q^b(z)
+ f^{acd} \pa (Q^b K^d)(z),
\nonu\\
\left[K^a, \pa(Q^b K^c) \right](z) & = & 
\frac{1}{6}k_2 \delta^{ac} \pa^3 Q^b(z)
+ \frac{1}{2} f^{acd} \pa^2 (Q^b K^d)(z),
\nonu\\
\left[J^a, R^b J^c \right](z) & = & 
\left( \left[K^a, Q^b K^c \right](z) \right)  |_{k_1 \leftrightarrow k_2, J^a \leftrightarrow K^a , Q^a \leftrightarrow R^a},
\nonu\\
\left[J^a, \pa(R^b J^c) \right](z)& = & \left( \left[K^a, \pa(Q^b K^c) \right](z) \right)  |_{k_1 \leftrightarrow k_2, J^a \leftrightarrow K^a , Q^a \leftrightarrow R^a},
\nonu\\
\left[K^a, R^b J^c \right](z) & = & \left( \left[J^a, Q^b K^c \right](z) \right)  |_{k_1 \leftrightarrow k_2, J^a \leftrightarrow K^a , Q^a \leftrightarrow R^a},
\nonu\\
\left[K^a, \pa(R^b J^c) \right](z) & = & 
\left( \left[J^a, \pa(Q^b K^c) \right](z) \right)  |_{k_1 \leftrightarrow k_2, J^a \leftrightarrow K^a , Q^a \leftrightarrow R^a}.
\label{comm2}
\eea

Finally, the commutators $[J^a,T](z)$ and $[K^a,T](z)$ 
to rearrange the 
normal ordered product $TW^{(3)}(w)$ in (\ref{spin5construction}) are needed. 
The OPE $J^a(z) \, T(w)$ are calculated as follows:
\bea
J^a(z) \, T(w) &=&\frac{1}{(z-w)^2} \, \frac{1}{(k_1+k_2+N)} \,(
k_2 J^a -k_1 K^a)(w)
\nonu \\
& - & 
\frac{1}{(z-w)} \, \frac{1}{(k_1+k_2+N)} \, f^{abc} J^b K^c(w) + \cdots .
\label{lastOPE}
\eea
The following commutators can be obtained from (\ref{lastOPE})
\bea
&&\left[J^a, T \right](z)=-\frac{1}{2(k_1+k_2+N)} 
\left( k_2 \pa^2 J^a-k_1 \pa^2 K^a \right)(z)
- \frac{1}{(k_1+k_2+N)} f^{abc} \pa (J^b K^c)(z),
\nonu\\
&&\left[K^a, T \right](z)=\left( [J^a, T ](z) 
\right) |_{k_1 \leftrightarrow k_2, J^a \leftrightarrow K^a }.
\label{comm3}
\eea
All the commutators in (\ref{comm1}), (\ref{comm2}) and (\ref{comm3})
are used to simplify the normal ordered product in Appendix $B$.

Although this paper did not present all the fully normal ordered products 
leading to the $52$ terms for the spin-$5$ current 
(\ref{spin5detail}) (due to the space limit of this paper), 
the complicated calculations from Appendices $B$ and $C$ to the final 
$52$ terms 
can be performed.

\section{ Explicit
$52$ coefficient functions of spin-$5$ Casimir operator }

The $52$ 
relative coefficient functions appearing in the spin-$5$ current 
are given by
\bea
a_1&=&-\frac{6 k N (k+N) (2 k+N)^2}{(N-2) (N+2)^2 (2 k+3 N+2)} ,
\nonu\\
a_2&=&\left[ 6 (k+N) (2 k+N) 
(30 k^3 N^4+204 k^3 N^3+382 k^3 N^2+184 k^3 N-120 k^3+65 k^2 N^5 \right. 
\nonu\\
&+& 518 k^2 N^4+1217 k^2
   N^3+960 k^2 N^2-188 k^2 N-240 k^2+10 k N^6+329 k N^5+1114 k N^4
\nonu\\
&+&  1265 k N^3+90 k N^2-452 k N-120 k+70
   N^6+264 N^5+374 N^4+108 N^3-152 N^2
\nonu\\
&-& \left.  80 N ) \right]
/
\left[ (N-2) (N+1) (N+2)^2 (2 k+3 N+2) d(N,k) \right],
\nonu\\
a_3&=& \left[ 6 (k+N) (2 k+N) (20 k^3 N^4+66 k^3 N^3+8 k^3 N^2-14 k^3 N+120 k^3+60 k^2 N^5 \right. 
\nonu\\
&+&  207 k^2 N^4+28 k^2
   N^3-145 k^2 N^2+358 k^2 N+240 k^2+40 k N^6+186 k N^5+71 k N^4
\nonu\\
&-&  240 k N^3+215 k N^2+452 k N+120 k+40
   N^6+66 N^5-44 N^4+2 N^3+152 N^2
\nonu\\
&+&  \left. 80 N) \right]
/
\left[ (N-2) (N+1) (N+2)^2 (2 k+3 N+2) d(N,k) \right],
\nonu\\
a_4&=&-\frac{12 N (N+1) (k+N) 
(10 k^2 N+8 k^2+14 k N^2+14 k N+5 N^3+6 N^2)}
{(N-2) (N+2)^2 (2 k+N) (2 k+3 N+2)},
\nonu\\
a_5&=&-\frac{24 N (N+1) (k+N) (2 k+N)}{(N-2) (N+2) (2 k+3 N+2)},
\nonu\\
a_6&=&-\frac{24 N (N+1) (k+N) 
(3 k N+4 k+2 N^2+3 N)}{(N-2) (N+2)^2 (2 k+3 N+2)},
\nonu\\
a_7&=&\frac{12 N (N+1) (k+N) 
(6 k N^2+14 k N+8 k+5 N^3+14 N^2+10 N)}{(N-2) (N+2)^2 (2 k+N) (2 k+3 N+2)},
\nonu\\
a_8&=&\frac{24 N (N+1) (k+N) (3 k N+4 k+2 N^2+3 N)}
{(N-2) (N+2) (2 k+N) (2 k+3 N+2)},
\nonu\\
a_9&=&\frac{24 N (N+1) (k+N)}{(N-2) (2 k+3 N+2)},
\nonu\\
a_{10}&=& 
\left[ -6 (N+1) (N+2) 
(10 k^3 N^2+118 k^3 N-60 k^3+25 k^2 N^3+347 k^2 N^2+22 k^2 N-120 k^2  \right.
\nonu\\
&+&\left.  10 k N^4+299 k N^3+217
   k N^2-72 k N-60 k+70 N^4+94 N^3+48 N^2+24 N) \right]
/ \nonu\\
&\times&\left[ 
(N-2) (2 k+N) (2 k+3 N+2) d(N,k)\right],
\nonu\\
a_{11}&=& \left[ 
-6 (N+1) (N+2) (15 k^3 N^2-33 k^3 N+60 k^3+50 k^2 N^3-67 k^2 N^2+63 k^2 N+120 k^2
\right.  \nonu\\
&+& \left. 40 k N^4+6 k N^3-22
   k N^2+72 k N+60 k+40 N^4+16 N^3-48 N^2-24 N) \right] / \nonu\\
&\times&
\left[ 
(N-2) (2 k+N) (2 k+3 N+2) d(N,k) \right],
\nonu\\
a_{12}&=& \frac{6 N (N+1) (N+2)}{(N-2) (2 k+N) (2 k+3 N+2)},
\nonu\\
a_{13}&=& \left[  
96 k^2 (k+N)^2 
(29 k^2 N-71 k^2+58 k N^2-113 k N-71 k+25 N^3-42 N^2-42 N) \right] /
\nonu\\
&\times& \left[ 
(N-2) (N+1) (2 k+3 N+2) p(N,k) \right],
\nonu\\
a_{14}&=& \left[ -24 (k+N) (1160 k^6 N^5+4724 k^6 N^4-1996 k^6 N^3-29228 k^6 N^2-30028 k^6 N \right. 
\nonu\\
&+&  12840 k^6+5800 k^5 N^6+26080 k^5
   N^5+1636 k^5 N^4-150572 k^5 N^3-221780 k^5 N^2
   \nonu\\
   &+& 6268 k^5 N+38520 k^5+10280 k^4 N^7+55241 k^4 N^6+34215 k^4 N^5-285673
   k^4 N^4
   \nonu\\
   &-&   585707 k^4 N^3-155620 k^4 N^2+131280 k^4 N+38520 k^4+7640 k^3 N^8+54988 k^3 N^7
   \nonu\\
   &+&  70558 k^3 N^6-237134 k^3
   N^5-710602 k^3 N^4-375346 k^3 N^3+156916 k^3 N^2+123644 k^3 N
   \nonu\\
   &+&  12840 k^3+2000 k^2 N^9+25468 k^2 N^8+53820 k^2 N^7-75987
   k^2 N^6-404325 k^2 N^5
   \nonu\\
   &-&  324821 k^2 N^4+79929 k^2 N^3+140224 k^2 N^2+28660 k^2 N+4400 k N^9+13880 k N^8
   \nonu\\
   &-&  1340 k
   N^7-89940 k N^6-106644 k N^5+18140 k N^4+65904 k N^3+20240 k N^2
   \nonu\\
   &+& \left.  
2052 N^8-2052 N^7-8436 N^6+2052 N^5+10944 N^4+4560 N^3)\right] 
/ \left[ (N-2) N (N+1) \right.
   \nonu\\
   & \times & \left. (N+2) (2 k+3 N+2) d(N,k) p(N,k) \right],
\nonu\\
a_{15}&=& \left[ 
-24 (k+N) 
(1740 k^6 N^5+6736 k^6 N^4-8064 k^6 N^3-37672 k^6 N^2-18252 k^6 N \right.
\nonu\\
&-& 12840 k^6+8700 k^5 N^6+37020 k^5
   N^5-29016 k^5 N^4-204048 k^5 N^3-153420 k^5 N^2
   \nonu\\
   &-& 102828 k^5 N-38520 k^5+15420 k^4 N^7+78399 k^4 N^6-23815 k^4 N^5-420707
   k^4 N^4
   \nonu\\
   &-&  422473 k^4 N^3-292000 k^4 N^2-179560 k^4 N-38520 k^4+11460 k^3 N^8+78632 k^3 N^7
   \nonu\\
   &+&  20162 k^3 N^6-406866
   k^3 N^5-513898 k^3 N^4-368274 k^3 N^3-296236 k^3 N^2-123644 k^3 N
   \nonu\\
   &-&  12840 k^3+3000 k^2 N^9+37152 k^2 N^8+34700 k^2
   N^7-186753 k^2 N^6-283235 k^2 N^5
   \nonu\\
   &-&  197099 k^2 N^4-207929 k^2 N^3-140224 k^2 N^2-28660 k^2 N+6600 k N^9+11640 k N^8
   \nonu\\
   &-&  36060
   k N^7-54380 k N^6-22716 k N^5-55100 k N^4-65904 k N^3-20240 k N^2
   \nonu\\
   &-&   \left. 2052 N^8+2052 N^7+8436 N^6-2052 N^5-10944 N^4-4560 N^3
) \right] / \left[ (N-2) N (N+1) \right. 
   \nonu\\
   &\times &  
\left. (N+2) (2 k+3 N+2) d(N,k) p(N,k) \right],
\nonu\\
a_{16}&=& \frac{96 k (k+N) 
(29 k^2 N-71 k^2+58 k N^2-113 k N-71 k+25 N^3-42 N^2-42 N)}
{(N-2) (2 k+3 N+2) p(N,k)},
   \nonu\\
a_{17}&=& \frac{576 (N+1) (k+N)^2 
(29 k^2 N-71 k^2+58 k N^2-113 k N-71 k+25 N^3-42 N^2-42 N)}
{(N-2) (2 k+3 N+2) p(N,k)},
   \nonu\\
a_{18}&=&  \left[ 
288 k (N+2) (k+N)^2 
(29 k^2 N-71 k^2+58 k N^2-113 k N-71 k+25 N^3-42 N^2 \right. 
\nonu\\
&-& \left.   42 N) \right] / 
\left[ (N-2) (2 k+N) (2 k+3 N+2) p(N,k) \right],
\nonu\\
a_{19}&=& -\frac{288 (N+1) (k+N) (29 k^2 N-71 k^2+58 k N^2-113 k N-71 k+25 N^3-42 N^2-42 N)}{(N-2) (2 k+3 N+2) p(N,k)},
\nonu\\
a_{20}&=& \left[ -96 k (N+2) (k+N) 
(29 k^2 N-71 k^2+58 k N^2-113 k N-71 k+25 N^3-42 N^2 \right. 
\nonu\\
&-& \left.  42 N) \right] 
/ \left[ (N-2) (2 k+N) (2 k+3 N+2) p(N,k) \right],
\nonu\\
a_{21}&=& \left[ 
-576 (N+1) (N+2) (k+N)^2 
(29 k^2 N-71 k^2+58 k N^2-113 k N-71 k+25 N^3 \right. 
\nonu\\
&-& \left.  42 N^2 -  42 N) \right] 
/ \left[ (N-2) (2 k+N) (2 k+3 N+2) p(N,k) \right],
\nonu\\
a_{22}&=& 
\left[ 24 (N+1) (N+2) 
(1670 k^5 N^3+936 k^5 N^2-9554 k^5 N-6420 k^5+8315 k^4 N^4 \right. 
\nonu\\
&+&  7366 k^4 N^3-47129 k^4 N^2-42640 k^4 N-19260
   k^4+14720 k^3 N^5+23946 k^3 N^4
   \nonu\\
   &-&  92706 k^3 N^3-106362 k^3 N^2-60890 k^3 N-19260 k^3+11040 k^2 N^6+35816 k^2
   N^5
   \nonu\\
   &-& 87917 k^2 N^4-132278 k^2 N^3-65641 k^2 N^2-32076 k^2 N-6420 k^2+3000 k N^7
   \nonu\\
   &+&   24900 k N^6-39648 k N^5-85680 k
   N^4-28068 k N^3-4608 k N^2-4272 k N+6600 N^7
   \nonu\\
   &-& \left.  6540 N^6-23544 N^5-4932 N^4+8208 N^3+2736 N^2) \right] / 
   \left[ (N-2) N (2 k+N) \right.
   \nonu\\
   &\times& \left.  (2 k+3 N+2) d(N,k) p(N,k) \right],
\nonu\\
a_{23}&=& 
\left[ 24 (N+1) (N+2) (1230 k^5 N^3+1824 k^5 N^2-14586 k^5 N+6420 k^5+6185 k^4 N^4 \right. 
\nonu\\ 
&+&  12234 k^4 N^3-68051 k^4 N^2-5640 k^4
   N+19260 k^4+10980 k^3 N^5+32594 k^3 N^4
   \nonu\\
   &-&  117914 k^3 N^3-81238 k^3 N^2+36750 k^3 N+19260 k^3+8060 k^2 N^6+40504
   k^2 N^5
   \nonu\\
   &-&  88523 k^2 N^4-135042 k^2 N^3-4019 k^2 N^2+32076 k^2 N+6420 k^2+2000 k N^7+22720 k N^6
   \nonu\\
   &-&  24692 k N^5-79280
   k N^4-35932 k N^3+4608 k N^2+4272 k N+4400 N^7-940 N^6
   \nonu\\
   &-& \left.  13416 N^5-13548 N^4-8208 N^3-2736 N^2) \right]
/
\left[ (N-2) N (2 k+N) (2 k+3 N+2)  \right.
\nonu\\
&\times&  
\left. d(N,k) p(N,k) \right],
   \nonu\\
a_{24}&=&
\left[ -96 (N+1) (N+2) 
(29 k^2 N-71 k^2+58 k N^2-113 k N-71 k+25 N^3-42 N^2 \right. 
\nonu\\
&-& \left.  42 N) \right] 
/
\left[ (N-2) (2 k+N) (2 k+3 N+2) p(N,k) \right],
\nonu\\
a_{25}&=&
\left[ -6 k (k+N) (2 k+N) (5 k^2 N^2+102 k^2 N+177 k^2+10 k N^3+113 k N^2+168 k N-15 k \right. 
\nonu\\
&+& \left.  6 N^3-4 N^2-10
   N) \right] 
/
\left[ (N+1) (N+2) d(N,k) \right],
\nonu\\
a_{26}&=&
\left[ -6 (k+N) (28 k^7 N^5+9980 k^7 N^4-15092 k^7 N^3-76428 k^7 N^2+47656 k^7 N+79840 k^7  \right.
\nonu\\
&-&  84 k^6 N^6+67148 k^6
   N^5-38204 k^6 N^4-518708 k^6 N^3-46512 k^6 N^2+493272 k^6 N
   \nonu\\
   &+&  184112 k^6-1085 k^5 N^7+183099 k^5 N^6+87751 k^5 N^5-1308927
   k^5 N^4-1204222 k^5 N^3
   \nonu\\
   &+&  301304 k^5 N^2+370520 k^5 N+15088 k^5-2625 k^4 N^8+260028 k^4 N^7+439584 k^4 N^6
   \nonu\\
   &-&  1470698
   k^4 N^5-3320543 k^4 N^4-2793274 k^4 N^3-1868128 k^4 N^2-979368 k^4 N
   \nonu\\
   &-&  202800 k^4-2408 k^3 N^9+203242 k^3 N^8+654939
   k^3 N^7-531605 k^3 N^6-3877529 k^3 N^5
   \nonu\\
   &-&  6792005 k^3 N^4-6991174 k^3 N^3-4226988 k^3 N^2-1240000 k^3 N-113616 k^3-756
   k^2 N^{10}
   \nonu\\
   &+&  83104 k^2 N^9+463719 k^2 N^8+321718 k^2 N^7-2089910 k^2 N^6-6378852 k^2 N^5
   \nonu\\
   &-&  8686421 k^2 N^4-6360778 k^2
   N^3-2357168 k^2 N^2-335728 k^2 N+13896 k N^{10}
   \nonu\\
   &+&  157824 k N^9+322196 k N^8-424092 k N^7-2686788 k N^6-4661748 k
   N^5-4028904 k N^4
   \nonu\\
   &-&  1759680 k N^3-308096 k N^2+20652 N^{10}+73512 N^9-1656 N^8-420384 N^7-912084 N^6
   \nonu\\
   &-& \left.  899496 N^5-439488 N^4-86208 N^3) \right]
/
\left[ (N-2) (N+1) (N+2) (k+N+1)  \right. 
\nonu\\
&\times&  \left. (2 k+3 N+2) d(N,k) p(N,k) \right],
\nonu\\
a_{27}&=&
\left[ 3 (k+N) (1764 k^6 N^7+9372 k^6 N^6-7844 k^6 N^5-183340 k^6 N^4-494352 k^6 N^3 \right. 
\nonu\\
&-&  342560 k^6 N^2+74144 k^6 N+205440
   k^6+8904 k^5 N^8+31028 k^5 N^7-76772 k^5 N^6
   \nonu\\
   &-&  809716 k^5 N^5-2334324 k^5 N^4-2397008 k^5 N^3-472096 k^5 N^2+1263328
   k^5 N+616320 k^5
   \nonu\\
   &+& 15071 k^4 N^9+12703 k^4 N^8-223001 k^4 N^7-1151167 k^4 N^6-3856922 k^4 N^5
   \nonu\\
   &-&  5914876 k^4 N^4-3482104
   k^4 N^3+2027248 k^4 N^2+2762784 k^4 N+616320 k^4+9884 k^3 N^{10}
   \nonu\\
   &-&  44374 k^3 N^9-268638 k^3 N^8-428282 k^3 N^7-2398410
   k^3 N^6-6576776 k^3 N^5
   \nonu\\
   &-&  6927340 k^3 N^4+117120 k^3 N^3+4306112 k^3 N^2+2032160 k^3 N+205440 k^3+2492 k^2 N^{11}
   \nonu\\
   &-&  48100
   k^2 N^{10}-131589 k^2 N^9+337475 k^2 N^8-52573 k^2 N^7-3524867 k^2 N^6
   \nonu\\
   &-&  6036286 k^2 N^5-1801944 k^2 N^4+3166328 k^2
   N^3+2473168 k^2 N^2+458560 k^2 N
   \nonu\\
   &-&  13624 k N^{11}-3348 k N^{10}+313836 k N^9+404028 k N^8-937572 k N^7-2354008 k
   N^6
   \nonu\\
   &-&  1046816 k N^5+1257632 k N^4+1312320 k N^3+323840 k N^2+9084 N^{11}+65436 N^{10}
   \nonu\\
   &+& \left.  84780 N^9-111084 N^8-314616 N^7-114528 N^6+248544 N^5+260928
   N^4+72960 N^3) \right] /
\nonu\\   
&\times&
\left[ (N-2) N (N+1) (N+2)^2 (2 k+3 N+2) d(N,k) p(N,k) \right],
\nonu\\
a_{28}&=&
\left[ -3 (k+N) (2 k+N) 
(138 k^4 N^4-180 k^4 N^3-862 k^4 N^2+1120 k^4 N+1088 k^4+621 k^3 N^5  \right.
\nonu\\
&-&  396 k^3 N^4-4419
   k^3 N^3+2454 k^3 N^2+8256 k^3 N+3264 k^3+897 k^2 N^6+52 k^2 N^5
   \nonu\\
   &-&  7441 k^2 N^4-1182 k^2 N^3+14934 k^2 N^2+12528
   k^2 N+2720 k^2+414 k N^7+307 k N^6
   \nonu\\
   &-&  4715 k N^5-4213 k N^4+8413 k N^3+12674 k N^2+5200 k N+544 k-30
   N^7-878 N^6
   \nonu\\
   &-& \left.  1110 N^5+1542 N^4+3244 N^3+1632 N^2+192 N) \right]
/
\left[ (N-2) (N+1) (N+2) \right.
\nonu\\
&\times& \left.  (k+N+1) (2 k+3 N+2) d(N,k) \right],
\nonu\\
a_{29}&=&
\left[ 3 (k+N) (2 k+N) 
(110 k^4 N^5+1172 k^4 N^4-26 k^4 N^3-4976 k^4 N^2-2544 k^4 N-960 k^4  \right.
\nonu\\
&+&  375 k^3 N^6+4348
   k^3 N^5+1951 k^3 N^4-18462 k^3 N^3-18440 k^3 N^2-8336 k^3 N-2880 k^3
   \nonu\\
   &+&  355 k^2 N^7+5468 k^2 N^6+5061 k^2 N^5-22922
   k^2 N^4-33494 k^2 N^3-18032 k^2 N^2
   \nonu\\
   &-&  9136 k^2 N-2880 k^2+90 k N^8+2425 k N^7+3551 k N^6-10535 k N^5-20129
   k N^4
   \nonu\\
   &-&  10402 k N^3-5048 k N^2-3984 k N-960 k+198 N^8+342 N^7-1554 N^6-2526 N^5
   \nonu\\
   &+& \left.  404 N^4+1008 N^3-864 N^2-640 N) \right]
/
\left[ (N-2) N (N+1) (N+2) (k+N+1)  \right.
\nonu\\
&\times& \left. (2 k+3 N+2) d(N,k) \right],
\nonu\\
a_{30}&=&
\left[ 6 (k+N) (40 k^4 N^4-120 k^4 N^3-168 k^4 N^2+888 k^4 N+512 k^4+160 k^3 N^5-360 k^3 N^4  \right.
\nonu\\
&-&  816 k^3 N^3+2840
   k^3 N^2+2848 k^3 N+480 k^3+195 k^2 N^6-338 k^2 N^5-1165 k^2 N^4
   \nonu\\
   &+&  3252 k^2 N^3+4804 k^2 N^2+1368 k^2 N-32
   k^2+70 k N^7-161 k N^6-492 k N^5+2011 k N^4
   \nonu\\
   &+& \left. 
3316 k N^3+932 k N^2-208 k N-38 N^7-20 N^6+546 N^5+808 N^4+184 N^3-96
   N^2) \right] /
\nonu\\
&\times&
\left[ (N-2) (N+2) (k+N+1) (2 k+3 N+2) d(N,k) \right],
\nonu\\
a_{31}&=&
\left[ -6 (k+N) (280 k^6 N^5-16056 k^6 N^4+10008 k^6 N^3+140488 k^6 N^2+32240 k^6 N \right. 
\nonu\\
&-&  120304 k^6+1540 k^5 N^6-98252 k^5
   N^5+3388 k^5 N^4+893796 k^5 N^3+724848 k^5 N^2
   \nonu\\
   &-&  512776 k^5 N-331808 k^5+3045 k^4 N^7-230305 k^4 N^6-120541 k^4
   N^5+2176321 k^4 N^4
   \nonu\\
   &+&  3000692 k^4 N^3-291612 k^4 N^2-1552216 k^4 N-448224 k^4+2660 k^3 N^8-257834 k^3 N^7
   \nonu\\
   &-&  275698 k^3
   N^6+2576254 k^3 N^5+5096606 k^3 N^4+1182380 k^3 N^3-2949896 k^3 N^2
   \nonu\\
   &-&  2066968 k^3 N-382240 k^3+980 k^2 N^9-137796 k^2
   N^8-234303 k^2 N^7+1547899 k^2 N^6
   \nonu\\
   &+&  4172067 k^2 N^5+1932785 k^2 N^4-3114796 k^2 N^3-3786148 k^2 N^2-1403128 k^2 N
   \nonu\\
   &-&  145520
   k^2-28456 k N^9-83612 k N^8+431852 k N^7+1650396 k N^6+1127164 k N^5
   \nonu\\
   &-&  1812528 k N^4-3230608 k N^3-1797664 k
   N^2-343360 k N-11436 N^9+34044 N^8
   \nonu\\
   &+& \left.  
262476 N^7+278244 N^6-417840 N^5-1059984 N^4-781536 N^3-200640 N^2) \right]
/
\nonu\\
&\times& 
\left[ (N-2) (N+2) (k+N+1) (2 k+3 N+2) d(N,k) p(N,k) \right],
\nonu\\
a_{32}&=&
\left[ 6 (k+N) (80 k^4 N^4-32 k^4 N^3-1056 k^4 N^2-928 k^4 N+1168 k^4+300 k^3 N^5-16 k^3 N^4  \right.
\nonu\\
&-&   3980 k^3 N^3-4888
   k^3 N^2+3144 k^3 N+2064 k^3+335 k^2 N^6+78 k^2 N^5-4681 k^2 N^4
   \nonu\\
   &-&  7692 k^2 N^3+2076 k^2 N^2+5072 k^2 N+1168
   k^2+110 k N^7-29 k N^6-1952 k N^5
   \nonu\\
   &-&  3641 k N^4+840 k N^3+3844 k N^2+1736 k N+272 k-46 N^7-220 N^6-374 N^5+320
   N^4
   \nonu\\
   &+&  \left. 1176 N^3+1008 N^2+352 N) \right]
/
\left[ (N-2) (N+2) (k+N+1) (2 k+3 N+2)  d(N,k) \right],
\nonu\\
a_{33}&=&
\left[ -6 (k+N) (1680 k^7 N^6-18016 k^7 N^5-137568 k^7 N^4-158976 k^7 N^3+362000 k^7 N^2 \right. 
\nonu\\
&+&   847840 k^7 N+367936 k^7+10080
   k^6 N^7-152464 k^6 N^6-1141312 k^6 N^5-1483104 k^6 N^4
   \nonu\\
   &+&  3199776 k^6 N^3+9279728 k^6 N^2+6833344 k^6 N+1526528 k^6+22470
   k^5 N^8-506358 k^5 N^7
   \nonu\\
   &-&  3834166 k^5 N^6-5740530 k^5 N^5+10503784 k^5 N^4+38139904 k^5 N^3+37670784 k^5 N^2
   \nonu\\
   &+&  14508128 k^5
   N+1673472 k^5+23205 k^4 N^9-848375 k^4 N^8-6738131 k^4 N^7
   \nonu\\
   &-&  11948949 k^4 N^6+16361850 k^4 N^5+78904344 k^4 N^4+97622512
   k^4 N^3+51571600 k^4 N^2
   \nonu\\
   &+&  10408288 k^4 N+239104 k^4+11340 k^3 N^{10}-761834 k^3 N^9-6657412 k^3 N^8
   \nonu\\
   &-&  14318048 k^3
   N^7+12041840 k^3 N^6+89502858 k^3 N^5+135816408 k^3 N^4+91329120 k^3 N^3
   \nonu\\
   &+&  26271120 k^3 N^2+1532224 k^3 N-275776 k^3+2100
   k^2 N^{11}-350556 k^2 N^{10}-3650887 k^2 N^9
   \nonu\\
   &-&  9812615 k^2 N^8+2602561 k^2 N^7+55433623 k^2 N^6+103916830 k^2 N^5+86470824
   k^2 N^4
   \nonu\\
   &+&  33199440 k^2 N^3+3984928 k^2 N^2-353440 k^2 N-64872 k N^{11}-998436 k N^{10}
   \nonu\\
   &-&  3549284 k N^9-1346868 k
   N^8+16948828 k N^7+40948040 k N^6+42003808 k N^5
   \nonu\\
   &+&  20918528 k N^4+4254336 k N^3+71552 k N^2-96972 N^{11}-520140 N^{10}-595836 N^9
   \nonu\\
   &+& \left.  1833180
   N^8+6422712 N^7+8244384 N^6+5247552 N^5+1577088 N^4+157056 N^3)\right]
/
\nonu\\
&\times& 
\left[ (N-2) (N+2)^2 (k+N+1) (2 k+N) (2 k+3 N+2) 
d(N,k) p(N,k) \right],
\nonu\\
a_{34}&=&
\left[ 6 (k+N) 
(120 k^4 N^5+344 k^4 N^4-648 k^4 N^3-1880 k^4 N^2+1264 k^4 N+3424 k^4  \right.
\nonu\\
&+&  440 k^3 N^6+1420 k^3 N^5-2008
   k^3 N^4-8516 k^3 N^3+520 k^3 N^2+13568 k^3 N+5216 k^3
   \nonu\\
   &+&  475 k^2 N^7+1840 k^2 N^6-1681 k^2 N^5-11834 k^2 N^4-5716
   k^2 N^3+14032 k^2 N^2+10768 k^2 N
   \nonu\\
   &+&  1248 k^2+150 k N^8+731 k N^7-386 k N^6-5129 k N^5-5002 k N^4+2152 k
   N^3+1776 k N^2
   \nonu\\
   &-&  1472 k N-544 k+42 N^8+40 N^7-286 N^6-780 N^5-1616 N^4-2976 N^3-2560 N^2
   \nonu\\
   &-& \left.  704 N) \right]
/
\left[ (N-2) (N+2)^2 (k+N+1) (2 k+3 N+2) d(N,k) \right],
\nonu\\
a_{35}&=&
\left[ -6 (k+N) 
(1400 k^7 N^6+16552 k^7 N^5+49800 k^7 N^4-54536 k^7 N^3-355040 k^7 N^2 
\right. 
\nonu\\
&-&  281216 k^7 N+23040 k^7+9660
   k^6 N^7+129604 k^6 N^6+456388 k^6 N^5-159988 k^6 N^4
   \nonu\\
   &-& 2903632 k^6 N^3-3759584 k^6 N^2-1014720 k^6 N+260864 k^6+25550
   k^5 N^8+406618 k^5 N^7
   \nonu\\
   &+& 1707322 k^5 N^6+472718 k^5 N^5-9573936 k^5 N^4-17420336 k^5 N^3-9085568 k^5 N^2
   \nonu\\
   &+&  481952 k^5
   N+877184 k^5+32165 k^4 N^9+650705 k^4 N^8+3312645 k^4 N^7+2838979 k^4 N^6
   \nonu\\
   &-&  15910142 k^4 N^5-39616552 k^4 N^4-30318608 k^4
   N^3-3526128 k^4 N^2+4227680 k^4 N
   \nonu\\
   &+&  1063936 k^4+19180 k^3 N^{10}+557942 k^3 N^9+3547076 k^3 N^8+5152928 k^3 N^7
   \nonu\\
   &-&  13677080
   k^3 N^6-48638126 k^3 N^5-50275896 k^3 N^4-14542440 k^3 N^3+6419200 k^3 N^2
   \nonu\\
   &+&  3974752 k^3 N+424576 k^3+4340 k^2
   N^{11}+243108 k^2 N^{10}+2058345 k^2 N^9+4414657 k^2 N^8
   \nonu\\
   &-&  5344003 k^2 N^7-32154533 k^2 N^6-43762318 k^2 N^5-20922244 k^2
   N^4+2409472 k^2 N^3
   \nonu\\
   &+&  4756976 k^2 N^2+962528 k^2 N+42200 k N^{11}+579740 k N^{10}+1805340 k N^9-335524 k N^8
   \nonu\\
   &-&  10393396 k
   N^7-18936904 k N^6-13222992 k N^5-1794048 k N^4+1928384 k N^3
   \nonu\\
   &+&  646272 k N^2+55860 N^{11}+280212 N^{10}+216372 N^9-1195860 N^8-3176952 N^7
   \nonu\\
   &-& \left.  3101712
   N^6-1138176 N^5+83904 N^4+109440 N^3) \right]
/
\left[ (N-2) (N+2)^2 (k+N+1) \right.
\nonu\\
&\times& \left. (2 k+N) (2 k+3 N+2) 
d(N,k) p(N,k) \right],
\nonu\\
a_{36}&=&
\left[ -3 (N+1) (60 k^5 N^4-308 k^5 N^3-160 k^5 N^2+1504 k^5 N+240 k^4 N^5-1428 k^4 N^4 \right. 
\nonu\\
&-&  756 k^4 N^3+7840 k^4 N^2+736
   k^4 N-3840 k^4+315 k^3 N^6-2543 k^3 N^5-746 k^3 N^4
   \nonu\\
   &+&  15900 k^3 N^3+4048 k^3 N^2-13856 k^3 N-11520 k^3+165
   k^2 N^7-2175 k^2 N^6+644 k^2 N^5
   \nonu\\
   &+&  15416 k^2 N^4+4516 k^2 N^3-13888 k^2 N^2-23456 k^2 N-11520 k^2+30 k N^8-863
   k N^7
   \nonu\\
   &+&  1464 k N^6+6911 k N^5-2494 k N^4-6040 k N^3-5328 k N^2-8832 k N-3840 k-126 N^8
   \nonu\\
   &+& \left.  640 N^7+1290 N^6-3236 N^5-3344
   N^4+4256 N^3+5376 N^2+1536 N) \right]
/
\left[ (N-2) N \right. 
\nonu\\
&\times&\left.  (k+N+1) (2 k+N) (2 k+3 N+2) d(N,k) \right],
\nonu\\
a_{37}&=&
\left[ 
3 (N+1) (784 k^6 N^4-8912 k^6 N^3+18368 k^6 N^2+61232 k^6 N-24432 k^6+4494 k^5 N^5  \right.
\nonu\\
&-&  56220 k^5 N^4+77494 k^5 N^3+439928
   k^5 N^2+49256 k^5 N-91664 k^5+9121 k^4 N^6
   \nonu\\
   &-&  136172 k^4 N^5+108325 k^4 N^4+1139978 k^4 N^3+654248 k^4 N^2-165744
   k^4 N-110032 k^4
   \nonu\\
   &+&  7644 k^3 N^7-158302 k^3 N^6+29770 k^3 N^5+1370726 k^3 N^4+1530866 k^3 N^3+244264 k^3 N^2
   \nonu\\
   &-&  245592
   k^3 N-42800 k^3+2212 k^2 N^8-88688 k^2 N^7-66315 k^2 N^6+770768 k^2 N^5
   \nonu\\
   &+&  1563901 k^2 N^4+789102 k^2 N^3-138440
   k^2 N^2-91824 k^2 N-19208 k N^8-64748 k N^7
   \nonu\\
   &+&  156824 k N^6+758660 k N^5+637384 k N^4+7872 k N^3-73216 k N^2-17820
   N^8-4416 N^7
   \nonu\\
   &+& \left.  145716 N^6+167928 N^5+10080 N^4-25536 N^3) \right]
/
\left[ (N-2) (2 k+N) (2 k+3 N+2)  \right. 
\nonu\\
&\times&\left. d(N,k) p(N,k) \right],
\nonu\\
a_{38}&=&
\left[ -3 (N+1) 
(116 k^5 N^3-476 k^5 N^2-704 k^5 N+1088 k^5+640 k^4 N^4-2252 k^4 N^3  \right.
\nonu\\
&-&   5084 k^4 N^2+5504 k^4 N+5440
   k^4+1305 k^3 N^5-3797 k^3 N^4-12518 k^3 N^3+7628 k^3 N^2
   \nonu\\
   &+&  21248 k^3 N+6528 k^3+1175 k^2 N^6-2693 k^2 N^5-13116
   k^2 N^4+1392 k^2 N^3+26700 k^2 N^2
   \nonu\\
   &+&  17600 k^2 N+2176 k^2+394 k N^7-725 k N^6-5192 k N^5-2259 k N^4+11094
   k N^3
   \nonu\\
   &+& \left.  
12416 k N^2+3328 k N-74 N^7-256 N^6-258 N^5+564 N^4+896 N^3+256 N^2) 
\right]
/
\nonu\\
&\times& 
\left[ (N-2) (k+N+1) (2 k+N) (2 k+3 N+2) d(N,k) \right],
\nonu\\
a_{39}&=&
\left[ 6 (N+1) 
(392 k^7 N^4+8120 k^7 N^3+30568 k^7 N^2-37688 k^7 N-118128 k^7+2604 
k^6 N^5 \right. 
\nonu\\
&+&   56100 k^6 N^4+231620 k^6
   N^3-162756 k^6 N^2-986912 k^6 N-492432 k^6+6545 k^5 N^6
   \nonu\\
   &+&  156248 k^5 N^5+742237 k^5 N^4-139386 k^5 N^3-3227336 k^5
   N^2-3091736 k^5 N-710320 k^5
   \nonu\\
   &+&  7665 k^4 N^7+223735 k^4 N^6+1292109 k^4 N^5+419449 k^4 N^4-5413350 k^4 N^3
   \nonu\\
   &-&  7793520 k^4
   N^2-3445416 k^4 N-415856 k^4+4088 k^3 N^8+172238 k^3 N^7+1309507 k^3 N^6
   \nonu\\
   &+&  1111550 k^3 N^5-5028613 k^3 N^4-10097078 k^3
   N^3-6453952 k^3 N^2-1470320 k^3 N
   \nonu\\
   &-&  79840 k^3+756 k^2 N^9+66812 k^2 N^8+767677 k^2 N^7+1069161 k^2 N^6-2579087 k^2
   N^5
   \nonu\\
   &-&  7097909 k^2 N^4-5792518 k^2 N^3-1818252 k^2 N^2-167416 k^2 N+10104 k N^9+241368 k N^8
   \nonu\\
   &+&  468452 k N^7-689864 k
   N^6-2584068 k N^5-2447784 k N^4-876176 k N^3-85408 k N^2
   \nonu\\
   &+&  
\left. 32148 N^9+77844 N^8-83052 N^7-390708 N^6-377496 N^5-111408 N^4+4128 N^3) \right]
/
\nonu\\
&\times& 
\left[ (N-2) (k+N+1) (2 k+N) (2 k+3 N+2) d(N,k) p(N,k) \right],
\nonu\\
a_{40}&=&
\frac{6 (N+1) (N+2) 
(5 k^2 N-15 k^2+10 k N^2-9 k N+177 k+6 N^2+102 N)}{(2 k+N) d(N,k)},
\nonu\\
a_{41}&=&
\left[ 12 k (k+N) (20 k^4 N^2-140 k^4 N+240 k^4+80 k^3 N^3-520 k^3 N^2+680 k^3 N+480 k^3 \right. 
\nonu\\
&+&  97 k^2 N^4-663
   k^2 N^3+660 k^2 N^2+1160 k^2 N+240 k^2+34 k N^5-349 k N^4+297 k N^3
   \nonu\\
   &+&  \left. 920 k N^2+340 k N-66 N^5+54 N^4+240 N^3+120
   N^2) \right]
/
\left[ (N-2) (2 k+3 N+2) p(N,k) \right],
\nonu\\
a_{42}&=&
\left[ -12 k (k+N) 
(220 k^6 N^5+2268 k^6 N^4+16220 k^6 N^3-11004 k^6 N^2-60936 k^6 N \right. 
\nonu\\
&+&  12240 k^6+1320 k^5 N^6+12924
   k^5 N^5+82236 k^5 N^4-32532 k^5 N^3-311076 k^5 N^2
   \nonu\\
   &-&  27192 k^5 N+36720 k^5+2820 k^4 N^7+27611 k^4 N^6+163705 k^4
   N^5-31147 k^4 N^4
   \nonu\\
   &-&  605101 k^4 N^3-154156 k^4 N^2+161604 k^4 N+36720 k^4+2480 k^3 N^8+27284 k^3 N^7
   \nonu\\
   &+&  158370 k^3 N^6-16750
   k^3 N^5-562158 k^3 N^4-121966 k^3 N^3+366676 k^3 N^2+161040 k^3 N
   \nonu\\
   &+&  12240 k^3+720 k^2 N^9+12212 k^2 N^8+72264 k^2
   N^7-23945 k^2 N^6-248831 k^2 N^5
   \nonu\\
   &+&  91009 k^2 N^4+459263 k^2 N^3+249320 k^2 N^2+33180 k^2 N+1968 k N^9+10176 k N^8
   \nonu\\
   &-&  25460
   k N^7-41604 k N^6+140596 k N^5+282372 k N^4+162400 k N^3+28560 k N^2
   \nonu\\
   &-&  \left. 1152 N^9-7956 N^8-60 N^7+41220 N^6+64476 N^5+37920 N^4+7920
   N^3) \right]
/
\left[ (N-2)  \right.
\nonu\\
&\times&  
\left. (N+1) (N+2) (2 k+3 N+2) d(N,k) p(N,k) \right],
\nonu\\
a_{43}&=&   
\left[ -3 (k+N) (2620 k^6 N^7+26236 k^6 N^6+63996 k^6 N^5-112412 k^6 N^4-543112 k^6 N^3 \right. 
\nonu\\
&-&  362960 k^6 N^2+177120 k^6 N+130560
   k^6+14460 k^5 N^8+138372 k^5 N^7+339876 k^5 N^6
   \nonu\\
   &-&  533524 k^5 N^5-2889680 k^5 N^4-2314752 k^5 N^3+1097120 k^5 N^2+1605760
   k^5 N
   \nonu\\
   &+&  391680 k^5+27885 k^4 N^9+268627 k^4 N^8+713991 k^4 N^7-914923 k^4 N^6-6097016 k^4 N^5
   \nonu\\
   &-&  5500484 k^4 N^4+3582440
   k^4 N^3+6954080 k^4 N^2+3034080 k^4 N+391680 k^4+21860 k^3 N^{10}
\nonu\\
&+&  235542 k^3 N^9+730862 k^3 N^8-731386 k^3 N^7-6477154
   k^3 N^6-6103364 k^3 N^5
   \nonu\\
   &+&  6939512 k^3 N^4+14493168 k^3 N^3+8536480 k^3 N^2+1959360 k^3 N+130560 k^3+5940 k^2 N^{11}
   \nonu\\
   &+&  89116
   k^2 N^{10}+351593 k^2 N^9-344817 k^2 N^8-3702509 k^2 N^7-3203103 k^2 N^6
   \nonu\\
   &+&  7528196 k^2 N^5+15700872 k^2 N^4+11056848 k^2
   N^3+3347120 k^2 N^2+353920 k^2 N
   \nonu\\
   &+&  10680 k N^{11}+60692 k N^{10}-154756 k N^9-1177780 k N^8-691044 k N^7+4178704 k
   N^6
   \nonu\\
   &+&  8510880 k N^5+6657344 k N^4+2347840 k N^3+304640 k N^2-1260 N^{11}-43764 N^{10}
   \nonu\\
   &-&  182292 N^9-35868 N^8+900240 N^7+1802016 N^6+1502016 N^5+580800
   N^4
   \nonu\\
   &+& \left.  84480 N^3) \right]
 /
\left[ (N-2) (N+1) (N+2)^2 (2 k+3 N+2) d(N,k) p(N,k) \right],
\nonu\\
a_{44}&=&   
\left[ -12 (N+1) (k+N) 
(20 k^4 N^2-140 k^4 N+240 k^4+115 k^3 N^3-765 k^3 N^2+1100 k^3 N \right. 
\nonu\\
&+&  480 k^3+195 k^2 N^4-1435
   k^2 N^3+2182 k^2 N^2+1152 k^2 N+240 k^2+90 k N^5-985 k N^4
   \nonu\\
   &+&  \left. 2044 k N^3+631 k N^2-88 k N-210 N^5+720 N^4-6 N^3-336
   N^2) \right]
/
\left[ (N-2) \right. 
\nonu\\
&\times& \left. (2 k+3 N+2) 
p(N,k) \right],
\nonu\\
a_{45}&=&   
\left[ 6 (k+N) 
(468 k^7 N^6-2100 k^7 N^5-82852 k^7 N^4-35468 k^7 N^3+374368 k^7 N^2 
\right. 
\nonu\\
&+&  251792 k^7 N+75840 k^7+3696 k^6
   N^7-2012 k^6 N^6-516100 k^6 N^5-516244 k^6 N^4
   \nonu\\
   &+&  2236500 k^6 N^3+2822784 k^6 N^2+1344592 k^6 N+303360 k^6+11115 k^5
   N^8+31369 k^5 N^7
   \nonu\\
   &-&  1303945 k^5 N^6-2141755 k^5 N^5+5125466 k^5 N^4+10391038 k^5 N^3+7137992 k^5 N^2
   \nonu\\
   &+&  2645816 k^5 N+455040
   k^5+15675 k^4 N^9+98058 k^4 N^8-1702470 k^4 N^7-4108438 k^4 N^6
   \nonu\\
   &+&  5521993 k^4 N^5+18212204 k^4 N^4+16761034 k^4
   N^3+8306672 k^4 N^2+2446024 k^4 N
   \nonu\\
   &+&  303360 k^4+10152 k^3 N^{10}+114482 k^3 N^9-1222297 k^3 N^8-4146939 k^3 N^7
   \nonu\\
   &+&  2596823 k^3
   N^6+16947359 k^3 N^5+20080134 k^3 N^4+12365574 k^3 N^3+4852232 k^3 N^2
   \nonu\\
   &+&  1074008 k^3 N+75840 k^3+2364 k^2 N^{11}+57944 k^2
   N^{10}-482005 k^2 N^9-2237644 k^2 N^8
   \nonu\\
   &+&  153636 k^2 N^7+8477520 k^2 N^6+12681099 k^2 N^5+9344316 k^2 N^4+4479258 k^2 N^3
   \nonu\\
   &+&  1373216
   k^2 N^2+181000 k^2 N+10536 k N^{11}-101248 k N^{10}-604332 k N^9-238644 k N^8
   \nonu\\
   &+&  2132108 k N^7+3987884 k N^6+3440456
   k N^5+1944024 k N^4+751328 k N^3+138080 k N^2
   \nonu\\
   &-&  9828 N^{11}-63840 N^{10}-51120 N^9+208560 N^8+490476 N^7+494160 N^6+326568 N^5
   \nonu\\
   &+& \left.  151296 N^4+34080 N^3)\right]
/
\left[ (N-2) (N+1) (N+2) (k+N+1) (2 k+3 N+2) \right.
\nonu\\
&\times& \left. d(N,k) p(N,k) \right],
\nonu\\
a_{46}&=&   
\left[ 6 (k+N) (1256 k^8 N^7+22952 k^8 N^6-14744 k^8 N^5-290984 k^8 N^4-147760 k^8 N^3 \right. 
\nonu\\
&+&  799584 k^8 N^2+850240 k^8 N+195840
   k^8+8580 k^7 N^8+188300 k^7 N^7+114524 k^7 N^6
   \nonu\\
   &-&  1939436 k^7 N^5-2426320 k^7 N^4+4268400 k^7 N^3+8530304 k^7 N^2+4627584
   k^7 N
   \nonu\\
   &+&  783360 k^7+23026 k^6 N^9+655646 k^6 N^8+1239430 k^6 N^7-4695070 k^6 N^6
   \nonu\\
   &-&  11912464 k^6 N^5+2898856 k^6 N^4+25688576
   k^6 N^3+23977792 k^6 N^2+8791296 k^6 N
   \nonu\\
   &+&  1175040 k^6+30995 k^5 N^{10}+1253145 k^5 N^9+4004649 k^5 N^8-4086181 k^5
   N^7
   \nonu\\
   &-&  27397424 k^5 N^6-23829864 k^5 N^5+19724976 k^5 N^4+44324592 k^5 N^3+28204416 k^5 N^2
   \nonu\\
   &+&  7809280 k^5 N+783360 k^5+21999
   k^4 N^{11}+1413912 k^4 N^{10}+6447090 k^4 N^9
   \nonu\\
   &+&  2294196 k^4 N^8-33065497 k^4 N^7-67129404 k^4 N^6-41813680 k^4 N^5+13265224
   k^4 N^4
   \nonu\\
   &+&  31384928 k^4 N^3+16477280 k^4 N^2+3503808 k^4 N+195840 k^4+7640 k^3 N^{12}
   \nonu\\
   &+&  939938 k^3 N^{11}+5751067 k^3
   N^{10}+7728037 k^3 N^9-20304955 k^3 N^8-77635215 k^3 N^7
   \nonu\\
   &-&  101155164 k^3 N^6-58732732 k^3 N^5-5633280 k^3 N^4+10914208 k^3
   N^3+5252608 k^3 N^2
   \nonu\\
   &+&  708480 k^3 N+1004 k^2 N^{13}+342128 k^2 N^{12}+2831143 k^2 N^{11}+6441570 k^2 N^{10}
   \nonu\\
   &-&  4505918 k^2
   N^9-44856020 k^2 N^8-86247029 k^2 N^7-81411198 k^2 N^6-38586456 k^2 N^5
   \nonu\\
   &-&  5465256 k^2 N^4+2291152 k^2 N^3+732672 k^2 N^2+52808
   k N^{13}+690048 k N^{12}
   \nonu\\
   &+&  2383148 k N^{11}+985116 k N^{10}-12130796 k N^9-33522668 k N^8-42209368 k N^7
   \nonu\\
   &-&  28755104 k
   N^6-9874528 k N^5-1005632 k N^4+170496 k N^3+57804 N^{13}+332760 N^{12}
   \nonu\\
   &+&  475944 N^{11}-1055904 N^{10}-4895412 N^9-7985016 N^8-6966624 N^7-3346848 N^6
   \nonu\\
   &-& \left.  782016
   N^5-55296 N^4) \right]
/
\left[ (N-2) (N+1) (N+2)^2 (k+N+1) (2 k+N) (2 k+3 N+2) \right. 
\nonu\\
&\times& \left. d(N,k) p(N,k) \right],
\nonu\\   
a_{47}&=&   
\left[ 12 (k+N) (188 k^4 N^6+872 k^4 N^5-1196 k^4 N^4-7000 k^4 N^3+960 k^4 N^2+10400 k^4 N  \right.
\nonu\\
&+&  1920 k^4+654 k^3 N^7+2754
   k^3 N^6-4286 k^3 N^5-23506 k^3 N^4+2440 k^3 N^3+45640 k^3 N^2
   \nonu\\
   &+&  25120 k^3 N+3840 k^3+657 k^2 N^8+2938 k^2 N^7-4231
   k^2 N^6-26996 k^2 N^5-916 k^2 N^4
   \nonu\\
   &+&  63288 k^2 N^3+56240 k^2 N^2+17440 k^2 N+1920 k^2+202 k N^9+1137 k N^8-1446
   k N^7
   \nonu\\
   &-&  12233 k N^6-2130 k N^5+35970 k N^4+43848 k N^3+18920 k N^2+2720 k N+102 N^9
   \nonu\\
   &-& \left.  120 N^8-1830 N^7-636 N^6+7308 N^5+11376
   N^4+6000 N^3+960 N^2) \right]
/
\left[ (N-2) \right.
\nonu\\
&\times& \left. (N+2)^2 (2 k+N) (2 k+3 N+2) p(N,k) \right],
\nonu\\
a_{48}&=&   
\left[ 6 (N+1) (N+2) (400 k^7 N^3-1440 k^7 N^2-4720 k^7 N+16320 k^7+3150 k^6 N^4 \right. 
\nonu\\
&-& 10140 k^6 N^3-41490 k^6 N^2+114360 k^6
   N+48960 k^6+9605 k^5 N^5-28416 k^5 N^4
   \nonu\\
   &-&  155131 k^5 N^3+347094 k^5 N^2+315884 k^5 N+23280 k^5+14110 k^4 N^6-38456
   k^4 N^5
   \nonu\\
   &-&  311476 k^4 N^4+562184 k^4 N^3+851094 k^4 N^2+175680 k^4 N-60720 k^4+9980 k^3 N^7
   \nonu\\
   &-&  22836 k^3 N^6-359779
   k^3 N^5+505816 k^3 N^4+1173921 k^3 N^3+462242 k^3 N^2-132340 k^3 N
   \nonu\\
   &-&  77040 k^3+2760 k^2 N^8-1936 k^2 N^7-238178
   k^2 N^6+255502 k^2 N^5+841998 k^2 N^4
   \nonu\\
   &+&  506154 k^2 N^3-31084 k^2 N^2-128304 k^2 N-25680 k^2+2160 k N^8-83652 k
   N^7
   \nonu\\
   &+&  81080 k N^6+287596 k N^5+203488 k N^4+67040 k N^3-18432 k N^2-17088 k N
   \nonu\\
   &-&  \left. 12600 N^8+18264 N^7+37512 N^6+7176 N^5+22416 N^4+32832
   N^3+10944 N^2) \right]
/
\nonu\\
&\times& 
\left[ (N-2) (2 k+N) (2 k+3 N+2) d(N,k) p(N,k) \right],
\nonu\\
a_{49}&=&    
\left[ -6 
(880 k^8 N^5+23472 k^8 N^4-80848 k^8 N^3-124080 k^8 N^2+181920 k^8 N+48960 k^8 \right. 
\nonu\\
&+&  6088 k^7 N^6+162880 k^7 N^5-599264
   k^7 N^4-1150160 k^7 N^3+1245208 k^7 N^2+1143952 k^7 N
   \nonu\\
   &+&  195840 k^7+16266 k^6 N^7+456126 k^6 N^6-1941482 k^6 N^5-4458174
   k^6 N^4+3679432 k^6 N^3
   \nonu\\
   &+&  6971784 k^6 N^2+2776480 k^6 N+293760 k^6+21065 k^5 N^8+658199 k^5 N^7-3657717 k^5 N^6
   \nonu\\
   &-&  9599335
   k^5 N^5+6131424 k^5 N^4+20594612 k^5 N^3+13282768 k^5 N^2+3110304 k^5 N
   \nonu\\
   &+&  195840 k^5+13595 k^4 N^9+512546 k^4 N^8-4423504
   k^4 N^7-12800418 k^4 N^6+6061833 k^4 N^5
   \nonu\\
   &+&  34745808 k^4 N^4+31310524 k^4 N^3+11379328 k^4 N^2+1557376 k^4 N+48960 k^4
   \nonu\\
   &+&  4032
   k^3 N^{10}+201162 k^3 N^9-3500243 k^3 N^8-10995929 k^3 N^7+3165031 k^3 N^6
   \nonu\\
   &+&  35051735 k^3 N^5+40792120 k^3 N^4+19883084
   k^3 N^3+4113432 k^3 N^2+261520 k^3 N
   \nonu\\
   &+&  444 k^2 N^{11}+29176 k^2 N^{10}-1738853 k^2 N^9-5977480 k^2 N^8+353060 k^2
   N^7
   \nonu\\
   &+&  20792514 k^2 N^6+29840905 k^2 N^5+17942726 k^2 N^4+4764956 k^2 N^3+414376 k^2 N^2
   \nonu\\
   &-&  984 k N^{11}-485480 k N^{10}-1869108
   k N^9-377396 k N^8+6674980 k N^7+11423548 k N^6
   \nonu\\
   &+&  7942360 k N^5+2417904 k N^4+231328 k N^3-57828 N^{11}-255024 N^{10}-114288
   N^9
   \nonu\\
   &+& \left. 
906360 N^8+1771428 N^7+1322328 N^6+401904 N^5+27552 N^4) \right]
/
\left[ (N-2) (k+N+1) \right. 
\nonu\\
&\times&  \left. (2 k+N) (2 k+3 N+2) d(N,k) p(N,k) \right],
\nonu\\
a_{50}&=&      
\left[ -12 (N+1) (150 k^8 N^4-240 k^8 N^3-2850 k^8 N^2+2580 k^8 N+12240 k^8+1249 k^7 N^5 \right. 
\nonu\\
&-&  718 k^7 N^4-26483 k^7 N^3+1600
   k^7 N^2+102676 k^7 N+48960 k^7+4343 k^6 N^6+3823 k^6 N^5
   \nonu\\
   &-&  104259 k^6 N^4-88667 k^6 N^3+351244 k^6 N^2+415372 k^6
   N+124800 k^6+8145 k^5 N^7
   \nonu\\
   &+&  23742 k^5 N^6-219920 k^5 N^5-421350 k^5 N^4+562947 k^5 N^3+1436832 k^5 N^2+892052 k^5 N
   \nonu\\
   &+&  254400
   k^5+8745 k^4 N^8+51233 k^4 N^7-262388 k^4 N^6-876902 k^4 N^5+286289 k^4 N^4
   \nonu\\
   &+&  2581741 k^4 N^3+2482290 k^4 N^2+1235264
   k^4 N+320400 k^4+5096 k^3 N^9+55166 k^3 N^8
   \nonu\\
   &-&  171993 k^3 N^7-971946 k^3 N^6-325141 k^3 N^5+2578794 k^3 N^4+3459476
   k^3 N^3
   \nonu\\
   &+&  2191720 k^3 N^2+992136 k^3 N+205440 k^3+1252 k^2 N^{10}+29716 k^2 N^9-54803 k^2 N^8
   \nonu\\
   &-&  579053 k^2 N^7-512091
   k^2 N^6+1442035 k^2 N^5+2571930 k^2 N^4+1703922 k^2 N^3
   \nonu\\
   &+&  887596 k^2 N^2+367824 k^2 N+51360 k^2+6328 k N^{10}-6856
   k N^9-167380 k N^8
   \nonu\\
   &-&  232872 k N^7+441324 k N^6+992072 k N^5+523128 k N^4+83120 k N^3+66240 k N^2
   \nonu\\
   &+&  34176 k N-924
   N^{10}-17076 N^9-30156 N^8+66300 N^7+163656 N^6+32904 N^5
   \nonu\\
   &-& \left.  127056 N^4-98496 N^3-21888 N^2) \right]
/
\left[ (N-2) (k+N+1) (2 k+N) (2 k+3 N+2)  \right.
\nonu\\
&\times& \left. d(N,k) p(N,k) \right],
\nonu\\
a_{51}&=&     
\left[ -12 (N+1) (N+2) (20 k^4 N^2-140 k^4 N+240 k^4+80 k^3 N^3-520 k^3 N^2+680 k^3 N \right. 
\nonu\\
&+&  480 k^3+97 k^2 N^4-663 k^2
   N^3+660 k^2 N^2+1160 k^2 N+240 k^2+34 k N^5-349 k N^4
   \nonu\\
   &+& \left.  297 k N^3+920 k N^2+340 k N-66 N^5+54 N^4
+240 N^3+120 N^2) \right]
/
\left[ (N-2) (2 k+N)  \right.
\nonu\\
&\times& \left. (2 k+3 N+2) p(N,k) \right],
\nonu\\
a_{52}&=&    
\left[ 12 (N+1) (N+2) (150 k^6 N^3-540 k^6 N^2-1770 k^6 N+6120 k^6+935 k^5 N^4-2430 k^5 N^3 \right. 
\nonu\\
&-&   14705 k^5 N^2+34620 k^5 N+18360
   k^5+2120 k^4 N^5-2762 k^4 N^4-43742 k^4 N^3
   \nonu\\
   &+&  87242 k^4 N^2+53262 k^4 N+18360 k^4+2060 k^3 N^6+1608 k^3 N^5-59061
   k^3 N^4
   \nonu\\
   &+&  129870 k^3 N^3+47143 k^3 N^2-4416 k^3 N+6120 k^3+720 k^2 N^7+4448 k^2 N^6-37640 k^2 N^5
   \nonu\\
   &+&  124048 k^2
   N^4+10136 k^2 N^3-102168 k^2 N^2-21288 k^2 N+1968 k N^7-10908 k N^6
   \nonu\\
   &+&  69624 k N^5+612 k N^4-115392 k N^3-47904 k
   N^2-1152 N^7+16704 N^6+1152 N^5
   \nonu\\
   &-& \left.  36000 N^4-19296 N^3) \right]
/
\left[ (N-2) (2 k+N) (2 k+3 N+2) d(N,k) p(N,k) \right],
\label{spin5coeff}   
\eea
where
\bea
d(N,k)&\equiv& 5 k^2 N+17 k^2+10 k N^2+39 k N+17 k +22 N^2+22 N,
\nonu\\
p(N,k)&\equiv& 7 k^2 N+107 k^2+14 k N^2+221 k N+107 k+114 N^2+114 N.
\eea
Although each coefficient function is rather complicated, as before, 
the exact (or large $N$ limit) expressions for the zero mode eigenvalue equations
are rather simple. One of the reasons why (\ref{spin5coeff})
is so complicated comes from the simplification for the independent terms.
Originally, after collecting the results from Appendices $B$ and $C$, 
there are approximately one hundred fifty terms. 
By rewriting the dependent terms 
in terms of the above $52$ terms using Mathematica, 
rather simple coefficient functions 
can combine and become quite complicated. In particular, the coefficient 
functions appearing in the derivative terms 
in spin-$5$ current are most complicated.    

Furthermore, the large $N$ 't Hooft limit (\ref{limit}) on these 
coefficient functions leads to
\bea
&&a_1 \rightarrow \frac{6 (\lambda -2)^2 (\lambda -1) N}{\lambda ^3 (\lambda +2)}, \qquad a_2 \rightarrow \frac{6 (\lambda -2) (5 \lambda -6) N}{\lambda ^2 (\lambda +2)},
\qquad a_3 \rightarrow -\frac{24 (\lambda -2) N}{\lambda ^2 (\lambda +2)},
\nonu\\
&&a_4 \rightarrow \frac{12 \left(\lambda ^2-6 \lambda +10\right) N}{(\lambda -2) \lambda  (\lambda +2)}, \qquad
a_5 \rightarrow  \frac{24 (\lambda -2) N}{\lambda  (\lambda +2)},
\qquad a_6 \rightarrow  \frac{24 (\lambda -3) N}{\lambda  (\lambda +2)},
\nonu\\
&&a_7 \rightarrow \frac{12 (\lambda -6) N}{(\lambda -2) (\lambda +2)},
\qquad a_8 \rightarrow \frac{24 (\lambda -3) N}{(\lambda -2) (\lambda +2)}, \qquad a_9 \rightarrow \frac{24 N}{\lambda +2},
\nonu\\
&&a_{10} \rightarrow -\frac{6 \lambda  N}{\lambda +2}, \qquad 
a_{11} \rightarrow \frac{6 \lambda  (\lambda +3) N}{(\lambda -2) (\lambda +2)}, \qquad
a_{12} \rightarrow -\frac{6 \lambda ^2}{(\lambda -2) (\lambda +2)},
\nonu\\
&&
a_{13} \rightarrow  \frac{96 (\lambda -1) \left(4 \lambda ^2-29\right) N}{7 \lambda ^3 (\lambda +1) (\lambda +2)}, \qquad
a_{14} \rightarrow \frac{192 \left(4 \lambda ^2-29\right) N}{7 \lambda ^2 (\lambda +1) (\lambda +2)}, \qquad
a_{15} \rightarrow \frac{288 \left(4 \lambda ^2-29\right) N}{7 \lambda ^2 (\lambda +1) (\lambda +2)},
\nonu\\
&&
a_{16} \rightarrow -\frac{96 \left(4 \lambda ^2-29\right)}{7 \lambda  (\lambda +1) (\lambda +2)}, \qquad
a_{17} \rightarrow \frac{576 \left(4 \lambda ^2-29\right) N}{7 (\lambda -1) \lambda  (\lambda +1) (\lambda +2)}, 
\nonu\\
&&a_{18} \rightarrow \frac{288 \left(4 \lambda ^2-29\right) N}{7 (\lambda -2) \lambda  (\lambda +1) (\lambda +2)}, \qquad
a_{19} \rightarrow -\frac{288 \left(4 \lambda ^2-29\right)}{7 (\lambda -1) (\lambda +1) (\lambda +2)}, 
\nonu\\
&& a_{20} \rightarrow  -\frac{96 \left(4 \lambda ^2-29\right)}{7 (\lambda -2) (\lambda +1) (\lambda +2)}, \qquad
 a_{21} \rightarrow \frac{576 \left(4 \lambda ^2-29\right) N}{7 (\lambda -2) (\lambda -1) (\lambda +1) (\lambda +2)},
 \nonu\\
&& a_{22} \rightarrow  \frac{24 \lambda  \left(7 \lambda ^3-34 \lambda ^2-7 \lambda +334\right)}{7 (\lambda -2) (\lambda -1) (\lambda +1) (\lambda +2)}, \qquad
a_{23} \rightarrow -\frac{24 \lambda  (\lambda +3) \left(7 \lambda ^2+25 \lambda -82\right)}{7 (\lambda -2) (\lambda -1) (\lambda +1) (\lambda +2)},
\nonu\\
&& a_{24} \rightarrow \frac{96 \lambda ^2 \left(4 \lambda ^2-29\right)}{7 (\lambda -2) (\lambda -1) (\lambda +1) (\lambda +2) N}, \qquad
a_{25} \rightarrow -\frac{6 (\lambda -2) (\lambda -1) N^2}{\lambda ^3},
\nonu\\
&& a_{26} \rightarrow -\frac{6 \left(9 \lambda ^2-40 \lambda +4\right) N^2}{5 \lambda ^2 (\lambda +2)}, \qquad
a_{27} \rightarrow \frac{3 \left(77 \lambda ^2-240 \lambda +252\right) N^2}{5 \lambda ^2 (\lambda +2)}, 
\nonu\\
&& a_{28} \rightarrow \frac{207 (\lambda -2) N^2}{5 \lambda ^2}, \qquad
a_{29} \rightarrow \frac{3 (\lambda -2) (13 \lambda -22) N^2}{\lambda ^2 (\lambda +2)}, \qquad
a_{30} \rightarrow -\frac{6 \left(\lambda ^2-8\right) N^2}{\lambda  (\lambda +2)},
\nonu\\
&&a_{31} \rightarrow -\frac{6 \left(3 \lambda ^2-4 \lambda +8\right) N^2}{\lambda  (\lambda +2)}, \qquad
a_{32} \rightarrow  -\frac{6 \left(\lambda ^2+4 \lambda -16\right) N^2}{\lambda  (\lambda +2)}, 
\nonu\\
&&a_{33} \rightarrow -\frac{18 \left(\lambda ^2-4 \lambda +8\right) N^2}{\lambda  (\lambda +2)}, \qquad
a_{34} \rightarrow -\frac{6 \left(\lambda ^2+8 \lambda -24\right) N^2}{\lambda  (\lambda +2)},
\nonu\\
&&a_{35} \rightarrow \frac{6 \left(\lambda ^3-6 \lambda ^2-4 \lambda +40\right) N^2}{(\lambda -2) \lambda  (\lambda +2)}, \qquad
a_{36} \rightarrow \frac{9 (\lambda -2) N^2}{\lambda +2}, 
\nonu\\
&&a_{37} \rightarrow \frac{3 \left(3 \lambda ^2+30 \lambda -112\right) N^2}{5 (\lambda -2) (\lambda +2)}, \qquad
a_{38} \rightarrow \frac{3 \left(21 \lambda ^2+60 \lambda +116\right) N^2}{5 (\lambda -2) (\lambda +2)},
\nonu\\
&&a_{39} \rightarrow \frac{6 \left(9 \lambda ^2+20 \lambda -56\right) N^2}{5 (\lambda -2) (\lambda +2)}, \qquad
a_{40} \rightarrow -\frac{6 \lambda  N}{\lambda -2}, \qquad
a_{41} \rightarrow \frac{12 (\lambda -1) \left(3 \lambda ^2-20\right) N^3}{7 \lambda ^3 (\lambda +2)},
\nonu\\
&&a_{42} \rightarrow -\frac{48 (\lambda -1) \left(2 \lambda ^2-11\right) N^3}{7 \lambda ^3 (\lambda +2)}, \qquad
a_{43} \rightarrow  -\frac{3 \left(25 \lambda ^2-252 \lambda +524\right) N^3}{7 \lambda ^2 (\lambda +2)},
\nonu\\
&&a_{44} \rightarrow \frac{60 (\lambda -4) (2 \lambda +1) N^3}{7 \lambda ^2 (\lambda +2)}, \qquad
a_{45} \rightarrow -\frac{18 \left(99 \lambda ^2-140 \lambda -156\right) N^3}{35 \lambda ^2 (\lambda +2)}, 
\nonu\\
&&a_{46} \rightarrow -\frac{6 \left(70 \lambda ^3-113 \lambda ^2+420 \lambda -628\right) N^3}{35 \lambda ^2 (\lambda +2)}, \qquad
a_{47} \rightarrow -\frac{12 \left(11 \lambda ^2-98 \lambda +188\right) N^3}{7 (\lambda -2) \lambda  (\lambda +2)},
\nonu\\
&&a_{48} \rightarrow -\frac{6 \left(7 \lambda ^3-19 \lambda ^2+70 \lambda +80\right) N^3}{7 (\lambda -2) \lambda  (\lambda +2)}, \qquad
a_{49} \rightarrow \frac{6 \left(133 \lambda ^3+50 \lambda ^2-952 \lambda +880\right) N^3}{35 (\lambda -2) \lambda  (\lambda +2)}, 
\nonu\\
&&a_{50} \rightarrow \frac{12 \left(14 \lambda ^3+100 \lambda ^2+49 \lambda +150\right) N^3}{35 (\lambda -2) \lambda  (\lambda +2)}, \qquad
a_{51} \rightarrow -\frac{12 \left(3 \lambda ^2-20\right) N^2}{7 (\lambda -2) (\lambda +2)},
\nonu\\
&&a_{52} \rightarrow \frac{12 (\lambda -10) (\lambda +3) N^2}{7 (\lambda -2) (\lambda +2)}.
\label{limitspin5coeff}
\eea 
In this case, each coefficient function has a very simple form after taking 
the large $N$ limit.  
The large $N$ behavior of these coefficient functions can be classified as
follows:
\bea
&& a_{24} \rightarrow \frac{1}{N},
\nonu \\
&& a_{12}, \, a_{16}, \, a_{19}, \, a_{20}, \, a_{22}, \, a_{23}, \,
 \rightarrow \mbox{const}, \nonu \\
&& a_1, \, a_2, \, a_3, \, a_4, \,
a_5, \, a_6, \, a_7, \, a_8, \,
a_9, \, a_{10}, \, a_{11}, \, a_{13}, \,
a_{14}, \, a_{15}, \, a_{17}, \, a_{18}, \,
a_{21}, \, a_{40} \rightarrow N,
\nonu \\
&& a_{25}, \,  a_{26}, \, a_{27}, \, a_{28}, \, a_{29}, \, a_{30}, \, a_{31}, 
\, a_{32},
\, a_{33}, \, a_{34}, \, a_{35}, \, a_{36}, \, a_{37}, \, a_{38}, \, a_{39},
\, a_{51}, \, a_{52}
\rightarrow N^2,
\nonu \\
&& a_{41}, \, a_{42}, \, a_{43}, \, a_{44}, \, a_{45}, \, a_{46}, \, a_{47}, 
\, a_{48}, \, a_{49}, 
\, a_{50} \rightarrow N^3. 
\label{Nbehavior}
\eea
In section $4$, the eigenvalue equations and three-point functions are 
analyzed using (\ref{limitspin5coeff}) and (\ref{Nbehavior}).

\section{Eigenvalue equations for the spin-$5$ current 
on the perturbative state}


Appendix $E$ presents the zero mode eigenvalue equations on the $52$ terms of
spin-$5$ current with perturbative state $| (f;0)>$.
At each expression, the exact $N$-dependent expressions appearing 
in (\ref{ddd1})-(\ref{dtrace}) are assumed. The only large $N$ limit is
presented. 
\bea
d^{abf} d^{fcg} d^{gde} (J^a J^b J^c J^d J^e )_0 |(f;0)>&=&
-\frac{1}{N} d^{abf} d^{fcg} d^{gde}  \mbox{Tr} ( T^a T^b T^c T^d T^e) 
|(f;0)>
\nonu \\
& \rightarrow &i N^4  |(f;0)>,
\label{1}
\\
d^{abf} d^{fcg} d^{gde} (J^a J^b J^c J^d K^e )_0 |(f;0)>&=&
\frac{1}{N} d^{abf} d^{fcg} d^{gde}  \mbox{Tr} ( T^a T^b T^c T^d T^e) 
|(f;0)>
\nonu\\
& \rightarrow &-i N^4  |(f;0)>,
\label{2}
\\
d^{abf} d^{feg} d^{gcd} (J^a J^b J^c J^d K^e )_0 |(f;0)>&=&
\frac{1}{N} d^{abf} d^{fcg} d^{gde}  \mbox{Tr} ( T^a T^b T^c T^d T^e) 
|(f;0)>
\nonu\\
&\rightarrow &-i N^4  |(f;0)>,
\label{3}
\\
d^{abf} d^{fcg} d^{gde} (J^a J^b J^c K^d K^e )_0 |(f;0)>&=&
 -\frac{1}{N} d^{abf} d^{fcg} d^{gde}   \mbox{Tr} (T^a T^b T^c T^d T^e )
|(f;0)>
\nonu\\
& \rightarrow &i N^4 |(f;0)>,
\label{4}
\\
d^{bdf} d^{fag} d^{gce} (J^a J^b J^c K^d K^e )_0 |(f;0)>&=&
 -\frac{1}{N} d^{acf} d^{fbg} d^{gde}   \mbox{Tr} (T^a T^b T^c T^d T^e )
|(f;0)>
\nonu\\
& \rightarrow &-4i N^2 |(f;0)>,
\label{5}
\\
d^{abf} d^{fdg} d^{gce} (J^a J^b J^c K^d K^e )_0 |(f;0)>&=&
 -\frac{1}{N} d^{abf} d^{fcg} d^{gde} \mbox{Tr} (T^a T^b T^c T^d T^e) |(f;0)>
\nonu\\
& \rightarrow &i N^4 |(f;0)>,
\label{6}
\\
d^{abf} d^{fcg} d^{gde} (J^a J^b K^c K^d K^e )_0 |(f;0)>&=&
 \frac{1}{N} d^{abf} d^{fcg}d^{gde}  \mbox{Tr} (T^a T^b T^c T^d T^e) |(f;0)>
\nonu\\
& \rightarrow  &-i N^4 |(f;0)>,
\label{7}
\\
d^{bcf} d^{fag} d^{gde} (J^a J^b K^c K^d K^e )_0 |(f;0)>&=&
 \frac{1}{N} d^{acf} d^{fbg} d^{gde} \mbox{Tr} (T^a T^b T^c T^d T^e) |(f;0)>
\nonu\\
& \rightarrow &4i N^2 |(f;0)>,
\label{8}
\\
d^{acf} d^{fdg} d^{gbe} (J^a J^b K^c K^d K^e )_0 |(f;0)>&=&
 \frac{1}{N} d^{abf} d^{fcg} d^{gde}  \mbox{Tr} ( T^a T^b T^c T^d T^e) 
|(f;0)>
\nonu\\
& \rightarrow &-i N^4 |(f;0)>,
\label{9}
\\
d^{abf} d^{fcg} d^{gde} (J^a K^b K^c K^d K^e )_0 |(f;0)>&=&
-\frac{1}{N} d^{abf} d^{fcg}d^{gde}  \mbox{Tr} (T^a T^b T^c T^d T^e) |(f;0)>
\nonu\\
& \rightarrow &i N^4 |(f;0)>,
\label{10}
\\
d^{bcf} d^{fag} d^{gde} (J^a K^b K^c K^d K^e )_0 |(f;0)>&=&
-\frac{1}{N} d^{abf} d^{fcg}d^{gde}  \mbox{Tr} (T^a T^b T^c T^d T^e) |(f;0)>
\nonu\\
& \rightarrow &i N^4 |(f;0)>,
\label{11}
\\
d^{abf} d^{fcg} d^{gde} (K^a K^b K^c K^d K^e )_0 |(f;0)>&=&
\frac{1}{N} d^{abf}d^{fcg} d^{gde}   \mbox{Tr} (T^a T^b T^c T^d T^e) |(f;0)>
\nonu\\
& \rightarrow &-i N^4 |(f;0)>,
\label{12}
\\
d^{abc} \delta^{de} (J^a J^b J^c J^d J^e )_0 |(f;0)>&=&
-\frac{1}{N} d^{abc} \delta^{de}  \mbox{Tr} ( T^a T^b T^c T^d T^e) 
|(f;0)>
\nonu\\
& \rightarrow &i N^3  |(f;0)>,
\label{13}
\\
d^{abc} \delta^{de} (J^a J^b J^c J^d K^e )_0 |(f;0)>&=&
\frac{1}{N} d^{abc} \delta^{de}  \mbox{Tr} (  T^a T^b T^c T^d T^e) 
|(f;0)>
\nonu\\
& \rightarrow &-i N^3  |(f;0)>,
\label{14}
\\
d^{abe} \delta^{cd} (J^a J^b J^c J^d K^e )_0 |(f;0)>&=&
\frac{1}{N} d^{abc} \delta^{de}  \mbox{Tr} (  T^a T^b T^c T^d T^e) 
|(f;0)>
\nonu\\
& \rightarrow &-i N^3  |(f;0)>,
\label{15}
\\
d^{abc} \delta^{de} (J^a J^b J^c K^d K^e )_0 |(f;0)>&=&
-\frac{1}{N} d^{abc} \delta^{de}  \mbox{Tr} (  T^a T^b T^c T^d T^e) 
|(f;0)>
\nonu\\
& \rightarrow &i N^3  |(f;0)>,
\label{16}
\\
d^{abd} \delta^{ce} (J^a J^b J^c K^d K^e )_0 |(f;0)>&=&
-\frac{1}{N} d^{abc} \delta^{de}  \mbox{Tr} ( T^a T^b T^c T^d T^e) 
|(f;0)>
\nonu\\
& \rightarrow &i N^3  |(f;0)>,
\label{17}
\\
\delta^{ab} d^{cde} (J^a J^b J^c K^d K^e )_0 |(f;0)>&=&
-\frac{1}{N} d^{abc} \delta^{de}   \mbox{Tr} ( T^a T^b T^c T^d T^e ) 
|(f;0)>
\nonu\\
& \rightarrow &i N^3  |(f;0)>,
\label{18}
\\
d^{abc} \delta^{de} (J^a J^b K^c K^d K^e )_0 |(f;0)>&=&
\frac{1}{N} d^{abc} \delta^{de}  \mbox{Tr} ( T^a T^b T^c T^d T^e) 
|(f;0)>
\nonu\\
& \rightarrow &-i N^3  |(f;0)>,
\label{19}
\\
\delta^{ab}d^{cde}  (J^a J^b K^c K^d K^e )_0 |(f;0)>&=&
\frac{1}{N} d^{abc} \delta^{de}  \mbox{Tr} ( T^a T^b T^c T^d T^e) 
|(f;0)>
\nonu\\
& \rightarrow &-i N^3  |(f;0)>,
\label{20}
\\
\delta^{ac}d^{bde}  (J^a J^b K^c K^d K^e )_0 |(f;0)>&=&
\frac{1}{N} d^{abc} \delta^{de}  \mbox{Tr} ( T^a T^b T^c T^d T^e ) 
|(f;0)>
\nonu\\
& \rightarrow &-i N^3  |(f;0)>,
\label{21}
\\
d^{abc} \delta^{de} (J^a K^b K^c K^d K^e )_0 |(f;0)>&=&
-\frac{1}{N} d^{abc} \delta^{de}  \mbox{Tr} ( T^a T^b T^c T^d T^e  ) 
|(f;0)>
\nonu\\
& \rightarrow &i N^3  |(f;0)>,
\label{22}
\\
\delta^{ab} d^{cde} (J^a K^b K^c K^d K^e )_0 |(f;0)>&=&
-\frac{1}{N} d^{abc} \delta^{de}  \mbox{Tr} ( T^a T^b T^c T^d T^e  ) 
|(f;0)>
\nonu\\
& \rightarrow &i N^3  |(f;0)>,
\label{23}
\\
d^{abc} \delta^{de} (K^a K^b K^c K^d K^e )_0 |(f;0)>&=&
\frac{1}{N} d^{abc} \delta^{de}  \mbox{Tr} ( T^a T^b T^c T^d T^e) 
|(f;0)>
\nonu\\
& \rightarrow &-i N^3  |(f;0)>,
\label{24}
\\
f^{ade} d^{bce} (J^a J^b J^c \pa J^d )_0 |(f;0)>&=&
\frac{1}{N} d^{abe} f^{cde}   \mbox{Tr} ( T^a T^b T^c T^d)  |(f;0)> \nonu\\
& \rightarrow &i N^3  |(f;0)>,
\label{25}
\\
d^{abe} f^{cde}  (J^a J^b J^c \pa K^d )_0 |(f;0)>&=&
\frac{1}{N} d^{abe} f^{cde}  \mbox{Tr} ( T^a T^b T^c T^d)  |(f;0)> \nonu\\
&\rightarrow &i N^3  |(f;0)>,
\label{26}
\\
d^{abe} f^{cde}  (J^a J^b \pa J^c  K^d )_0 |(f;0)>&=&
\frac{1}{N} d^{abe} f^{cde}  \mbox{Tr} ( T^a T^b T^c T^d)  |(f;0)> \nonu\\
& \rightarrow &i N^3  |(f;0)>,
\label{27}
\\
d^{ace} f^{bde}  (J^a J^b \pa J^c  K^d )_0 |(f;0)>&=&
\frac{1}{N} d^{ace} f^{bde}  \mbox{Tr} ( T^a T^b T^c T^d)  |(f;0)> \nonu\\
&=&0,
\label{28}
\\
d^{ade} f^{bce}  (J^a J^b \pa J^c  K^d )_0 |(f;0)>&=&
\frac{1}{N} d^{abe} f^{cde}  \mbox{Tr} ( T^a T^b T^c T^d)  |(f;0)> \nonu\\
& \rightarrow &i N^3  |(f;0)>,
\label{29}
\\
d^{abe} f^{cde}  (J^a J^b  K^c \pa K^d )_0 |(f;0)>&=&
\frac{1}{N} d^{abe} f^{cde}  \mbox{Tr} ( T^a T^b T^c T^d)  |(f;0)> \nonu\\
&\rightarrow & i N^3  |(f;0)>,
\label{30}
\\
d^{ace} f^{bde}  (J^a J^b  K^c \pa K^d )_0 |(f;0)>&=&
-\frac{1}{N} d^{abe} f^{cde}   \mbox{Tr} ( T^a T^b T^c T^d)  |(f;0)> \nonu\\
& \rightarrow &-i N^3  |(f;0)>,
\label{31}
\\
d^{ade} f^{bce}  (J^a J^b  K^c \pa K^d )_0 |(f;0)>&=&
\frac{1}{N} d^{ace} f^{bde}  \mbox{Tr} ( T^a T^b T^c T^d)  |(f;0)> \nonu\\
&=&0,
\label{32}
\\
d^{ace} f^{bde}  (J^a \pa J^b  K^c K^d )_0 |(f;0)>&=&
-\frac{1}{N} d^{abe} f^{cde}  \mbox{Tr} ( T^a T^b T^c T^d)  |(f;0)> \nonu\\
& \rightarrow &-i N^3  |(f;0)>,
\label{33}
\\
 f^{ade} d^{bce} (J^a \pa J^b  K^c K^d )_0 |(f;0)>&=&
-\frac{1}{N}  d^{ace} f^{bde}   \mbox{Tr} ( T^a T^b T^c T^d)  |(f;0)> \nonu\\
&=&0,
\label{34}
\\
f^{abe} d^{cde}   (J^a \pa J^b  K^c K^d )_0 |(f;0)>&=&
-\frac{1}{N} d^{abe} f^{cde}   \mbox{Tr} ( T^a T^b T^c T^d)  |(f;0)> \nonu\\
& \rightarrow &-i N^3  |(f;0)>,
\label{35}
\\
d^{abe} f^{cde}   (J^a K^b  K^c \pa K^d )_0 |(f;0)>&=&
-\frac{1}{N} d^{abe} f^{cde}   \mbox{Tr} (  T^a T^b T^c T^d)  |(f;0)> \nonu\\
& \rightarrow &-i N^3  |(f;0)>,
\label{36}
\\
f^{ade} d^{bce}   (J^a K^b  K^c \pa K^d )_0 |(f;0)>&=&
\frac{1}{N} d^{abe} f^{cde}    \mbox{Tr} (  T^a T^b T^c T^d)  |(f;0)> \nonu\\
& \rightarrow &i N^3  |(f;0)>,
\label{37}
\\
f^{ace} d^{bde}   (J^a K^b  K^c \pa K^d )_0 |(f;0)>&=&
-\frac{1}{N}d^{ace} f^{bde}    \mbox{Tr} ( T^a T^b T^c T^d)  |(f;0)> \nonu\\
&=&0,
\label{38}
\\
f^{ade} d^{bce}   (\pa J^a K^b  K^c  K^d )_0 |(f;0)>&=&
\frac{1}{N} d^{abe} f^{cde}    \mbox{Tr} (  T^a T^b T^c T^d)  |(f;0)> \nonu\\
& \rightarrow &i N^3  |(f;0)>,
\label{39}
\\
f^{ade} d^{bce}   ( K^a K^b  K^c \pa K^d )_0 |(f;0)>&=&
-\frac{1}{N} d^{abe} f^{cde}    \mbox{Tr} (  T^a T^b T^c T^d)  |(f;0)> \nonu\\
& \rightarrow &-i N^3  |(f;0)>,
\label{40}
\\
d^{abc} (J^a J^b \pa^2 J^c )_0 |(f;0)>
&=&-\frac{2}{N}d^{abc}     \mbox{Tr} ( T^a T^b T^c) |(f;0)> \nonu\\
& \rightarrow &-2i N^2  |(f;0)>,
\label{41}
\\
d^{abc} (J^a \pa J^b \pa J^c )_0 |(f;0)>
&=&-\frac{1}{N}d^{abc}     \mbox{Tr} ( T^a T^b T^c) |(f;0)> \nonu\\
& \rightarrow &-i N^2  |(f;0)>,
\label{42}
\\
d^{abc} (J^a \pa^2 J^b K^c )_0 |(f;0)>
&=&\frac{2}{N}d^{abc}     \mbox{Tr} ( T^a T^b T^c) |(f;0)> \nonu\\
& \rightarrow &2i N^2  |(f;0)>,
\label{43}
\\
d^{abc} (J^a  J^b \pa^2 K^c )_0 |(f;0)>
&=&\frac{2}{N}d^{abc}     \mbox{Tr} ( T^a T^b T^c) |(f;0)> \nonu\\
& \rightarrow &2i N^2  |(f;0)>,
\label{44}
\\
d^{abc} (\pa J^a \pa J^b  K^c )_0 |(f;0)>
&=&\frac{1}{N}d^{abc}     \mbox{Tr} ( T^a T^b T^c) |(f;0)> \nonu\\
& \rightarrow &i N^2  |(f;0)>,
\label{45}
\\
d^{abc} ( J^a \pa J^b  \pa K^c )_0 |(f;0)>
&=&\frac{1}{N}d^{abc}     \mbox{Tr} ( T^a T^b T^c) |(f;0)> \nonu\\
& \rightarrow &i N^2  |(f;0)>,
\label{46}
\\
d^{abc} ( \pa^2 J^a K^b K^c )_0 |(f;0)>
&=&-\frac{2}{N}d^{abc}     \mbox{Tr} ( T^a T^b T^c) |(f;0)> \nonu\\
& \rightarrow &-2i N^2  |(f;0)>,
\label{47}
\\
d^{abc} (J^a K^b \pa^2 K^c )_0 |(f;0)>
&=&-\frac{2}{N}d^{abc}     \mbox{Tr} ( T^a T^b T^c) |(f;0)> \nonu\\
& \rightarrow &-2i N^2  |(f;0)>,
\label{48}
\\
d^{abc} ( \pa J^a K^b  \pa K^c )_0 |(f;0)>
&=&-\frac{1}{N}d^{abc}     \mbox{Tr} ( T^a T^b T^c) |(f;0)> \nonu\\
& \rightarrow &-i N^2  |(f;0)>,
\label{49}
\\
d^{abc} (  J^a \pa K^b  \pa K^c )_0 |(f;0)>
&=&-\frac{1}{N}d^{abc}     \mbox{Tr} ( T^a T^b T^c) |(f;0)> \nonu\\
& \rightarrow &-i N^2  |(f;0)>,
\label{50}
\\
d^{abc} (K^a K^b \pa^2 K^c )_0 |(f;0)>
&=&\frac{2}{N}d^{abc}     \mbox{Tr} ( T^a T^b T^c) |(f;0)> \nonu\\
& \rightarrow &2i N^2  |(f;0)>,
\label{51}
\\
d^{abc} ( K^a \pa K^b  \pa K^c )_0 |(f;0)>
&=&\frac{1}{N}d^{abc}     \mbox{Tr} ( T^a T^b T^c) |(f;0)> \nonu\\
& \rightarrow &i N^2  |(f;0)>.
\label{52}
\eea
From these results, the large $N$ behavior 
for these $52$ eigenvalue equations can be classified as follows:
\bea
&& (\mbox{5th}, \mbox{8th}, \mbox{41st}, \mbox{42nd},
\mbox{43rd}, \mbox{44th}, \mbox{45th},
\mbox{46th}, \mbox{47th}, \mbox{48th},
\mbox{49th}, \mbox{50th}, \mbox{51st}, \mbox{52nd})-
\mbox{terms} 
\rightarrow N^2,
\nonu \\
&& (\mbox{13th}, \mbox{14th}, \mbox{15th}, \mbox{16th},
\mbox{17th}, \mbox{18th}, \mbox{19th},
\mbox{20th}, \mbox{21st}, 
\mbox{22nd}, \mbox{23rd},
\mbox{24th}, \mbox{25th},
\mbox{26th}, \mbox{27th},
\mbox{29th}, \nonu \\
&& \mbox{30th}, \mbox{31st}, \mbox{33rd}, \mbox{35th},\mbox{36th}, \mbox{37th}, \mbox{39th},\mbox{40th})-\mbox{terms} 
\rightarrow N^3,
\nonu \\ 
&& (\mbox{1st}, \mbox{2nd}, \mbox{3rd}, \mbox{4th}, \mbox{6th}, 
\mbox{7th}, \mbox{9th}, \mbox{10th}, \mbox{11th}, 
\mbox{12th})-
\mbox{terms} \rightarrow N^4.
\label{Nbehavior2}
\eea
In section $4$, the leading $N^5$ behavior in the spin-$5$ current 
is analyzed using (\ref{Nbehavior}) and (\ref{Nbehavior2}).

\section{Eigenvalue equations for other higher spin currents of spins 
$s=2,3,4$}

The eigenvalue equations for the zero mode of the higher spin-$4$
current can be analyzed as follows: 
\bea
&& d^{abcd} (J^a J^b J^c J^d )_0 |(f;0)>
=\frac{1}{N} d^{abcd}  \mbox{Tr} ( T^a T^b T^c T^d) |(f;0)>
\nonu\\
&& =\frac{2(N^2-1)(N^2-4)(N^2-9)}{N(N^2+1)}  |(f;0)>
 \rightarrow  2 N^3|(f;0)>,
\label{4-1} \\
&& d^{abcd} (J^a J^b J^c K^d )_0 |(f;0)>
=-\frac{1}{N} d^{abcd}  \mbox{Tr} ( T^a T^b T^c T^d) |(f;0)>
\nonu\\
& &=-\frac{2(N^2-1)(N^2-4)(N^2-9)}{N(N^2+1)}  |(f;0)>
\rightarrow -2 N^3|(f;0)>,
\label{4-2} \\
&& d^{abcd} (J^a J^b K^c K^d )_0 |(f;0)>
=\frac{1}{N} d^{abcd}  \mbox{Tr} ( T^a T^b T^c T^d) |(f;0)>
\nonu\\
&& =\frac{2(N^2-1)(N^2-4)(N^2-9)}{N(N^2+1)}  |(f;0)>
\rightarrow 2 N^3|(f;0)>,
\label{4-3} \\
&& d^{abcd} (J^a K^b K^c K^d )_0 |(f;0)>
=-\frac{1}{N} d^{abcd}  \mbox{Tr} ( T^a T^b T^c T^d) |(f;0)>
\nonu\\
&& =-\frac{2(N^2-1)(N^2-4)(N^2-9)}{N(N^2+1)}  |(f;0)>
\rightarrow - 2 N^3|(f;0)>,
\label{4-4} \\
&& d^{abcd} (K^a K^b K^c K^d )_0 |(f;0)>
=\frac{1}{N} d^{abcd}  \mbox{Tr} ( T^a T^b T^c T^d) |(f;0)> \nonu\\
&& =\frac{2(N^2-1)(N^2-4)(N^2-9)}{N(N^2+1)}  |(f;0)> \rightarrow 2 N^3|(f;0)>,
\label{4-5} \\
&& d^{abe}d^{cde} (J^a J^b J^c J^d )_0 |(f;0)>
=\frac{1}{N} d^{abe}d^{cde}  \mbox{Tr} ( T^a T^b T^c T^d) |(f;0)>
\nonu\\
&& =\frac{1}{N^3}(N^2-4)^2(N^2-1)  |(f;0)>
\rightarrow  N^3|(f;0)>,
\label{4-6} \\
&& d^{abe}d^{cde} (J^a J^b J^c K^d )_0 |(f;0)>
=-\frac{1}{N} d^{abe}d^{cde}  \mbox{Tr} ( T^a T^b T^c T^d) |(f;0)>
\nonu\\
&& =-\frac{1}{N^3}(N^2-4)^2(N^2-1)  |(f;0)>
\rightarrow  -N^3|(f;0)>,
\label{4-7} \\
&& d^{abe}d^{cde} (J^a J^b K^c K^d )_0 |(f;0)>
=\frac{1}{N} d^{abe}d^{cde}  \mbox{Tr} ( T^a T^b T^c T^d) |(f;0)>
\nonu\\
&& =\frac{1}{N^3}(N^2-4)^2(N^2-1)  |(f;0)>
\rightarrow   N^3|(f;0)>,
\label{4-8} \\
&& d^{abe}d^{cde} (J^a K^b K^c K^d )_0 |(f;0)>
-\frac{1}{N} d^{abe}d^{cde}  \mbox{Tr} ( T^a T^b T^c T^d) |(f;0)>
\nonu\\
&& =-\frac{1}{N^3}(N^2-4)^2(N^2-1)  |(f;0)>
\rightarrow  -N^3|(f;0)>,
\label{4-9} \\
&& d^{abe}d^{cde} (K^a K^b K^c K^d )_0 |(f;0)>
=\frac{1}{N} d^{abe}d^{cde}  \mbox{Tr} ( T^a T^b T^c T^d) |(f;0)>
\nonu\\
&& =\frac{1}{N^3}(N^2-4)^2(N^2-1)  |(f;0)>
 \rightarrow  N^3|(f;0)>,
\label{4-10} \\
&& \delta^{ab} \delta^{cd} (J^a J^b J^c J^d )_0 |(f;0)>
=\frac{1}{N} \delta^{ab} \delta^{cd}  \mbox{Tr} ( T^a T^b T^c T^d) |(f;0)>
\nonu\\
&& =\frac{1}{N^2} (N^2-1)^2  |(f;0)>
\rightarrow  N^2|(f;0)>,
\label{4-11} \\
&& \delta^{ab} \delta^{cd} (J^a J^b J^c K^d )_0 |(f;0)>
=-\frac{1}{N} \delta^{ab} \delta^{cd}  \mbox{Tr} (  T^a T^b T^c T^d) |(f;0)>
\nonu\\
&&=-\frac{1}{N^2} (N^2-1)^2  |(f;0)>
\rightarrow  -N^2|(f;0)>,
\label{4-12} \\
&& \delta^{ab} \delta^{cd} (J^a J^b K^c K^d )_0 |(f;0)>
=\frac{1}{N} \delta^{ab} \delta^{cd}  \mbox{Tr} ( T^a T^b T^c T^d) |(f;0)>
\nonu\\
&&=\frac{1}{N^2} (N^2-1)^2  |(f;0)> \rightarrow  N^2|(f;0)>,
\label{4-13} \\
&& \delta^{ab} \delta^{cd} (J^a K^b K^c K^d )_0 |(f;0)>
=-\frac{1}{N} \delta^{ab} \delta^{cd}  \mbox{Tr} ( T^a T^b T^c T^d) |(f;0)>
\nonu\\
&&=-\frac{1}{N^2} (N^2-1)^2  |(f;0)>
\rightarrow  -N^2|(f;0)>,
\label{4-14} \\
&& \delta^{ab} \delta^{cd} (K^a K^b K^c K^d )_0 |(f;0)>
=\frac{1}{N} \delta^{ab} \delta^{cd}  \mbox{Tr} ( T^a T^b T^c T^d) |(f;0)>
\nonu\\
&&=\frac{1}{N^2} (N^2-1)^2  |(f;0)>
\rightarrow  N^2|(f;0)>,
\label{4-15} \\
&& (\pa^2 J^a K^a )_0 |(f;0)>
= -\frac{2}{N} \delta^{ab}  \mbox{Tr} ( T^a T^b)|(f;0)> 
 =\frac{2}{N}(N^2-1) |(f;0)> \nonu \\
&& \rightarrow  2N|(f;0)>,
\label{4-17} \\
&& (\pa J^a \pa K^a )_0 |(f;0)>
=-\frac{1}{N} \delta^{ab}  \mbox{Tr} ( T^a T^b)|(f;0)> 
 =\frac{1}{N}(N^2-1) |(f;0)> \nonu \\
&& \rightarrow  N|(f;0)>,
\label{4-20} \\
&& (J^a \pa^2 K^a )_0 |(f;0)>
=-\frac{2}{N} \delta^{ab}  \mbox{Tr} ( T^a T^b)|(f;0)> 
 =  \frac{2}{N}(N^2-1) |(f;0)>
\nonu \\
&& \rightarrow  2N|(f;0)>,
\label{4-22} \\
&& \delta^{ac} \delta^{bd} (J^a J^b K^c K^d )_0 |(f;0)>
=\frac{1}{N} \delta^{ab} \delta^{cd} \mbox{Tr} ( T^a T^b T^c T^d)|(f;0)> \nonu\\
&& =\frac{1}{N^2} (N^2-1)^2  |(f;0)>
\rightarrow  N^2|(f;0)>.
\label{4-23}
\eea
In particular, the following identity with (\ref{dabcd})
can be  used to simplify
(\ref{4-1})-(\ref{4-5}):
\bea
d^{abcd} \mbox{Tr} (T^a T^b T^c T^d) 
&= & 3  d^{abe}d^{cde} \mbox{Tr} (T^a T^b T^c T^d)
-\frac{12(N^2-4)}{N(N^2+1)} \delta^{ab} \delta^{cd} \mbox{Tr} (T^a T^b T^c T^d)
\nonu\\
&+& \left(-\frac{3(N^2-4)(N^2-3)}{N^2+1}+\frac{4(N^2-4)(N^2-3)}{N^2+1} \right) \delta^{ab} \mbox{Tr} (T^a T^b )
\nonu\\
&=&\frac{2(N^2-1)(N^2-4)(N^2-9)}{N^2+1},
\label{fourdidentity}
\eea
where the quartic product of the generators with (\ref{TT1}) 
has the following identities 
\bea
d^{abe}d^{cde} \mbox{Tr} (T^a T^b T^c T^d) 
&= & \frac{1}{N^2}(N^2-4)^2(N^2-1), 
\nonu \\
\delta^{ab} \delta^{cd} \mbox{Tr} (T^a T^b T^c T^d) 
&= & \delta^{ab} \delta^{cd} \left( \frac{1}{N}  \delta^{ab} \delta^{cd}\right)
=  \frac{1}{N} (N^2-1)^2. 
\label{ididentity}
\eea

The following eigenvalue equations for other state can be obtained:
\bea
&& d^{abcd} (J^a J^b J^c J^d )_0 |(0;f)>
=\frac{1}{N} d^{abcd}  \mbox{Tr} ( T^a T^b T^c T^d) |(0;f)>
\nonu\\
&& =\frac{2(N^2-1)(N^2-4)(N^2-9)}{N(N^2+1)}  |(0;f)>
\rightarrow  2N^3|(0;f)>,
\label{44-1} \\
&& d^{abe}d^{cde} (J^a J^b J^c J^d )_0  |(0;f)>
=\frac{1}{N} d^{abe}d^{cde}  \mbox{Tr} ( T^a T^b T^c T^d)  |(0;f)>
\nonu\\
&& =\frac{1}{N^3}(N^2-4)^2(N^2-1)   |(0;f)>
\rightarrow  N^3|(0;f)>,
\label{44-6} \\
&& \delta^{ab} \delta^{cd} (J^a J^b J^c J^d )_0 |(0;f)>
=\frac{1}{N} \delta^{ab} \delta^{cd}  \mbox{Tr} ( T^a T^b T^c T^d) |(0;f)>
\nonu\\
&&=\frac{1}{N^2} (N^2-1)^2  |(0;f)>
\rightarrow  N^2|(0;f)>.
\label{44-11}
\eea
In addition, the identities in (\ref{fourdidentity}) and (\ref{ididentity}) can be used.

Combine the equations (\ref{4-1})-(\ref{4-23}) with the coefficients in Appendix $A$ 
for the first eigenvalue equation  
 and (\ref{44-1}), (\ref{44-6}) and (\ref{44-11}) with the coefficients
for the second eigenvalue equation.
The resulting eigenvalue equations have very simple 
factorized forms as follows:
\bea
W^{(4)}_0 |(f;0)>&=&\left[ \frac{2(N^2-1)(N^2-4)(N^2-9)}{N(N^2+1)}(c_1-c_2+c_3-c_4+c_5) \right.
\nonu\\
&+&\frac{1}{N^3}(N^2-4)^2(N^2-1) (c_6-c_7+c_8-c_9+c_{10}) 
\nonu\\
&+&\frac{1}{N^2} (N^2-1)^2 (c_{11}-c_{12}+c_{13}-c_{14}+c_{15}) 
\nonu\\
&+& \left. \frac{1}{N}(N^2-1)(2 c_{17}+c_{20}+2 c_{22}) + 
\frac{1}{N^2} (N^2-1)^2 c_{23} \right] |(f;0)>
\nonu\\
&=&
 \left[
\frac{2 (k+1) (N-3) (N^2-1) (k+2 N) (k+2 N+1) (3 k+4 N+3)}{N^2 
(k+N) (2 k+N) \, d(N,k)}\right] 
|(f;0)>,
\nonu\\
W^{(4)}_0 |(0;f)>&=&\left[\frac{2(N^2-1)(N^2-4)(N^2-9)}{N(N^2+1)} c_1
+\frac{1}{N^3}(N^2-4)^2(N^2-1)c_6 \right.
\nonu\\
&+& \left. \frac{1}{N^2} (N^2-1)^2 c_{11}\right]|(0;f)>
\nonu\\
&=&
\left[ \frac{2 k (k+1) (N-3) (N^2-1) (k+2 N) (3 k+2 N)}{
N^2 (k+N+1)  (2 k+3 N+2) \, d(N,k)} \right]
|(0;f)>,
\label{spin4result}
\eea
where
 $d(N,k )  \equiv  
17 k+17 k^2+22 N+39 k N+5 k^2 N+22 N^2+10 k N^2$.

The spin-$3$ current case can be analyzed further as follows:
\bea
W^{(3)}_0 |(f;0)>&=&\left(-A_1+A_2-A_3+A_4 \right)d^{abc}K_0^a K_0^b K_0^c |(f;0)>
\nonu\\
&=&\left(-A_1+A_2-A_3+A_4 \right) \left[ 
\frac{i}{N^2}(N^2-4)(N^2-1) \right] |(f;0)>
\nonu\\
&=&- \left[ \frac{i}{N^2} 
(N^2-4) (N^2-1) (k+N+1) (k+2 N+1) (2 k+3 N+2) \right]
|(f;0)>,
\nonu\\
W^{(3)}_0 |(0;f)>&=&A_1 d^{abc}J_0^a J_0^b J_0^c |(0;f)>
=A_1 \left[ \frac{i}{N^2}(N^2-4)(N^2-1) \right] |(0;f)>
\nonu\\
&=& \left[ \frac{i}{N^2} 
k (N^2-4)(N^2-1) (k+N) (2 k+N) \right] |(0;f)>,
\label{spin3result}
\eea
where the property (\ref{dtrace}) is used and the coefficients 
in (\ref{BA}) are substituted.

Finally, the eigenvalue equations for the spin-$2$ current 
can be summarized as
\bea
T_0 |(f;0)>
&=&\left[ -\frac{k}{2 (N+1) (k+N+1)}J_0^a J_0^a
-\frac{1}{2 (k+N) (k+N+1)} K_0^a K_0^a \right.
\nonu\\
&+&\left. \frac{1}{k+N+1} J_0^a K_0^a \right] |(f;0)>
\nonu\\
&=&\left( -\frac{k+2 N+1}{2 (N+1) (k+N)} \right) 
\left[ -\frac{1}{N}(N^2-1) \right]
|(f;0)>
\nonu\\
&=& \left[\frac{(N-1) (k+2 N+1)}{2 N (k+N)} \right] |(f;0)>,
\nonu\\
T_0 |(0;f)>
&=&-\frac{k}{2 (N+1) (k+N+1)} J_0^a J_0^a |(0;f)>
\nonu\\
&=& -\frac{k}{2 (N+1) (k+N+1)} \left[ -\frac{1}{N}(N^2-1) \right] 
|(0;f)>
\nonu\\
&=& \left[ \frac{k (N-1)}{2 N (k+N+1)} \right] |(0;f)>,
\label{spin2result}
\eea
where the property 
$ \delta^{ab}  \mbox{Tr} (T^a T^b )  = -(N^2-1)$ is used.


\end{document}